\documentclass[preprint,12pt]{aastex}

\def\nodatah{--}
\def\nodatae{ }

\def\ho{76.9}
\def\houp{3.9}
\def\hodown{3.4}
\def\houpsys{10.0}
\def\hodownsys{8.0}

\def\HO{$H_0=\ho\pm^{\houp}_{\hodown}\pm^{\houpsys}_{\hodownsys}$ km s$^{-1}$ Mpc$^{-1}$}
\def\HOstat{$H_0=\ho\pm^{\houp}_{\hodown}$ km s$^{-1}$ Mpc$^{-1}$}
\def\hocut{77.6}
\def\hocutup{4.8}
\def\hocutdown{4.3}

\def\hocutupsys{10.1}
\def\hocutdownsys{8.2}
\def\HOcut{$H_0=\hocut\pm^{\hocutup}_{\hocutdown}\pm^{\hocutupsys}_{\hocutdownsys}$ km s$^{-1}$ Mpc$^{-1}$}

\def\hoall{73.7}
\def\hoallup{4.6}
\def\hoalldown{3.8}

\def\hoallupsys{9.5}
\def\hoalldownsys{7.6}
\def\HOall{$H_0=\hoall\pm^{\hoallup}_{\hoalldown}\pm^{\hoallupsys}_{\hoalldownsys}$ km s$^{-1}$ Mpc$^{-1}$}

\def\chandra{\it Chandra\rm}
\def\ovro{\it OVRO\rm}
\def\bima{\it BIMA\rm}
\def\rxj1347{\it RXJ~1347.5-1145\rm}
\def\nclu{38}
\def\chires{31.6}
\def\chirescut{53.9}
\def\chiresall{53.1}

\def\ndof{37}

\begin{document}
                                                                                
\title{
     Determination of the Cosmic Distance Scale from\\
     Sunyaev-Zel'dovich Effect and {\it Chandra} X-ray\\ 
     Measurements of High Redshift Galaxy Clusters}

\author{Massimiliano~Bonamente$\,^{1,2}$,
Marshall~K.~Joy$\,^{1}$, Samuel~J.~LaRoque$\,^{3}$, John~E.~Carlstrom$\,^{3,4}$,
Erik~D.~Reese$\,^{5}$ and Kyle~S.~Dawson$\,^{6}$
}
                                                                               
\affil{
\(^{\scriptstyle 1} \){NASA Marshall Space Flight Center, Huntsville, AL}\\
\(^{\scriptstyle 2} \){Department of Physics, University of Alabama,
Huntsville, AL}\\
\(^{\scriptstyle 3} \){Department of Astronomy and Astrophysics, University of
Chicago, Chicago, IL\\ and \\
Kavli Institute for Cosmological Physics,  University of Chicago, Chicago, IL}\\
\(^{\scriptstyle 4} \){Department of Physics,
 University of Chicago, Chicago, IL \\and \\
Enrico Fermi Institute, University of Chicago, Chicago, IL}\\
\(^{\scriptstyle 5} \){Physics Department, University of California, Davis, CA}\\
\(^{\scriptstyle 6} \){Physics Department, University of California, Berkeley,CA\\
(now at Lawrence Berkeley National Laboratory)}
}

\begin{abstract}

We determine the distance to \nclu\/ clusters of galaxies in the
redshift range 0.14$\leq$z$\leq$0.89 using X-ray data 
from \chandra\/ and
Sunyaev-Zeldovich Effect data from the {\it Owens Valley Radio
Observatory} and the {\it Berkeley-Illinois-Maryland Association}
interferometric arrays.  The cluster plasma and dark matter
distributions are analyzed using a hydrostatic equilibrium model that
accounts for radial variations in density, temperature and abundance,
and the statistical and systematic errors of this method are
quantified.  The analysis is performed via a Markov chain Monte Carlo
technique that provides simultaneous estimation of all model
parameters.  We measure a Hubble constant of \HO (statistical followed
by systematic uncertainty at 68\% confidence) for an $\Omega_{M}=0.3$,
$\Omega_{\Lambda}$=0.7 cosmology.  We also analyze the data using an
isothermal $\beta$ model that does not invoke the hydrostatic
equilibrium assumption, and find \HOall; to avoid effects from cool
cores in clusters, we repeated this analysis excluding the central 100
kpc from the X-ray data, and find \HOcut\/ (statistical followed by
systematic uncertainty at 68\% confidence).  The consistency between
the models illustrates the relative insensitivity of SZE/X-ray
determinations of $H_0$ to the details of the cluster model.  Our
determination of the Hubble parameter in the distant universe agrees
with the recent measurement from the {\it Hubble Space Telescope} key
project that probes the nearby universe.

\end{abstract}

\section{Introduction}
\label{sec:intro}

Combined analysis of radio and X-ray data provides a method to
determine directly the distances to galaxy clusters.  Galaxy clusters
are the largest gravitationally collapsed structures in the universe,
with a hot diffuse plasma ($T_e \sim 10^7 - 10^8$ K) that fills the
intergalactic space.  Cosmic microwave background (CMB) photons
passing through this hot intracluster medium (ICM) have a $\sim 1$\%
chance of inverse Compton scattering off the energetic electrons,
causing a small ($\sim 1$ mK) distortion of the CMB spectrum, known as
the Sunyaev-Zel'dovich Effect (SZE: Sunyaev \& Zel'dovich 1970, 1972;
for reviews see Birkinshaw 1999; Carlstrom, Holder, \& Reese 2002).
The same hot gas emits X-rays primarily through thermal
bremsstrahlung.  The SZE is a function of the integrated pressure,
$\Delta T \propto \int n_e T_e d\ell$, where $n_e$ and $T_e$ are the
electron number density and temperature of the hot gas, and the
integration is along the line-of-sight.  The X-ray emission scales as
$S_X \propto \int n_e^2 \Lambda_{ee} d\ell$, where $\Lambda_{ee}$ is the
X-ray cooling function.  The different dependences on density, along
with a model of the cluster gas, enable a direct distance
determination to the galaxy cluster.  This method is independent of
the extragalactic distance ladder and provides distances to high
redshift galaxy clusters.  

The $\sim 1$ mK SZE signal proved challenging for initial searches,
but recent improvements in both technology and observational
strategies have made observations of the SZE fairly routine.  High
signal-to-noise ratio (S/N) detections of the SZE have been made with
single dish observations at radio wavelengths (Birkinshaw and Hughes
1994; Herbig et al.\ 1995; Myers et al.\ 1997; Hughes and Birkinshaw
1998; Mason et al.\ 2001), millimeter wavelengths (Holzapfel et al.\
1997a,b; Pointecouteau et al.\ 1999, 2001) and submillimeter
wavelengths (Lamarre et al.\ 1998; Komatsu et al.\ 1999), while
interferometric observations at centimeter wavelengths have produced 
images of the SZE (Jones et al.\ 1993; Grainge et al.\ 1993;
Carlstrom et al.\ 1996, 2000;
Grainge et al.\ 2002; Reese et al.\ 2000, 2002; Grego et al.\ 2000, 2001;
La~Roque et al.\ 2003; Udomprasert et al.\ 2004).

SZE/X-ray distances provide a measure of the Hubble constant that is
independent of the extragalactic distance ladder and probe high
redshifts, well into the Hubble flow.  The SZE/X-ray determinations of
$H_0$ bridge the gap between observations of nearby objects (e.g. the
Hubble Space Telescope Key Project, Freedman et al.\ 2001) and
expansion values inferred from CMB anisotropy (Spergel et
al.\ 2003)  and supernova (Riess et al.\ 2005) measurements.  
Previous SZE/X-ray determinations of the Hubble parameter
have progressed from analysis of individual galaxy clusters, to
samples of a few (Myers et al.\ 1997; Mason et al.\ 2001; Jones et
al.\ 2005), up to a sample of 18 galaxy clusters using ROSAT X-ray data (Reese et al 2002;
for reviews see Reese 2004 and Carlstrom, Holder, \& Reese 2002).  In
most cases, simple isothermal $\beta$ models were adopted for the
cluster gas, since the data did not warrant a more sophisticated
treatment.

We present a Markov chain Monte Carlo joint analysis of
interferometric SZE observations and \chandra\ X-ray imaging
spectroscopy observations of a sample of \nclu\ galaxy clusters with
redshifts $0.14 \leq z \leq 0.89$.  The unprecedented spatial
resolution of \chandra\ combined with its simultaneous spectral
resolution allow more realistic modeling of the intracluster plasma
than previous studies, thus enabling a more accurate determination of
the Hubble constant.

\section{Observations of galaxy clusters \label{sample}}

\subsection{Interferometric Sunyaev-Zel'dovich effect data}
Interferometric radio observations of the \nclu\/ clusters in Table
\ref{tab:data} were performed at the Berkeley-Illinois-Maryland
Association observatory (\bima) and at the Owens Valley Radio
Observatory (\ovro).  The
arrays were equipped with 26-36 GHz receivers to obtain maps of the
Sunyaev-Zel'dovich Effect (SZE) toward the clusters (Carlstrom et al.\
1996, 2000; Reese et al.\ 2000).
These frequencies are on the Rayleigh-Jeans end of the microwave
spectrum, and the scattering with cluster electrons causes an
intensity decrease that, in terms of brightness temperature,
corresponds to a change in $T_{CMB}$ of order $-1$~mK.

Most of the \ovro\/ and \bima\/ telescopes were placed in a compact
configuration to maximize the sensitivity on angular scales
subtended by distant clusters (typically $\sim 1'$) and a few
telescopes were placed at longer baselines for simultaneous point
source imaging (Reese et al.\ 2002).
The SZE data consist of the position in the Fourier domain ($u$-$v$
plane) and the visibilities --- the real and imaginary Fourier
component pairs as functions of $u$ and $v$, which are the Fourier
conjugate variables to right ascension and declination.  The effective
resolution of the interferometer, the synthesized beam, depends on the
$u$-$v$ coverage and is therefore a function of the array
configuration and source position. A typical size for the synthesized
beam of our observations is $\sim 1^\prime$, as shown in Figure
\ref{5clusters}.
The SZE data were reduced using the MIRIAD (Sault et al.\ 1995) and
MMA (Scoville et al.\ 1993) software packages and images were made with
DIFMAP (Pearson et al.\ 1994) software.  Absolute flux calibration
was peformed using Mars observations adopting the brightness
temperature from the Rudy (1987) Mars model. The gain was monitored
with observations of phase calibrators, and remained stable at the 1\%
level over a period of months.  Data were excised when one telescope
was shadowed by another, when cluster observations were not bracketed
by two phase calibrators, when there were anomalous changes in the
instrumental response between calibrator observations, or when there
was spurious correlation.
Positions of point sources were identified using the long baseline
data; their fluxes are included as free parameters in the model, using
the same methodology as Reese et al.\ (2002).
Additional details of the SZE data analysis are provided in Reese et
al.\ (2002) and Grego et al.\ (2000).

\subsection{\chandra\/ \bf X-ray data \label{chandra}}
The \chandra\/ X-ray data for the \nclu\/ clusters in our sample were
obtained primarily through the Guaranteed Time program of Leon van
Speybroeck.  The observations were performed with the ACIS-I and
ACIS-S detectors. The two ACIS instruments provide spatially resolved
X-ray spectroscopy and imaging with an angular resolution of
$\sim0.5^{\prime\prime}$ and with energy resolution of $\sim 100-200$
eV.  Data analysis was performed with the CIAO software (version 3.2)
and the CALDB calibration information (version 3.1) provided by the
\chandra\/ calibration team (Chandra Interactive Analysis of
Observations, http://cxc.harvard.edu/ciao/).

The first step in the data analysis was to process the Level 1 data to
correct for the charge transfer inefficiency of the ACIS detectors.
We then generated a Level 2 event file applying standard filtering
techniques: we selected grade=0,2,3,4,6, status=0 events (as defined
in the Chandra Proposers Observatory Guide) and filtered the event
file for periods of poor aspect solution using the good time interval
(GTI) data provided with the observations.  Periods of high background
count rates were occasionally present, typically due to Solar flares
(Markevitch 2001).  We discarded these periods using an iterative
procedure in which we constructed a light-curve of a background region
in 500 second bins, and time intervals that were in excess of the
median count rate by more than 3$\sigma$ were discarded from the
dataset.  The Chandra instruments are affected by the buildup of a
contaminant on the optical blocking filter located along the optical
path to the ACIS detector; we accounted for this efficiency reduction
using CIAO and CALDB.  Spectra were accumulated in concentric annuli
centered at the peak of the X-ray emission, each containing approximately
the same number of source photons after removal of point sources.  
Both images and spectra were limited to 0.7-7 keV in order to exclude the
low-energy and high-energy data that are more strongly affected by
background and by calibration uncertainties.  
The X-ray images were binned in 1.97$^{\prime \prime}$ pixels;
this sets the limiting angular resolution of our X-ray data, as the
\chandra/\ point response function in the center of the X-ray image
is smaller than our adopted pixel size.
The X-ray background was measured for each cluster exposure, 
using peripheral regions of the
detector (ACIS-S) or adjacent detector chips (ACIS-I) that are source
free.  Additional details of the \chandra\/ X-ray data analysis are
provided in Section 2 of Bonamente et al. (2004).

Images of the X-ray surface brightness of selected clusters are shown
in Figure \ref{5clusters}, with SZE contours overlaid, and in Appendix
1 for all \nclu\/ clusters.
\chandra\/ also provides spatially-resolved spectroscopy that allows a
determination of the temperature and metal abundance of the hot
plasma. The spectral properties of the plasma are obtained by fits to
an optically thin emission model, with absorbing column $N_H$ fixed at
the Galactic value. The uncertainty in $N_H$ of $\sim10^{19}$
cm$^{-2}$ (Dickey and Lockman 1990) results in uncertainties in the
measured temperatures of less than 1\%, and therefore has a negligible
effect in the measurement of the cluster distances.
In Figure \ref{5clusters} we show the radial profiles of the X-ray
surface brightness and of the plasma temperature for several
representative clusters, along with their best fit curves as
determined from the modeling described in the next Section; radial
brightness and temperature profiles for the full cluster sample are
presented in Appendix 2 and 3.

In Figure \ref{tprofiles} we show the composite radial temperature
profiles for the \nclu\/ clusters.
The clusters containing plasma with a central cooling time $t_{cool}
\leq 0.5t_{Hubble}$, which we refer to as the``cool core" sample, are
shown on top, while clusters with longer cooling times are shown on
the bottom~\footnote{The clusters with $t_{cool} \leq 0.5t_{Hubble}$
are Abell~586, MACS~J0744.8+3927, ZW~3146, Abell~1413,
MACS~J1311.0-0310, Abell~1689, RX~J1347.5-1145, MS~1358.4+6245,
Abell~1835, MACS~J1423.8+2404, RX~J2129.7+0005, Abell~2163, Abell~2204
and Abell~2261. The Hubble time is approximately $t_{Hubble} \simeq
H_0^{-1}$ (Carroll, Press and Turner 1992) with $H_0$=72 km s$^{-1}$
Mpc$^{-1}$ and the cooling time ($t_{cool}\simeq 3 k_B T / 2
\Lambda_{ee} n_e$) is calculated using the central density and the
temperature from an isothermal $\beta$ model fit.}.  The temperature
profiles for the \nclu\ clusters lie within the envelope of the 21
clusters observed by BeppoSAX (see Fig. 3, Fig. 4 and Table 2 of De
Grandi and Molendi 2002).
The spectra were extracted in concentric annuli that contain a similar
number of photons in each annulus for each cluster. The radial
temperature profile data for all clusters, including the temperature,
the background-subtracted counts and the $\chi^2$, are reported in
Appendix 3 (Table \ref{table_tprof}).  Metal abundances of the hot cluster
plasma have a marginal effect on the X-ray cooling function (see
Section \ref{method}).  We assume the De Grandi et al.\ (2004)
abundance profile in our analysis, which is consistent with our
measured abundances.
\clearpage
\begin{figure}
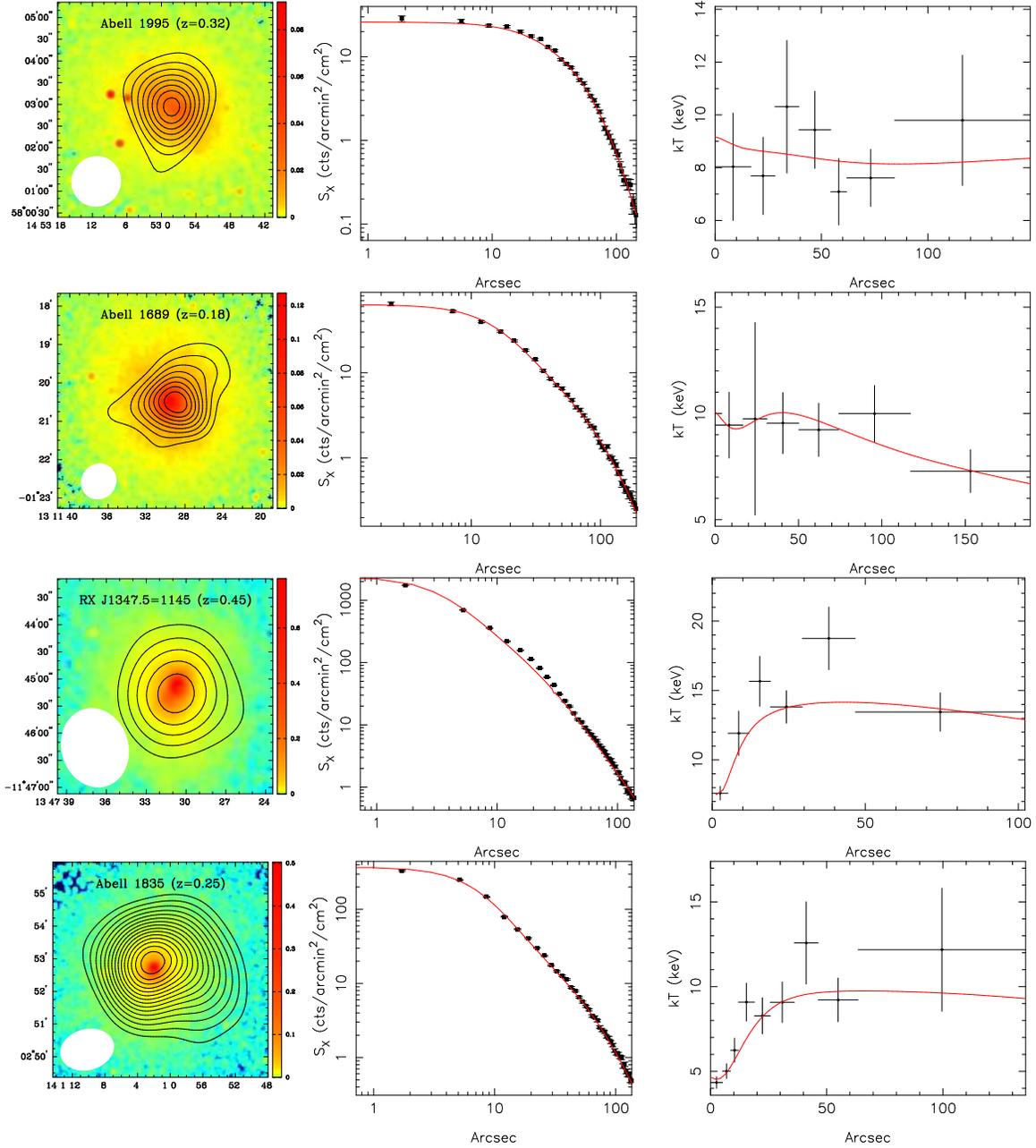

\begin{center}
\includegraphics[angle=-90,width=1.75in]{f1a.eps}
\includegraphics[angle=-90,width=4.2in]{f1b.eps}
\includegraphics[angle=-90,width=1.75in]{f1c.eps}
\includegraphics[angle=-90,width=4.2in]{f1d.eps}
\includegraphics[angle=-90,width=1.75in]{f1e.eps}
\includegraphics[angle=-90,width=4.2in]{f1f.eps}
\includegraphics[angle=-90,width=1.75in]{f1g.eps}
\includegraphics[angle=-90,width=4.2in]{f1h.eps}
\vspace{-5mm}
\end{center}
\caption{\small (Left) \chandra\/ images of the X-ray surface
brightness in 0.7-7 keV band in units of counts pixel$^{-1}$
(1.97$^{\prime\prime}$ pixels) for selected clusters.
 Overlaid are the SZE decrement contours, with contour levels
(+1,-1,-2,-3,-4,...) times the rms noise in each image; the
full-width-at-half-maximum of the SZE synthesized beam (effective
point-spread function) is shown in the lower left
corner.  The X-ray images were smoothed with a $\sigma=2^{\prime\prime}$
Gaussian kernel.  (Center) Radial profile of the background subtracted
X-ray surface brightness; the solid line is the best-fit model
obtained with the parameters of Table \ref{tab:results}.  (Right)
Radial profiles of the \chandra\/ temperatures, the solid line is the
best-fit hydrostatic equilibrium model with the parameters of Table
\ref{tab:results}.
\label{5clusters} 
}
\vspace{-0.5cm}
\end{figure}

\begin{figure}
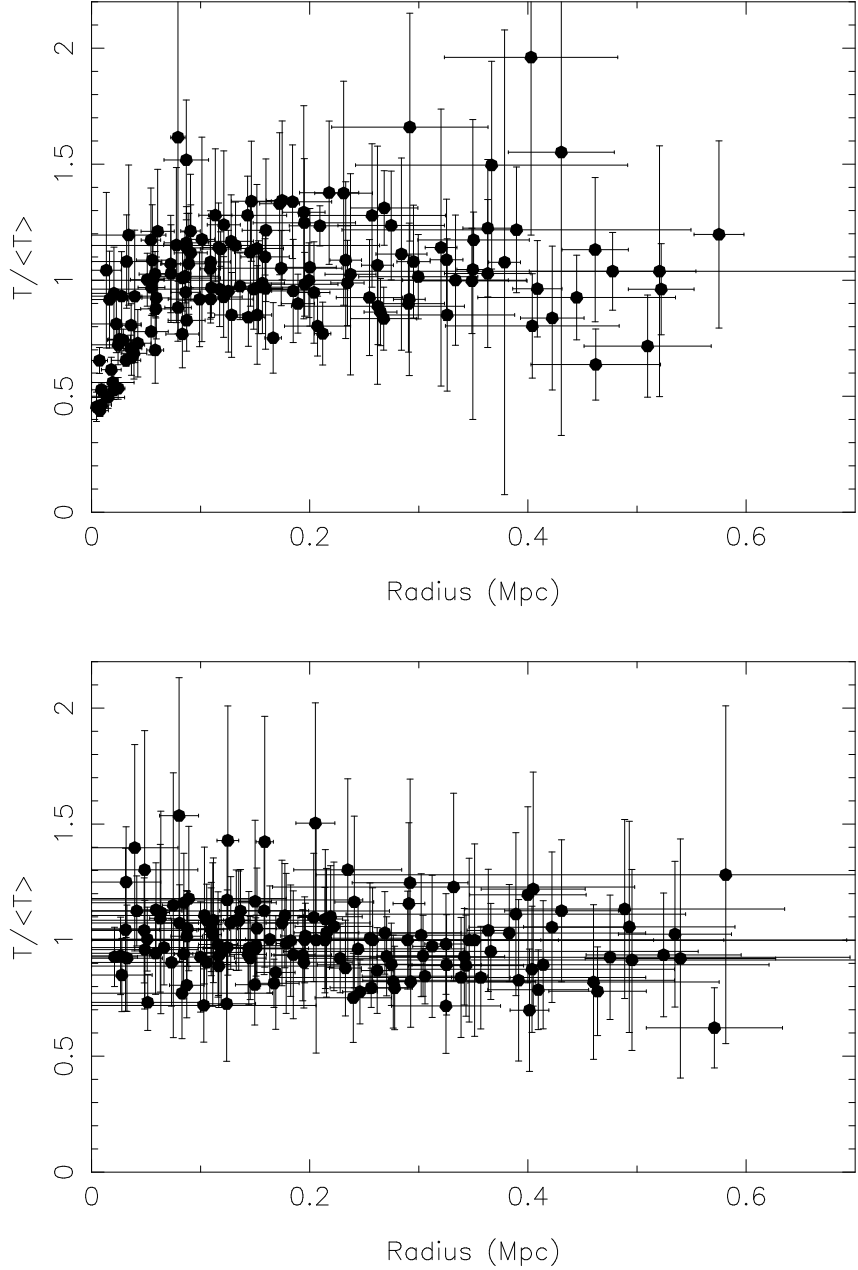

\begin{center}
\includegraphics[angle=-90,width=4.75in]{f2a.eps}
\includegraphics[angle=-90,width=4.75in]{f2b.eps}
\end{center}
\caption{\small Temperature profiles for clusters with central cooling time 
$t_{cool} \leq 0.5\times t_{Hubble}$  (top) and
for clusters with $t_{cool} > 0.5\times t_{Hubble}$ (bottom) \label{tprofiles}.
We assume the best-fit Hubble constant of Section \ref{results} and $\Omega_{M}=0.3$, 
$\Omega_{\Lambda}=0.7$. T/$<$T$>$ is plotted as function of projected radius.
}
\vspace{-0.5cm}
\end{figure}
\clearpage
\section{Measuring distances with X-ray and Sunyaev-Zel'dovich Effect data \label{method}}
\subsection{The hydrostatic equilibrium model \label{hse-method}}
To determine the distance to a cluster, we must first construct a
realistic model for the cluster gas distribution.  
At the center of clusters the density may be high enough that the
radiative cooling time-scale is less than the cluster's age, leading
to a reduction in temperature and an increase in central density.
This increases the central X-ray emissivity in the \chandra\/
passband, as shown in Figure \ref{5clusters} for the clusters
\rxj1347 and Abell~1835.  At large radii, the density of the gas is
sufficiently low that X-ray emission can be sustained for cosmological
periods without significant cooling.  Cool core clusters effectively
exhibit two components: a centrally concentrated gas peak and a broad,
shallower distribution of the gas.  This phenomenon motivates the
modelling of the gas density with a function of the form:
\begin{equation}
n_e(r) = n_{e0} \cdot
\left[f\left( 1+\frac{r^2}{r_{c1}^2} \right)^{-\frac{3\beta}{2}}+
(1-f)\left( 1+\frac{r^2}{r_{c2}^2} \right)^{-\frac{3\beta}{2}} \right]
\label{density}
\end{equation}
This shape generalizes the single $\beta$-model profile, introduced by
Cavaliere and Fusco-Femiano (1976) and commonly used to fit X-ray
surface brightness profiles, to a double $\beta$-model of the density
that has the freedom of following both the central spike in density
and the more gentle outer distribution.
A double $\beta$-model of the surface brightness was first used by
Mohr et al.\ (1999) to fit X-ray data of galaxy clusters; the density
model of Equation \ref{density} was further developed by La~Roque
(2005).  The quantity $n_{e0}$ is the central density, 
$f$ governs the fractional contributions of 
the narrow and broad components ($0\leq f \leq 1$),
$r_{c1}$ and $r_{c2}$ are the two core radii that describe the shape
of the inner and outer portions of the density distribution and
$\beta$ determines the slope at large radii (the same $\beta$ is used
for both the central and outer distribution in order to reduce the
total number of degrees of freedom).

The X-ray surface brightness is related to the gas density as
\begin{equation}
S_X=\frac{1}{4 \pi (1+z)^4} \int n_e^2 \Lambda_{ee} dl
\label{XSB}
\end{equation}
where $z$ is the cluster redshift, $n_e$ is the electron density of
the plasma (Equation \ref{density}), $\Lambda_{ee}$ is the X-ray
cooling function, and the integration is performed along the line of
sight $l$.  We calculate $\Lambda_{ee}$ as a function of plasma
temperature and energy in the rest frame of the cluster, including
contributions from relativistic electron-ion thermal bremsstrahlung,
electron-electron thermal bremsstrahlung, recombination, and two
photon processes; the cooling function is then redshifted to the
detector frame, convolved with the telescope and detector response,
and integrated over the 0.7-7 keV Chandra bandpass, following the 
method described in 
Reese et al.\ (2000).
The calculation of $\Lambda_{ee}$ requires a temperature profile in
order to perform the integration, which we determine from our
Chandra data (Figure \ref{tprofiles}).
The appropriate response for each image was generated from CIAO.

The SZE decrement is proportional to the integrated gas pressure as
\begin{equation}
\Delta T_{CMB}=f_{(x,T_e)} T_{CMB} \int \sigma_T n_e \frac{k_B T_e}{m_e c^2} dl
\label{SZE}
\end{equation}
where $f_{(x,T_e)}$ is the frequency dependence of the SZE, $x=h \nu/
k_B T_{CMB}$ and $f_{(x,T_e)}\simeq-2$ at our observing frequency of
30 GHz, $T_{CMB}=2.728$ K (Fixsen et al.\ 1996), $\sigma_T$ is the
Thomson cross section, $k_B$ is the Boltzmann constant, $c$ is the
speed of light in vacuum, $m_e$ is the electron mass, $T_e$ the
electron temperature, and the integration is along the line of sight.

Historically, the cluster distance has been solved for directly by
taking advantage of the different density dependences of the X-ray
emission and SZE decrement (e.g., Hughes, Birkinshaw and Arnaud 1991;
Reese et al.\ 2002; Bonamente et al.\ 2004):
\begin{eqnarray}
\label{eq-system}
S_X \propto \int n_e^2 \Lambda_{ee} dl=\int n_e^2 \Lambda_{ee} D_A d\theta\\\nonumber
\Delta T_{CMB} \propto \int n_e T_e dl=\int n_e T_e D_A d\theta 
\end{eqnarray}
The details of the plasma modelling, such as the numerical integration
of the density profile, are included in the proportionality constants
of Equations \ref{eq-system}.
The cluster angular diameter distance $D_A\equiv dl/d\theta$, where
$\theta$ is the line-of-sight angular size, can be inferred with a
joint analysis of SZE and X-ray data by assuming a cluster geometry to
relate the measured angular size in the plane of the sky to that along
the line of sight. For our adopted spherical geometry, these two sizes
are equal.

Our model includes the distribution of dark matter in clusters.
The baryonic matter reaches hydrostatic equilibrium in the potential well defined
by the baryonic and dark matter components, on a time scale that is shorter than the
cluster's age (Sarazin 1988).  Under spherical symmetry, this results
in the condition
\begin{equation} 
\frac{dP}{dr}=-\rho_g \frac{d\phi}{dr}
\end{equation}
where $P$ is the gas pressure, $\rho_g$ is the gas density and
$\phi=-G M(r)/r$ is the gravitational potential due to both dark
matter and the plasma.  Using the ideal gas equation of state for the
diffuse cluster plasma, $P=\rho_g k_B T / \mu m_p$ where $\mu$ is the
mean molecular weight and $m_p$ is the proton mass, one obtains a
relationship between the cluster temperature and the cluster mass
distribution:
\begin{equation}
\frac{dT}{dr}=-\left[ \frac{\mu m_p}{k_B} \frac{d\phi}{dr} + \frac{T}{\rho_g} \frac{d\rho_g}{dr} \right]=
-\left[ \frac{\mu m_p}{k_B} \frac{G M}{r^2} + \frac{T}{\rho_g} \frac{d\rho_g}{dr} \right]
\label{eq-hse}
\end{equation}

We combine these hydrostatic equilibrium equations with a dark matter
density distribution from Navarro, Frenk and White (1997):
\begin{equation}
\rho_{DM}(r)=\mathcal{N} \left[ \frac{1}{(r/r_s) (1+r/r_s)^2} \right]
\label{NFW}
\end{equation}
where $\mathcal{N}$ is a density normalization constant and $r_s$ is a
scale radius. These model equations are combined with the X-ray and SZE
data using a Markov chain Monte Carlo method, described in the
following Section.

In Figure \ref{5clusters} the best-fit line (in red) of the X-ray
surface brightness is obtained using
the density distribution of Equation \ref{density} and the hydrostatic
equilibrium model described in this Section.
For clusters in which a single $\beta$ model is an acceptable fit to
the X-ray surface brightness ($\chi^2_r <$1.5; see Appendix 2), we
simplify the density model of Equation \ref{density} by fixing the
parameter $f=0$.  We fit spherically symmetric models to all of the
clusters, including those which do not appear circular in X-ray and
radio observations, as this approach gives an unbiased estimator of
cluster distances when a large sample of clusters is used (Sulkanen
1999).

\subsection{Parameter estimation using the Markov chain Monte Carlo method \label{mcmc}}
Our model consists of five parameters that describe the gas density
($n_{e0}$, $f$, $r_{c1}$, $r_{c2}$ and $\beta$; Equation
\ref{density}), two parameters that describe the dark matter density
($\mathcal{N}$ and $r_s$ from Equation \ref{NFW}) and the angular diameter
distance $D_A$.
Additional parameters such as the cluster position, point source
positions, and point source fluxes are also included.  A detailed
discussion of radio point sources is provided in Reese et al.\ (2002).
By linking the central densities between the X-ray and SZE datasets,
and allowing $D_A$ to vary, the model can be integrated along the line
of sight and compared with the X-ray and SZE data simultaneously,
according to Equations \ref{XSB} and \ref{SZE}.  The model parameters
can also be used with Equation \ref{eq-hse} to solve for the cluster
temperature profile, which is integrated along the line of sight and
compared with the spectral data as described below.
The Markov chain Monte Carlo (MCMC) method used to estimate the model
parameters is described in Bonamente et al.\ (2004). In this Section
we provide a brief overview of the method, focusing on the changes we
applied to accommodate the new hydrostatic equilibrium model of Section
\ref{hse-method}.

The first step of the Markov chain Monte Carlo method is the
calculation of the joint likelihood $\mathcal{L}$ of the X-ray and SZE
data with the model.  This calculation follows three independent
steps, one for each of the datasets involved: SZE data, X-ray images,
and X-ray spectra.  The likelihood calculation for the SZE data is
performed directly in the Fourier plane, where the data are taken and
where we understand the noise properties of the data. The likelihood
is given by
\begin{equation}
ln(\mathcal{L}_{SZE})=\displaystyle \sum_i (-\frac{1}{2} \left(\Delta R_i^2+\Delta I_i^2)\right) W_i
\end{equation}
where $\Delta R_i$ and $\Delta I_i$ are the difference between model
and data for the real and imaginary components at each point $i$ in
the Fourier plane, and $W_i$ is a measure of the Gaussian noise ($1/\sigma^2$).

Since the X-ray counts are distributed according to Poisson
statistics, the likelihood is given by
\begin{equation}
ln(\mathcal{L}_{image})=\displaystyle \sum_i [ D_i ln(M_i) - M_i -\ln(D_i!) ]
\end{equation}
where $M_i$ is model prediction (including cluster and background
components), and $D_i$ is the number of counts detected in pixel
$i$. For details on the X-ray/SZE joint analysis, see Reese et al.\
(2000).

The spectral likelihood is calculated by comparing the predicted
temperature profile with the observed one:
\begin{equation}
ln(\mathcal{L}_{spectra})=-\frac{1}{2} \chi^2 -\frac{1}{2}
\displaystyle\sum_i ln({2 \pi \sigma_i^2})
\label{lik-spectra}
\end{equation}
where $i$ labels the bins in the temperature profile (Figure
\ref{tprofiles}), $\chi^2\equiv \displaystyle \sum_i
([T_i-M_i]/\sigma_i)^2 $, $T_i$ and $M_i$ are the measured and
model-predicted temperatures, and $\sigma_i$ is the measured temperature
uncertainty.  The last term on the right in Equation \ref{lik-spectra}
depends only on the data and will cancel when performing the
likelihood ratio test.  Likelihood evaluation for
the spectral data requires another numerical integration to
solve for $T_e(r)$, according to Equation \ref{eq-hse}.  This
temperature profile is weighted by the square of the density and the
cooling function (Equation \ref{XSB}) and then integrated along the
line of sight to determine the emission-weighted temperature
profiles, which can be directly compared with the measured temperature
profile.  The joint likelihood of the spatial and spectral models is
given by

$\mathcal{L}=\mathcal{L}_{SZE}\cdot\mathcal{L}_{image}\cdot\mathcal{L}_{spectra}$.
                                                                                                                                                 
A Markov chain is a sequence of model parameters constructed with the
property that the model parameters appear in the chain with a
frequency that is proportional to their {\it posterior} probability,
i.e., the probability of occurrence in the light of the current
observations.  We start by assuming vague {\it prior} probability
distributions for all parameters as top-hat functions between two
extreme values.  The first link of the MCMC is chosen as the midpoint
of the prior distributions.  We then select a candidate for the next
link in the chain using a {\it proposal} distribution, in our case a
simple top-hat function of constant width around the previous
parameter values.  These candidate parameter values are accepted into
the chain or rejected according to the Metropolis-Hastings criterion
(Metropolis 1953, Hastings 1970) that takes into account the
likelihood information.  This process is iterated for a large number
of steps, which we chose as 100,000. This number ensures that the MCMC
has reached convergence towards the posterior probability distribution
functions of the parameters (Bonamente et al. 2004). Convergence is
tested using the Raftery-Lewis test (Raftery and Lewis 1992; Gilks et
al.\ 1996), the Gelman-Rubin test (Gelman and Rubin 1992) and the
Geweke test (Geweke 1992).  Confidence intervals for the model
parameters are obtained by computing the cumulative distribution of
the occurrence for each model parameter.  We consider the median of
the distribution as the best-fit value and calculate 68\% confidence
intervals around the median.  

The results of the MCMC analysis are
shown in Table \ref{tab:results}, in which we report the best-fit
values of $\mathcal{N}$, $r_s$, $n_{e0}$, $r_{c1}$, $\beta$, $f$, $r_{c2}$ and
$D_A$ for each cluster.

\subsection{Uncertainty analysis \label{error}}

The uncertainties in Table \ref{tab:results} represent the
photon-counting statistical uncertainties of the X-ray images and
spectra, and the statistical uncertainty of the SZE observations, as
described in Section \ref{mcmc}.  Other sources of statistical and
systematic uncertainty that affect our measurements are discussed in
this Section and listed in Table \ref{table-sys}.  For comparison with
the uncertainties encountered in previous studies, see Reese et al.\
(2002).  We note that the Chandra and OVRO/BIMA sample allows us to
obtain a distance scale measurement averaged over a large number of
clusters; this ensemble average significantly reduces the impact of
the single-cluster statistical uncertainies shown in Tables
\ref{tab:results} and \ref{table-sys}.

In the uncertainty analysis we make use of the 
following relationship that follows from Equation \ref{eq-system},
as shown in Bonamente et al. (2004): 
\begin{equation}
D_A \propto \frac{\Delta T_{CMB}^2 \Lambda_{ee}}{S_X T_e^2}
\label{eq-DA}
\end{equation}
Note that $D_A$ is
proportional to $\Delta T_{CMB}^2$ and $T_e^{3/2}$ (since
$\Lambda_{ee} \propto T_e^{1/2}$), so the
distance determination is strongly dependent on the accuracy of the
SZE decrement and X-ray temperature measurements.

\subsubsection{Uncertainty in Galactic $N_H$ \label{nh}}
In the spectral fits of Section \ref{chandra} we used the HI column
densities of Dickey and Lockman (1990), which have an uncertainty of
$\sigma_{N_H}$= $1\times10^{19}$ cm$^{-2}$. A variation of the HI
column density will primarily affect the best-fit X-ray
temperature. The temperature, in turn, affects the measurement of
cluster distances through Equation \ref{eq-DA}, $D_A \propto
\Lambda_{e} T_e^{-2}$, in which the X-ray cooling function is
$\Lambda_{ee} \propto T^{0.5}$.  We obtained spectral fits of our
clusters using $N_H+\sigma_{N_H}$ and $N_H-\sigma_{N_H}$ as the HI
column densities, and found that the best-fit temperatures change by
less than 0.5\%. The uncertainty in Galactic $N_H$ therefore results
in a $D_A$ uncertainty of $\leq1$\% ($D_A \propto T_e^{-1.5}$).

\subsubsection{Cluster asphericity \label{asphericity}}
Most clusters do not appear circular in X-ray or radio observations
(e.g., Mohr et al.\ 1995).  Numerical simulations by Sulkanen (1999)
show that a spherical model fit to triaxial X-ray and SZE clusters
yields an unbiased estimate of cluster distance when a large ensemble
of clusters is used; the standard deviation of the measured distance
for one cluster is $\sim15$\%.

\subsubsection{Small scale clumps in the intracluster gas}
Clumping of the intracluster medium on scales smaller than the Chandra
resolution is a potential source of systematic error. The presence of
clumps enhances the measured X-ray emission ($S_X$) by the factor:
\begin{equation}
C\equiv\frac{<n_e^2>}{<n_e>^2} \geq 1.
\end{equation}
A factor C$>$1 results in a measured angular diameter distance
($D_A\propto S_X^{-1}$) that is lower than what one measures if no
clumping is present, and such $D_A$ should be increased by a factor of
$C$ if clumping occurs. In this case, the resulting best-fit $H_0$
would decrease.

Concurrent studies by LaRoque et al.\ (2006) suggest that as long
as the clumps of X-ray emission observed by \chandra\/ are excised from the 
data, the clusters in this sample are not affected by additional clumping of the hot gas.
We therefore do not include this source of uncertainty in our
error analysis.  There is indication that clumpiness of the gas may be
a factor for high-redshift clusters (Jeltema et al.\ 2005), and this
effect may in principle lead to increased scatter at large z in
Figure \ref{distances}.

\subsubsection{Point sources in the field \label{psources}}
Undetected radio point sources near the cluster center mask the central
decrement.  According to Equation \ref{eq-DA}, $D_A \propto \Delta
T_{CMB}^2$, and an underestimate of the SZE decrement will result in
an underestimate of the cluster distance.  The synthesized beam of the
SZE instrument also has negative sidelobes, and therefore
overestimates of the decrement are also possible. A detailed treatment
of the effect of point sources by La~Roque et al.\ (2006) using this
cluster sample result in a $\sim 8$\% uncertainty in the determination
of $D_A$.

For the X-ray data, the superior angular resolution of Chandra allows
one to locate the point sources and mask them from the analysis, so no
uncertainty from undetected X-ray point sources is introduced.

\subsubsection{Kinetic SZE effect and CMB anisotropies}
Peculiar velocities of clusters introduce a distortion in the CMB
spectrum, known as the kinetic SZE. For a typical line-of-sight
peculiar velocity of 300 km s$^{-1}$ (Watkins 1997, Colberg et al.\
2000) and a cluster of $T_e=8$ keV, the kinetic SZE is 4\% of the
thermal SZE. Since $D_A\propto \Delta T_{CMB}^2$, the kinetic SZE
effect introduces an uncertainty of 8\% to the determination of
cluster distances.
                                                                                                                                    
Limits on CMB anisotropies have been measured by Dawson et al.\
(2001, 2006) and Holzapfel et al.\ (2000) with BIMA at the frequency and
angular scales of
the observations presented in this paper. The 68\% confidence upper
limit is $\Delta T_{CMB}<19$ $\mu$K at $l\sim5500$ ($\sim$ 2
$^\prime$scales). This results in a 68\% uncertainty of $\leq$1\% in
the measurement of $\Delta T_{CMB}$, and $\leq$2\% in the measurement
of $D_A$.
Both of these effects are expected to average out for a sample of clusters
widely separated on the sky.

\subsubsection{Radio halos  and relics}
Extended steep-spectrum nonthermal radio halo sources have been
detected in the center of several clusters (Giovannini and Feretti
2000; Giovannini et al.\ 1999; Hanish 1982), and similar extended
sources (radio relics) have also been found in other clusters at large
radii from the cluster core (Giovannini and Feretti 2004; Feretti
2004).  The on-center radio halo sources can mask the SZE decrement,
resulting in an underestimate of $D_A$.  Reese et al.\ (2002)
determined that the average effect of a radio halo at the BIMA and
OVRO frequency is small: $\sim$1.5\% of the thermal decrement.  We
include a one-sided +3\% systematic uncertainty in the $D_A$
measurement to account for the possible presence of extended
non-thermal radio halo emission in the cluster core.  The frequency of
occurrence and the flux of radio relic and radio halo sources are
similar (Feretti 2004), but radio relics are attenuated by the
interferometer due to their large displacement from the cluster
center; we estimate that the systematic uncertainty due to radio relic
sources is negligible ($<$1\%).

\subsubsection{X-ray background \label{xray-background}}
The X-ray background is measured following the method described in
Section \ref{chandra} and in Bonamente et al.\ (2004).  For each
cluster, we measure the 0.7-7 keV counts in a background region, with
a statistical error equal to the square root of the number of counts,
and extract a background spectrum from the same region.  We assessed
the effect of the X-ray background subtraction on the distance
measurements by performing the following analysis.  For the spatial
analysis, we used the upper and lower limits of the measured
background level; in the spectral analysis, we used the temperatures
obtained by subtracting the background spectrum, rescaled by an amount
equal to $\pm$ the fractional error of the measured background level.
These additional MCMC runs resulted in $D_A$ measurements that are
within $\sim$2\% of those obtained with the nominal background. We
therefore add a $\pm2\%$ uncertainty in the measurement of $D_A$.

\subsubsection{X-ray calibration \label{xray-cal}}
The absolute calibration of the Chandra ACIS effective area is known
to $\sim$ 5\% in the 0.7-7 keV band of interest
(http:/asc.harvard.edu/cal).  This uncertainty affects the $D_A$
measurements directly through the surface brightness terms in Equation
\ref{eq-DA}, resulting in a 5\% systematic uncertainty on $S_X$.

Temperature measurements with Chandra may be subject to systematic
offsets caused by effective area and energy calibration errors, which
we estimate at 5\% (http:/asc.harvard.edu/cal).  According to Equation
\ref{eq-DA}, $D_A \propto \frac{\Lambda_{ee}}{S_X T_e^2}$, where
$\Lambda_{ee}\propto T_e^{0.5}$, therefore the effect of the
temperature measurement uncertainty results in a $\sim$7.5\%
uncertainty on the distance for one cluster.

\subsubsection{SZE calibration}
The absolute calibration of the interferometric observations is known
to about 4\%, resulting in an uncertainty of 8\% in the distance
measurement of one cluster.  Reese et al.\ (2002) also studied the
effect of imprecisions in the measurement of the BIMA and OVRO primary
beams, and conclude that the effect on distance measurements is
negligible.

\section{Measurement of the Hubble constant}
We now use the \nclu\/ cluster distances to estimate the Hubble
constant.  The angular diameter distance $D_A$ is a function of the
cluster redshift $z$, the mass density $\Omega_{M}$, the dark energy
density $\Omega_{\Lambda}$, and the Hubble constant $H_0$, which is
the overall normalization:
\begin{equation}
D_A(z)=\frac{1}{H_0} \cdot \frac{c}{|\Omega_{k}|^{1/2}(1+z)}\cdot sinn\left[ |\Omega_{k}|^{1/2}
\int_{0}^{z} [(1+\zeta)^2(1+\Omega_{M} \zeta) -\zeta(2+\zeta)\Omega_{\Lambda}]^{-1/2} d\zeta\right]
\end{equation}
where the function $sinn(x)$ is defined as $sinh(x)$ for
$\Omega_{k}>0$, $sinn(x)=x$ for $\Omega_{k}=0$, $sinn(x)=sin(x)$ for
$\Omega_{k}<0$, and $\Omega_{k}=1-\Omega_{M}-\Omega_{\Lambda}$
(Carroll, Press and Turner 1992)\footnote{Throughout this paper,
$\Omega_{M}$, $\Omega_{\Lambda}$, and $\Omega_{k}$ are defined at the
present epoch (cf. Carroll, Press, and Turner 1992).}.  Observations
of the CMB anisotropy (Spergel et al. 2003), high-redshift supernovae
(Riess et al.\ 2004; Knop et al.\ 2003; Tonry et al.\ 2003) and mass
measurements of galaxy clusters (e.g.,Grego et al.\ 2001; Vikhlinin et
al.\ 2003; Allen et al.\ 2004) indicate a flat, dark energy-dominated
universe with $\Omega_{M}\simeq0.3$ and $\Omega_{\Lambda}\simeq0.7$,
and these values are adopted in all subsequent analyses unless
otherwise specified.

\subsection{Measurement of the Hubble constant using the hydrostatic equilibrium model\label{results}}
We fit the theoretical $D_A(z)$ function to our sample of \nclu\/
cluster distances obtained with the hydrostatic equilibrium model.
For the fit, we combine the statistical errors given in Table
\ref{table-sys} with the data modelling $D_A$ errors in Table
\ref{tab:results}, and obtain \HOstat (68\% confidence interval,
statistical uncertainty only).  The fit uses the MCMC parameter
estimation method described in section \ref{mcmc}, with the likelihood
calculated using Equation \ref{lik-spectra}.  The $\chi^2$ statistic
of the best-fit model is \chires\/ for \ndof\/ degrees of freedom.

The total systematic errors in $D_A$ are calculated by combining the
individual systematic uncertainties of Table \ref{table-sys} in
quadrature, applying the resulting errors to all \nclu\/ cluster
distances, and repeating the fit.  We obtain a systematic uncertainty
in $H_0$ of (+\houpsys,-\hodownsys) km s$^{-1}$ Mpc $^{-1}$.

\begin{figure}
\begin{center}
\includegraphics[angle=-90,width=7in]{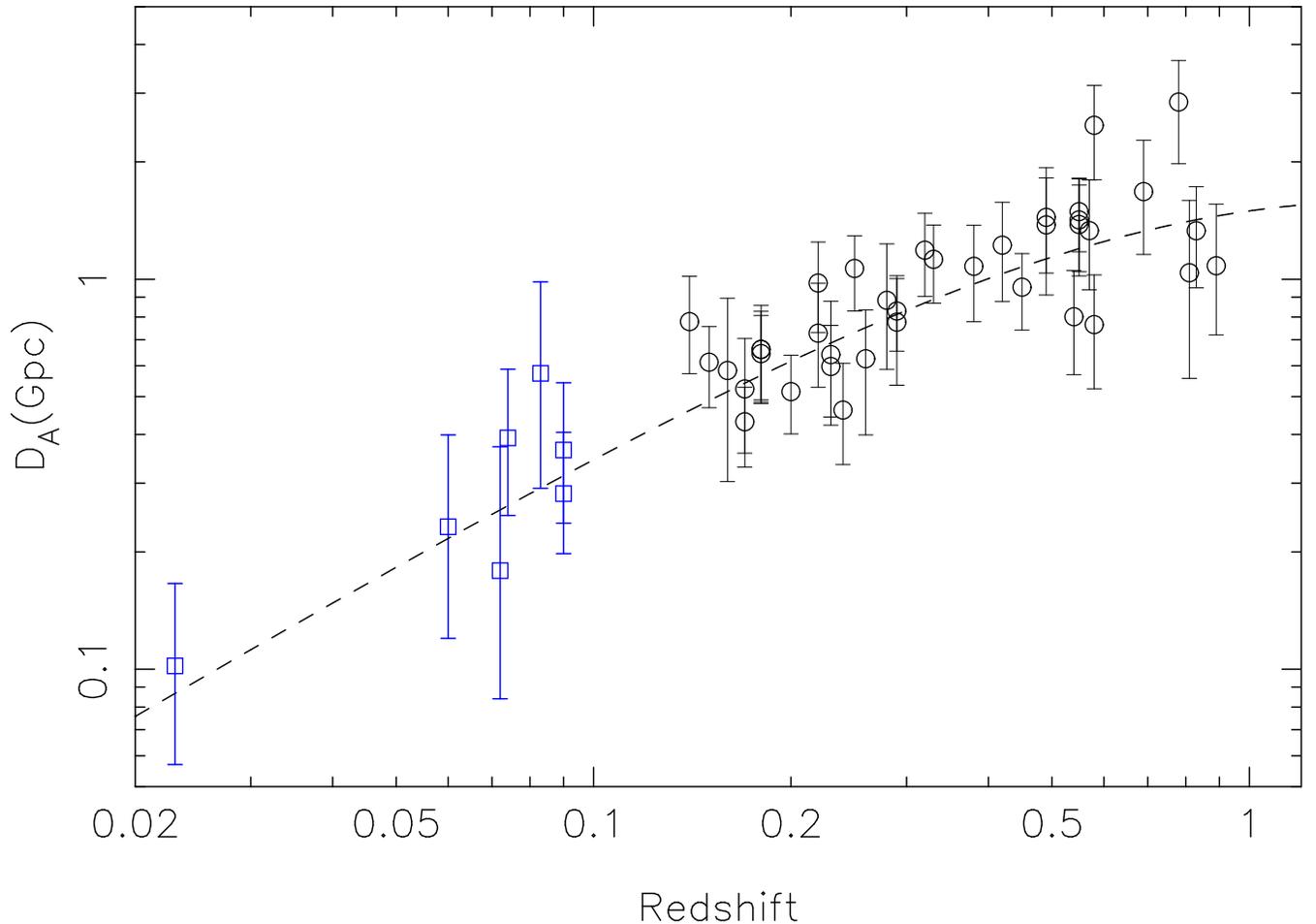}
\vspace{-5mm}
\end{center}
\caption{\small Angular diameter distances of the \nclu\/ clusters
(open circles).  The error bars are the total statistical
uncertainties, obtained by combining the X-ray and SZE data modelling
uncertainties (Table \ref{tab:results}) and the additional sources of
random error described in Section \ref{error} and Table
\ref{table-sys}. The systematic errors of Table \ref{table-sys} are
not shown. Dashed line is the angular diameter curve using the
best-fit Hubble constant $H_0$=\ho\/ km s$^{-1}$ Mpc$^{-1}$ and
$\Omega_{M}=0.3$, $\Omega_{\Lambda}=0.7$.  The open squares are from
the low redshift sample of Mason et al.\ (2001), and they are not
included in the fit.
\label{distances}}
\vspace{-0.5cm}
\end{figure}

Figure \ref{distances} shows the Chandra/SZE cluster distance
measurements, and the theoretical curve for the
best-fit Hubble constant $H_0$=\ho\/  km s$^{-1}$ Mpc$^{-1}$ and $\Omega_{M}=0.3$, $\Omega_{\Lambda}=0.7$.
\clearpage
\clearpage
We also show the angular diameter distances of nearby clusters from
Mason et al.\ (2001), to demonstrate that the best-fit curve is in
agreement with low-redshift X-ray/SZE measurements.  Our measurement
of $H_0$ in the distant universe is in agreement with the {\it Hubble
Space Telescope} Key Project measurement of $H_0=72\pm8$ km s$^{-1}$
Mpc$^{-1}$ (Freedman et al.\ 2001), which probes the nearby universe.

To address the effects of cosmology on the value of the
Hubble constant obtained from the SZE/X-ray method, we also repeat the
fit of our cluster distances varying the $\Omega_{M}$ and
$\Omega_{\Lambda}$ in a fiducial interval around the currently favored
$\Lambda$CDM model parameters, $\Omega_{M}=0.2-0.4$ and
$\Omega_{\Lambda}=0.6-0.8$.  The fits yield
$H_0=78.8\pm^{4.1}_{3.5}$ (68\% statistical error), with a
$\chi^2=31.9$ statistic for $\Omega_{M}=0.2$, $\Omega_{\Lambda}$=0.8,
and $H_0=74.9\pm^{3.8}_{3.2}$ ($\chi^2=31.5$) for $\Omega_{M}=0.4$,
$\Omega_{\Lambda}$=0.6.  Finally, we fit of our cluster distances with
the theoretical $D_A(z)$ function for a matter-dominated universe with
$\Omega_{M}=1.0$ and $\Omega_{\Lambda}=0.0$. The best-fit value of the
Hubble constant in this case is $H_0$=67.1$\pm^{4.5}_{3.6}$ (68\%
statistical error), with a $\chi^2$ statistic of 32.5 for \ndof\/
degrees of freedom. These fits have the same quality as that for the
currently favored $\Omega_{M}=0.3$, $\Omega_{\Lambda}=0.7$ cosmology,
indicating that cluster distances alone can not yet effectively
constrain the energy density parameters.

\subsection{Measurement of the Hubble constant using the isothermal $\beta$-model \label{nocenter}}
We compare the cluster distance results from the hydrostatic model of
Section \ref{results} to the results from other ICM models to
determine how sensitive the distance measurements are to the details
of the plasma modeling.
These models consist of a simple isothermal $\beta$-model, with a
density profile described by Equation \ref{density} with $f$=0, and
with a constant temperature.  Since cluster centers often feature a
sharp gradient in density and temperature, not consistent with this
simple $f$=0 model, we also excised the central $r < 100$ kpc of the X-ray
data from the analysis.  Figure \ref{tprofiles} and
Table \ref{table_tprof} show that, when the central 100 kpc are removed
from the X-ray data, the temperature profiles out to $\sim$600 kpc are
essentially flat.  With these simplifying assumptions, the X-ray
surface brightness and SZE decrement have simple analytical functions
(see, e.g., Birkinshaw et al.\ 1991), and numerical integrations are
no longer needed. Also, we do not enforce hydrostatic equilibrium, and
accordingly do not consider the dark matter distribution and the
spectral likelihood information in the MCMC procedure described in
Section \ref{mcmc}.

There is no simple way to mask the central 100 kpc from our
interferometric data, because these data are fit in the Fourier plane
(La~Roque 2005).  However, the SZE data are less sensitive to the
presence of a dense core than the X-ray data (Equation
\ref{eq-system}), and the X-ray data drive the fit for the density
shape parameters.  In addition, even clusters with X-ray structures in
the core are normally in pressure equilibrium (Markevitch et al.\
2000, 2001), and should therefore have smooth SZE profiles.  We
therefore use the entire SZE dataset and the 100 kpc-cut X-ray dataset
for this analysis.  The assumptions of the model outlined above are
described and tested in more detail in La~Roque et al.\ (2006).

\begin{figure}
\begin{center}
\includegraphics[angle=-90,width=7in]{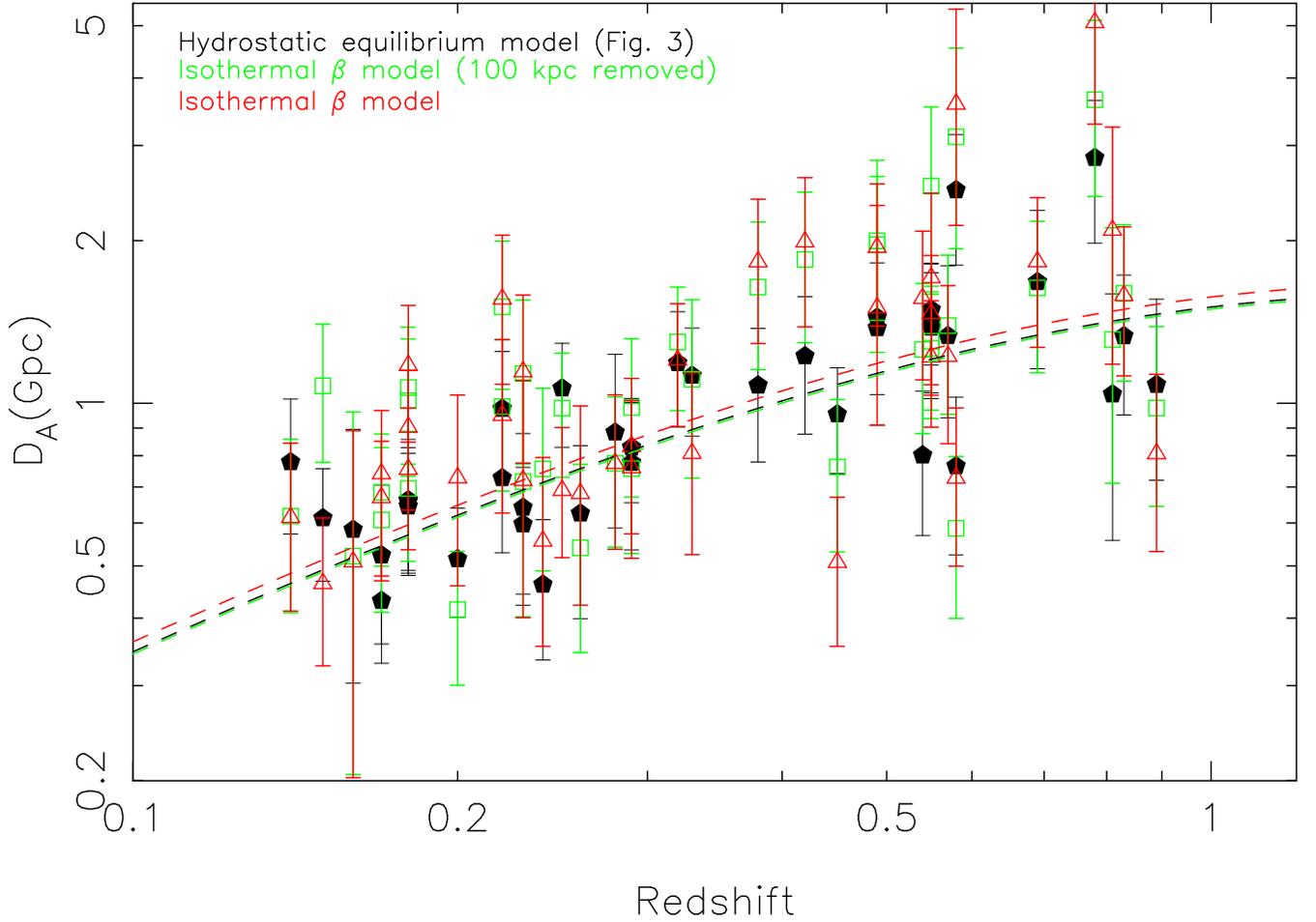}
\vspace{-5mm}
\end{center}
\caption{\small Angular diameter distances of the \nclu\/ clusters,
using the simple $r<$100 kpc-cut isothermal $\beta$ model (green) and
the isothermal $\beta$ model (red) described in Section
\ref{nocenter}.  The error bars are the total statistical
uncertainties, obtained by adding the X-ray and SZE data modelling
uncertainties (Table \ref{table:nocenter}, Table \ref{table:center})
and the additional sources of random error described in Section
\ref{error} and Table \ref{table-sys}. The systematic errors of Table
\ref{table-sys} are not shown.  Dashed lines are the best-fit angular
diameter curves using the best-fit Hubble constant $H_0$=\hocut\/ km
s$^{-1}$ Mpc$^{-1}$ (green) and $H_0$=\hoall\/ km s$^{-1}$ Mpc$^{-1}$
(red) and $\Omega_{M}=0.3$, for $\Omega_{\Lambda}=0.7$.  In black are
the distances obtained with the hydrostatic equilibrium model of
Section \ref{results} (Figure \ref{distances}).
\label{fig:nocenter}}
\vspace{-0.5cm}
\end{figure}

The model includes the following parameters: $S_{X0}$, $r_c$, $\beta$,
$\Delta T_0$, $kT$ and $A$, with the X-ray surface brightness $S_X$
and the SZE decrement $\Delta T$ following the equations
\begin{eqnarray}
S_X = S_{X0} \left( 1 + \frac{r^2}{r_c^2} \right) ^{(1-6\beta)/2}\\
\Delta T =  \Delta T_0 \left ( 1 + \frac{r^2}{r_c^2} \right)
^{(1-3\beta)/2}
\end{eqnarray}
The angular diameter distance $D_A$ is calculated according to
Equation \ref{eq-DA}, which is explained in detail in Bonamente et
al.\ (2004).  Applying this simple model to the data, we calculate the
model parameters (Table \ref{table:nocenter}) and the angular diameter
distances (Figure \ref{fig:nocenter}). The same fitting technique
employed in Section \ref{results} above yields a best-fit Hubble
constant of \HOcut (68\% confidence interval, statistical followed by
systematic errors), with a fit statistic of $\chi^2$=\chirescut\/ for
\ndof\/ degrees of freedom.
In this joint X-ray/SZE analysis, 
the spatial parameters
$S_{X0}$, $r_c$, and $\beta$ are constrained almost exclusively by the 
X-ray data, rather than by the SZE imaging, 
due to the high angular resolution
and the large number of counts in the \chandra/\ images.
Constraints on $\Delta T_0$ are obtained from the SZE data, 
while $kT$ and $A$ are obtained from the X-ray spectroscopy.  The value
of performing the joint MCMC analysis involving all three datasets
is that the full probability density
function for each model parameter can be obtained, with all statistically-allowable
parameter variations included.

Finally, we include the results for the standard isothermal $\beta$
model fit to the entire X-ray dataset, i.e., without the excision of
the central 100 kpc.  This exercise is provided for comparison with
earlier analyses that used such modelling (e.g., Reese et al.\ 2002),
and it is useful to assess the impact of the bright cluster cores on
the determination of the distance scale.  This model yields the
best-fit parameters in Table \ref{table:center}, the angular diameter
distances of Figure \ref{fig:nocenter} and a best-fit Hubble constant
of \HOall (68\% confidence interval, statistical followed by
systematic errors), with a fit statistic of $\chi^2$=\chiresall\/ for
\ndof\/ degrees of freedom.

These results indicate that the measurement of the cosmic distance
scale using X-ray and SZE observations of galaxy clusters is
insensitive to the details of the hot ICM model: the spread between
the three models explored here is 3 km s$^{-1}$ Mpc$^{-1}$.

\clearpage

\clearpage

\subsection{Comparison with recent distance measurements \label{comparison}}
We compare our results with recent X-ray/SZE measurements of the Hubble constant
using clusters in our sample.

Jones et al.\ (2005), Saunders et al.\ (2003) and Grainge et al.\
(2002) use ROSAT and ASCA X-ray data and SZE data from the Ryle
telescope for Abell~697, Abell~773, Abell~1413, Abell~1914 and
Abell~2218. They employ an ellipsoidal $\beta$ model to find
$H_0=66\pm11\pm^{9}_{8}$ km s$^{-1}$ Mpc$^{-1}$ for a $\Lambda$CDM
cosmology, in agreement with our hydrostatic equilibrium model and our
isothermal $\beta$ model results.

Schmidt, Allen and Fabian (2004) use Chandra X-ray data and several
published SZE measurements of RXJ1347.5-1145, Abell~1835 and Abell~478
(the latter not included in our sample), and employ a hydrostatic
equilibrium model similar to the one used in this paper. They obtain a
best-fit $H_0=69\pm8$ km s$^{-1}$ Mpc$^{-1}$ for a $\Lambda$CDM
cosmology, also in agreement with our hydrostatic equilibrium model
results.

Worrall and Birkinshaw (2003) use a $\beta$ model fit to the
XMM-Newton X-ray data and the Hughes and Birkinshaw (1998) SZE data of
CL0016+1609 to find a best-fit $D_A=1.36\pm0.15$ Gpc. This measurement
is in excellent agreement with our hydrostatic equilibrium model for
this cluster
($D_A=1.38\pm0.22$ Gpc, Table \ref{tab:results}) and isothermal
$\beta$ models ($D_A=1.22\pm^{0.21}_{0.19}$ Gpc, Table
\ref{table:nocenter}; $D_A=1.30\pm^{0.21}_{0.19}$ Gpc, Table
\ref{table:center}).

\section{Conclusions}
We analyzed \nclu\/ clusters of galaxies with \chandra\ X-ray imaging
spectroscopy and \ovro-\bima\/ SZE data, the largest sample to date
used to measure $H_0$.
We applied a hydrostatic equilibrium model that accounts for radial
variations in cluster temperature, and for sharp density gradients
caused by the cooling of the plasma in the cluster core.  The joint
analysis of X-ray and SZE data yields a direct measurement of the
cosmic distance scale in the redshift range 0.14$<$z$<$0.89.  We
measure a Hubble constant of \HO\ for an $\Omega_M=0.3$,
$\Omega_\Lambda=0.7$ cosmology (68 \% confidence interval, statistical
followed by systematic uncertainty), which is in agreement with the
{\it Hubble Space Telescope} Key Project results obtained at low
redshift.  We also analyze our measurements with a simple isothermal
model of the hot plasma without the hydrostatic equilibrium
assumption. The results from this simple model are in good agreement
with the hydrostatic equilibrium model, indicating that the X-ray/SZE
method used to determine the cosmic distance scale is largely
insensitive to the details of the hot plasma modeling.

\vspace{2cm}
Acknowledgments:

This paper is dedicated to Leon van Speybroeck and his colleagues on
the \chandra\/ project, including H.\ Ebeling, W.\ Forman, J.P.\ Hughes,
C.\ Jones, M.\ Markevitch, H.\ Tananbaum, A.\ Vikhlinin and M.\ Weisskopf; without their
effort to construct an exceptional observatory and to obtain deep
observations of a large number of clusters, this research would not
have been possible.  The support of the BIMA and OVRO staff over many
years is also gratefully acknowledged, including J.R.\ Forster,
C.\ Giovanine, R.\ Lawrence, S.\ Padin, R.\ Plambeck, S.\ Scott and
D.\ Woody.  We thank C.\ Alexander, K.\ Coble, A.\ Cooray, 
L.\ Grego, G.\ Holder, W.\ Holzapfel,
A.\ Miller, J.\ Mohr, D.\ Nagai, S.\ Patel and P.\ Whitehouse for their outstanding
contributions to the SZE instrumentation, observations, and analysis. 
We also thank J.\ Mohr and D.\ Nagai for contributions to the development 
of the hydrostatic equilibrium model.

This work was supported by NASA LTSA grant NAG5-7985 and also in part
by NSF grants PHY-0114422 and AST-0096913, the David and Lucile
Packard Foundation, the McDonnell Foundation, and a MSFC director's
discretionary award.  Research at the Owens Valley Radio Observatory
and the Berkeley-Illinois-Maryland Array was supported by National
Science Foundation grants AST 99-81546 and 02-28963.  Calculations
were performed at the Space Plasma Interactive Data Analysis and
Simulation Laboratory at the Center for Space Plasma and Aeronomy
Research of the University of Alabama at Huntsville.


\clearpage
\pagestyle{empty}
\begin{deluxetable}{lrcccccccccl}
\rotate
\tabletypesize{\scriptsize}
\tablewidth{0pt}
\tablecaption{Cluster Data\label{tab:data}}
\tablehead{
\colhead{} &
\colhead{} &
\multicolumn{5}{c}{\chandra\ X-ray Data} &
\multicolumn{4}{c}{Interferometric SZE Data}&
\colhead{}
\\
\colhead{} &
\colhead{} &
\multicolumn{5}{c}{\hrulefill} &
\multicolumn{4}{c}{\hrulefill} &
\colhead{}
\\
\colhead{Cluster} &
\colhead{z} &
\colhead{ObsID} &
\colhead{Chip} &
\colhead{(ks)} &
\colhead{(hh mm ss)} &
\colhead{($\circ\;\;$ $\prime\;\;$ $\prime\prime$ )} &
\colhead{BIMA (hr)} &
\colhead{OVRO (hr)} &
\colhead{(hh mm ss)} &
\colhead{($\circ\;\;$ $\prime\;\;$ $\prime\prime$ )} &
\colhead{$z$ reference}
}
\startdata
CL~0016+1609 & 0.541 
        & \phn520 & I3 & \phn67.4 & 00 18 33.5 & +16 26 12.5 & 43 & 100 & 00 18 33.3 & +16 26 04.0
&  \citet{stocke1991}
\\
Abell~68 & 0.255 
	& 3250 & I3 & \phn10.0 & 00 37 06.2 & +09 09 33.2 & 54 & \nodatah & 00 37 04.0 & +09 10 02.5
& \citet{struble1999}
\\
Abell~267 & 0.230 
	& 1448 & I3 & \phn\phn7.4 & 01 52 42.1 & +01 00 35.7 & 50 & \nodatah & 01 52 42.3 & +01 00 26.0
 & \citet{struble1999}
\\
Abell~370  & 0.375 
	& \phn515 & S3 & \phn65.3 & 02 39 53.2 & -01 34 35.0 &26 & 33 & 02 39 52.4 & -01 34 43.8
& \citet{struble1999}
\\
MS~0451.6-0305 & 0.550 
	& \phn902 & S3 & \phn42.2 & 04 54 11.4 & -03 00 52.7 & \nodatah & 30 & 04 54 11.6 & -03 01 01.3 
& \citet{gioia1994} \\
	& \nodatae
	& \phn529 & I3 & \phn13.9 & \nodatae& \nodatae& \nodatae& \nodatae& \nodatae& \nodatae
& \nodatae
\\
MACS~J0647.7+7015 & 0.584
	& 3196 & I3 & \phn19.3 & 06 47 50.2 & +70 14 54.6 & \nodatah & 23 & 06 47 50.2 & +70 14 56.1
& \citet{laroque2003}\\
	& \nodatae
	& 3584 & I3 & \phn20.0 & \nodatae& \nodatae & \nodatae& \nodatae& \nodatae& \nodatae
& \nodatae
\\
Abell~586 & 0.171
	& \phn530 & I3 & \phn10.0 & 07 32 20.2 & +31 37 55.6& 45 & \nodatah & 07 32 19.6 & +31 37 55.3
& \citet{struble1999}
\\
MACS~J0744.8+3927 & 0.686
	& 3197 & I3 & \phn20.2 & 07 44 52.8 & +39 27 26.7 & \phn8 & 17 & 07 44 52.4 & +39 27 33.2
& \citet{laroque2003}\\
	& \nodatae
	& 3585 & I3 & \phn19.4 & \nodatae& \nodatae& \nodatae& \nodatae& \nodatae& \nodatae
& \nodatae
\\
Abell~611 & 0.288
	& 3194 & S3 & \phn36.1 & 08 00 56.6 & +36 03 24.1 & \nodatah & 57 & 08 00 56.5 & +36 03 22.9
& \citet{struble1999}
\\
Abell~665 & 0.182
	& 3586 & I3 & \phn29.7 & 08 30 58.1 & +65 50 51.6 & 52 & 16  & 08 30 58.6 & +65 50 49.8 
 & \citet{struble1999}\\
	& \nodatae
	& \phn531 & I3 & \phn9.0 & \nodatae& \nodatae& \nodatae& \nodatae& \nodatae& \nodatae
& \nodatae
\\
Abell~697& 0.282 
	& 4217 & I3 & \phn19.5 & 08 42 57.5 & +36 21 56.2 & \nodatah & 47 & 08 42 57.8 & +36 21 54.5
& \citet{struble1999}
\\
Abell~773 & 0.217
	& \phn533 & I3 & \phn11.3 & 09 17 52.8 & +51 43 38.9 & 26 & 66 & 09 17 53.5 & +51 43 49.8 
& \citet{struble1999}\\
	& \nodatae
	& 3588 & I3 & \phn\phn9.4 & \nodatae& \nodatae& \nodatae& \nodatae& \nodatae& \nodatae
& \nodatae
\\
ZW~3146& 0.291
	& \phn909 & I3 & \phn46.0 & 10 23 39.7 & +04 11 09.5 & 25 & 15 & 10 23 37.8 & +04 11 17.8
& \citet{allen1992}
\\
MS~1054.5-0321& 0.826
        & 512 & S3  & \phn89.1 & 10 56 59.4 & -03 37 34.2 & \nodatah  & 43  & 10 56 59.1 & -03 37 34.0  
& \citet{gioia1995}
\\
MS~1137.5+6625 & 0.784
	& \phn536 & I3 & \phn77.0 & 11 40 22.3 & +66 08 16.0 & 88 & \nodatah & 11 40 23.1  & +66 08 05.3
& \citet{donahue1999}
\\
MACS~J1149.5+2223 & 0.544 
	& 1656 & I3 & \phn18.5 & 11 49 35.5 & +22 24 02.3 & 39 & \nodatah & 11 49 34.9 & +22 23 54.8 
& \citet{laroque2003}\\
	& \nodatae
	& 3589 & I3 & \phn20.0 & \nodatae& \nodatae& \nodatae& \nodatae& \nodatae& \nodatae
& \nodatae
\\
Abell~1413& 0.142
	& 1661 & I3 & \phn\phn9.7 & 11 55 18.0 & +23 24 17.0 & 28 & \nodatah &  11 55 17.7 & +23 24 39.5
& \citet{struble1999}\\
	& \nodatae
	& \phn537 & I3 & \phn\phn9.6 & \nodatae& \nodatae& \nodatae& \nodatae& \nodatae& \nodatae
& \nodatae
\\
CL~J1226.9+3332& 0.890 
	& 3180 & I3 & \phn31.7 & 12 26 57.9 & +33 32 47.4 & 33 & \nodatah & 12 26 58.0 & +33.32 57.9 
& \citet{ebeling2001a}\\
	& \nodatae
	& 932  & S3 & \phn9.9  &  \nodatae& \nodatae& \nodatae& \nodatae& \nodatae& \nodatae
& \nodatae
\\
MACS~J1311.0-0310 & 0.490
	&3258 & I3 & 14.9 & 13 11 01.7 & -03 10 38.5 & 39 & \nodatah &  13 11 02.2 &  -03 10 45.6 
& \citet{allen2004}
\\

Abell~1689& 0.183
	& 1663 & I3 & \phn10.7 & 13 11 29.5 & -01 20 28.2 & 16 & 26 & 13 11 29.1 & -01 20 29.7
& \citet{struble1999}\\
	& \nodatae
	& \phn540 & I3 & \phn10.3 &  \nodatae& \nodatae& \nodatae& \nodatae& \nodatae& \nodatae
& \nodatae
\\
RX~J1347.5-1145& 0.451
	& 3592 & I3 & \phn57.7 & 13 47 30.6 & -11 45 08.6 & 22 & \phn3 &13 47 30.6 & -11 45 12.3
& \citet{schindler1995}
\\

MS~1358.4+6245& 0.327
	& \phn516 & S3 & \phn48.1 & 35 59 50.6 & +62 31 04.1& 70 & \nodatah & 13 59 50.2 & +62 31 07.0
& \citet{gioia1994}
\\
Abell~1835& 0.252
	& \phn495 & S3 & \phn19.5 & 14 01 02.0 & +02 52 41.7 & 27 & 23 & 14 01 01.8 & +02 52 45.6
& \citet{struble1999}\\
	& \nodatae
	& \phn496 & S3 & \phn10.7 & \nodatae& \nodatae& \nodatae& \nodatae& \nodatae& \nodatae
& \nodatae
\\
MACS~J1423.8+2404& 0.545
	& 4195 & S3 & 115.6 & 14 23 47.9 & +24 04 42.6  & 35 & \nodatah & 14 23 47.7 & +24 04 37.3
& \citet{laroque2003}
\\
Abell~1914 & 0.171 
	& 3593 & I3 & \phn18.9 & 14 26 00.8 & +37 49 35.7 & 24 & \nodatah & 14 26 01.3 & 37 49 38.6 
& \citet{struble1999}\\
	& \nodatae
	& \phn542 & I3 & \phn\phn8.1 & \nodatae& \nodatae& \nodatae& \nodatae& \nodatae& \nodatae
& \nodatae
\\
Abell~1995& 0.322
	& \phn906 & S3 & \phn56.7 & 14 52 57.9 & +58 02 55.8 & 50 & 58 & 14 52 58.1 & +58 02 57.0
& \citet{patel2000}
\\
Abell~2111& 0.229
	& \phn544 & I3 & \phn10.3 &15 39 41.0 & +34 25 08.8 & 36 & \nodatah & 15 39 40.2  & +34 25 00.4
& \citet{struble1999}
\\
Abell~2163& 0.202
	& 1653 & I1 & \phn71.1 & 16 15 46.2 & -06 08 51.3 & 23 & 37 & 16 15 43.6 & -06 08 46.6
& \citet{struble1999}
\\
Abell~2204 & 0.152
       & 499 & S3   & \phn8.6 &  16 32 46.9 & +05 34 31.9  & 30 & \nodatah & 16 32 46.6 & +05 34 20.6
& \citet{struble1999}\\
	& \nodatae
       &  6104 & I3  & \phn9.6 &  \nodatae& \nodatae& \nodatae& \nodatae& \nodatae& \nodatae
& \nodatae
\\	
Abell~2218& 0.176 
	& 1666 & I0 & \phn41.7 & 16 35 51.9 & +66 12 34.5 & 32 & 70 & 16 35 48.7 & +66 12 28.1 
& \citet{struble1999}
\\
RX~J1716.4+6708& 0.813 
	& \phn548 & I3 & \phn51.7 & 17 16 48.8 & +67 08 25.3 & 37 & \nodatah & 17 16 51.2 & +67 07 49.6
& \citet{henry1997}
\\
Abell~2259& 0.164 
	& 3245 & I3 & \phn10.0 & 17 20 08.5 & +27 40 11.0 & 25 & \nodatah & 17 20 09.0 & +27 40 09.4 
& \citet{struble1999}
\\
Abell~2261& 0.224
	& \phn550 & I3 & \phn\phn9.1 & 17 22 27.1 & +32 07 57.8 & 23 & 40 & 17 22 26.9 & +32 07 59.9 
& \citet{struble1999}
\\
MS~2053.7-0449& 0.583
	& \phn551 & I3 & \phn44.3 & 20 56 21.2 & -04 37 47.8 & \nodatah & 154 & 20 56 21.0 & -04 37 47.2
& \citet{stocke1991}\\
	& \nodatae
        & 1667 & I3 & \phn44.5 & \nodatae& \nodatae& \nodatae& \nodatae& \nodatae& \nodatae
& \nodatae
\\
MACS~J2129.4-0741& 0.570
	& 3199 & I3 & \phn8.5 & 21 29 26.0 & -07 41 28.7 & \nodatah & 24 & 21 29 24.9 & -07 41 43.9
& \citet{laroque2003}\\
	& \nodatae
        & 3595 & I3 & \phn18.4 & \nodatae& \nodatae& \nodatae& \nodatae& \nodatae& \nodatae
& \nodatae
\\
RX~J2129.7+0005& 0.235
	& \phn552 & I3 & \phn10.0 & 21 29 39.9 &+00 05 19.8 & 47 & \nodatah & 21 29 38.1 & +00 05 12.4
& \citet{ebeling1998}
\\
MACS~J2214.9-1359& 0.483
	& 3259 & I3 & \phn19.5 & 22 14 57.3 & -14 00 12.3 &41 & 11 & 22 14 58.4 & -14 00 10.9
& Note (\tablenotemark{a}\ ) \\
	& \nodatae       
& 5011 & I3 & \phn16.1 & \nodatae& \nodatae & \nodatae& \nodatae& \nodatae& \nodatae
& \nodatae
\\
MACS~J2228.5+2036& 0.412
	& 3285 & I3 & \phn19.9 & 22 28 33.0 & +20 37 14.4 & 39 & \nodatah & 22 28 33.1 & +20 37 14.2
& \citet{bohringer2000}
\\
\enddata
\tablenotetext{a}{Redshift derived from the Fe lines in the Chandra x-ray spectrum, this paper.}
\end{deluxetable}

\clearpage
\pagestyle{plaintop}
\begin{deluxetable}{lclcccccc}
\tablecaption{Results of the hydrostatic equilibrium model\label{tab:results}}
\tabletypesize{\scriptsize}
\tablewidth{600pt}
\rotate
\tablehead{ Cluster  &$\mathcal{N} $ & $r_s$ &$ n_{e0} $  &$ r_{c1} $  &$ \beta $  &$ f $  &$ r_{c2} $  &$ D_A $  \\ 
 &  (10$^{-25}$ g cm$^{-3}$)   & (arcsec)    & (cm$^{-3}$)   &  (arcsec)   &     &     & (arcsec)   & Gpc }
\startdata
CL~0016+1609 & $ 0.10 ^{+0.14}_{-0.06}$ & $225 ^{+233}_{-96}$  & $ 1.40 ^{+0.18}_{-0.15}$ $ \times 10^{-2}$ & $ 10.3 ^{+4.4}_{-2.5}$ & $ 0.761 ^{+0.031}_{-0.036}$ & $ 0.48 ^{+0.05}_{-0.05}$ & $ 47.8 ^{+3.8}_{-3.7}$ & $ 1.38 ^{+0.22}_{-0.22}$  \\ 
Abell~0068 & $ 3.29 ^{+7.60}_{-2.51}$ & $70 ^{+62}_{-27}$  & $ 8.89 ^{+1.68}_{-1.18}$ $ \times 10^{-3}$ & -- & $ 0.693 ^{+0.026}_{-0.028}$ & -- & $ 47.8 ^{+2.8}_{-3.0}$ & $ 0.63 ^{+0.16}_{-0.19}$ \\ 
Abell~0267 & $2.02 ^{+3.04}_{-1.24}$ &  $ 75 ^{+50}_{-31}$ & $ 1.17 ^{+0.11}_{-0.10}$ $ \times 10^{-2}$ & -- & $ 0.698 ^{+0.031}_{-0.030}$ & -- & $ 40.9 ^{+2.8}_{-2.8}$ & $ 0.60 ^{+0.11}_{-0.09}$ \\ 
Abell~0370 & $ 1.63 ^{+1.80}_{-0.87}$  &$ 51 ^{+21}_{-13}$&  $ 5.33 ^{+0.58}_{-0.40}$ $ \times 10^{-3}$ & -- & $ 0.740 ^{+0.035}_{-0.028}$ & -- & $ 55.6 ^{+3.1}_{-2.6}$ & $ 1.08 ^{+0.19}_{-0.20}$ \\ 
MS~0451.6-0305  &$ 0.27 ^{+0.58}_{-0.16}$ &$ 110 ^{+75}_{-44}$ & $ 1.26 ^{+0.12}_{-0.09}$ $ \times 10^{-2}$ & -- & $ 0.777 ^{+0.019}_{-0.019}$ & -- & $ 34.5 ^{+1.1}_{-1.1}$ & $ 1.42 ^{+0.26}_{-0.23}$ \\ 
MACS~J0647.7+7015  & $12.01 ^{+16.67}_{-8.41}$ & $ 36 ^{+22}_{-13}$ & $ 2.19 ^{+0.34}_{-0.25}$ $ \times 10^{-2}$ & -- & $ 0.653 ^{+0.019}_{-0.017}$ & -- & $ 19.9 ^{+1.2}_{-1.2}$ & $ 0.77 ^{+0.21}_{-0.18}$ \\ 
Abell~0586 & $1.78 ^{+1.97}_{-1.05}$ & $ 102 ^{+40}_{-26}$  & $ 1.83 ^{+0.25}_{-0.21}$ $ \times 10^{-2}$ & -- & $ 0.627 ^{+0.017}_{-0.013}$ & -- & $ 32.0 ^{+1.7}_{-1.4}$ & $ 0.52 ^{+0.15}_{-0.12}$ \\ 
MACS~J0744.8+3927 & $ 0.27 ^{+0.84}_{-0.22}$ &$ 94 ^{+102}_{-51}$ &  $ 1.14 ^{+0.22}_{-0.15}$ $ \times 10^{-1}$ & $ 3.4 ^{+0.6}_{-0.7}$ & $ 0.635 ^{+0.049}_{-0.039}$ & $ 0.93 ^{+0.01}_{-0.01}$ & $ 25.8 ^{+4.7}_{-4.7}$ & $ 1.68 ^{+0.48}_{-0.38}$  \\ 
Abell~0611 &$ 1.73 ^{+1.87}_{-0.90}$& $ 64 ^{+15}_{-12}$  & $ 5.27 ^{+0.97}_{-1.00}$ $ \times 10^{-2}$ & $ 2.8 ^{+0.4}_{-0.3}$ & $ 0.600 ^{+0.014}_{-0.008}$ & $ 0.66 ^{+0.08}_{-0.07}$ & $ 22.5 ^{+1.6}_{-1.2}$ & $ 0.78 ^{+0.18}_{-0.18}$ \\ 
Abell~0665 & $ 0.18 ^{+0.14}_{-0.09}$ & $ 340 ^{+150}_{-86}$ & $ 9.13 ^{+1.34}_{-1.06}$ $ \times 10^{-3}$ & $ 3.2 ^{+0.8}_{-0.5}$ & $ 0.730 ^{+0.015}_{-0.016}$ & $ 0.11 ^{+0.10}_{-0.08}$ & $ 64.4 ^{+1.7}_{-1.8}$ & $ 0.66 ^{+0.09}_{-0.10}$ \\ 
Abell~0697 & $ 0.76 ^{+1.63}_{-0.59}$ &$ 93 ^{+66}_{-32}$   & $ 9.82 ^{+1.55}_{-1.28}$ $ \times 10^{-3}$ & -- & $ 0.584 ^{+0.014}_{-0.016}$ & -- & $ 41.6 ^{+1.6}_{-1.9}$ & $ 0.88 ^{+0.30}_{-0.23}$ \\ 
Abell~0773 & $ 1.22 ^{+1.98}_{-0.88}$ & $ 54 ^{+40}_{-19}$ & $ 8.04 ^{+0.68}_{-0.64}$ $ \times 10^{-3}$ & -- & $ 0.564 ^{+0.020}_{-0.022}$ & -- & $ 40.2 ^{+2.2}_{-2.3}$ & $ 0.98 ^{+0.17}_{-0.14}$ \\ 
ZW~3146 & $ 0.66 ^{+0.08}_{-0.05}$ & $ 121 ^{+4}_{-6}$   & $ 1.70 ^{+0.02}_{-0.03}$ $\times 10^{-1}$ & $ 4.4 ^{+0.1}_{-0.1}$ & $ 0.668 ^{+0.005}_{-0.004}$ & $ 0.881 ^{+0.004}_{-0.003}$ & $ 25.5 ^{+0.7}_{-0.4}$ & $ 0.83 ^{+0.02}_{-0.02}$ \\ 
MS~1054-0321 & $ 0.04 ^{+0.08}_{-0.02}$ &$ 666 ^{+571}_{-359}$  & $ 6.15 ^{+0.71}_{-0.56}$ $ \times 10^{-3}$ & -- & $ 1.791 ^{+0.148}_{-0.209}$ & -- & $ 83.7 ^{+4.9}_{-7.3}$ & $ 1.33 ^{+0.28}_{-0.26}$ \\ 
MS~1137.5+6625 & $ 1.73 ^{+7.31}_{-1.40}$ &$ 16 ^{+18}_{-9}$  & $ 1.26 ^{+0.16}_{-0.11}$ $ \times 10^{-2}$ & -- & $ 0.667 ^{+0.044}_{-0.043}$ & -- & $ 14.2 ^{+1.5}_{-1.3}$ & $ 2.85 ^{+0.52}_{-0.63}$ \\ 
MACS~J1149.5+2223  & $ 0.74 ^{+3.06}_{-0.50}$ &  $ 110 ^{+46}_{-29}$   & $ 8.53 ^{+1.04}_{-0.89}$ $ \times 10^{-3}$ & -- & $ 0.673 ^{+0.020}_{-0.022}$ & -- & $ 42.8 ^{+2.1}_{-2.4}$ & $ 0.80 ^{+0.19}_{-0.16}$ \\ 
Abell~1413 &  $ 0.47 ^{+0.58}_{-0.27}$ &$ 121 ^{+51}_{-47}$  & $ 3.66 ^{+0.65}_{-0.42}$ $ \times 10^{-2}$ & $ 6.5 ^{+1.5}_{-1.3}$ & $ 0.531 ^{+0.018}_{-0.014}$ & $ 0.76 ^{+0.02}_{-0.02}$ & $ 39.3 ^{+4.5}_{-3.7}$ & $ 0.78 ^{+0.18}_{-0.13}$ \\ 
CL~J1226.9+3332  & $ 4.09 ^{+9.41}_{-3.58}$ & $ 46 ^{+58}_{-19}$ & $ 3.01 ^{+0.47}_{-0.44}$ $ \times 10^{-2}$ & -- & $ 0.715 ^{+0.038}_{-0.038}$ & -- & $ 15.8 ^{+1.3}_{-1.4}$ & $ 1.08 ^{+0.42}_{-0.28}$ \\ 
MACS~J1311.0-0310  & $ 7.59 ^{+17.81}_{-7.09}$ & $ 19 ^{+47}_{-9}$ & $ 3.93 ^{+0.72}_{-0.55}$ $ \times 10^{-2}$ & -- & $ 0.613 ^{+0.022}_{-0.020}$ & -- & $ 9.3 ^{+0.7}_{-0.7}$ & $ 1.38 ^{+0.47}_{-0.37}$ \\ 
Abell~1689 & $2.68 ^{+1.20}_{-1.16}$ & $ 75 ^{+19}_{-10}$ & $ 4.054 ^{+0.36}_{-0.26}$ $ \times 10^{-2}$ & $ 21.7 ^{+0.9}_{-1.0}$ & $ 0.873 ^{+0.039}_{-0.041}$ & $ 0.87 ^{+0.01}_{-0.01}$ & $ 104.9 ^{+5.1}_{-5.5}$ & $ 0.65 ^{+0.09}_{-0.09}$ \\ 
RX~J1347.5-1145 &  $ 4.57 ^{+1.06}_{-0.86}$ & $ 47 ^{+5}_{-5}$ & $ 2.81 ^{+0.16}_{-0.12} \times 10^{-1}$ & $ 3.9 ^{+0.2}_{-0.1}$ & $ 0.631 ^{+0.009}_{-0.008}$ & $ 0.942 ^{+0.004}_{-0.004}$ & $ 22.9 ^{+1.8}_{-1.4}$ & $ 0.96 ^{+0.06}_{-0.08}$ \\ 
MS~1358.4+6245 & $ 0.58 ^{+0.21}_{-0.19}$ &$ 90 ^{+26}_{-18}$  & $ 9.62 ^{+0.79}_{-0.78}$ $ \times 10^{-2}$ & $ 3.3 ^{+0.2}_{-0.2}$ & $ 0.675 ^{+0.017}_{-0.016}$ & $ 0.934 ^{+0.003}_{-0.003}$ & $ 37.2 ^{+1.7}_{-1.9}$ & $ 1.13 ^{+0.09}_{-0.10}$ \\ 
Abell~1835 & $0.28 ^{+0.10}_{-0.03}$ & $ 150 ^{+11}_{-11}$ & $ 1.10 ^{+0.05}_{-0.02} \times 10^{-1}$ & $ 9.3 ^{+0.2}_{-0.2}$ & $ 0.798 ^{+0.013}_{-0.017}$ & $ 0.940 ^{+0.001}_{-0.001}$ & $ 63.7 ^{+1.5}_{-1.6}$ & $ 1.07 ^{+0.02}_{-0.08}$ \\ 
MACS~J1423.8+2504  & $ 1.83 ^{+0.02}_{-0.07}$ & $ 33 ^{+1}_{-1}$  & $ 1.60 ^{+0.02}_{-0.08} \times 10^{-1}$ & $ 4.2 ^{+0.1}_{-0.1}$ & $ 0.721 ^{+0.012}_{-0.008}$ & $ 0.975 ^{+0.001}_{-0.001}$ & $ 36.7 ^{+0.9}_{-0.7}$ & $ 1.49 ^{+0.06}_{-0.03}$ \\ 
Abell~1914 & $ 5.79 ^{+2.60}_{-1.85}$ & $ 81 ^{+14}_{-11}$ & $ 1.72 ^{+0.13}_{-0.08}$ $ \times 10^{-2}$ & $6.6^{+0.6}_{-0.8}$ & $ 0.899 ^{+0.007}_{-0.012}$ & $0.008 ^{+0.018}_{-0.008}$ & $ 68.3 ^{+0.7}_{-1.0}$ & $ 0.44 ^{+0.04}_{-0.05}$ \\ 
Abell~1995 & $ 0.07^{+0.06}_{-0.04}$ & $359^{+205}_{-117}$ & $9.35^{+0.74}_{-0.56} \times 10^{-3}$ & $31.2^{+3.0}_{-3.5}$ & $1.298^{+0.062}_{-0.096}$ & $0.462^{+0.033}_{-0.033}$ & $83.5^{+3.7}_{-7.1}$ & $1.19^{+0.15}_{-0.14}$ \\
Abell~2111 &  $ 0.47 ^{+2.74}_{-0.38}$ & $ 172 ^{+354}_{-107}$  & $ 5.99 ^{+1.05}_{-0.79}$ $ \times 10^{-3}$ & -- & $ 0.600 ^{+0.026}_{-0.025}$ & -- & $ 50.4 ^{+3.8}_{-3.5}$ & $ 0.64 ^{+0.20}_{-0.17}$ \\ 
Abell~2163 & $ 0.26 ^{+0.12}_{-0.09}$ & $ 390 ^{+87}_{-52}$  & $ 1.09 ^{+0.07}_{-0.04}$ $ \times 10^{-2}$ & $4.0^{+1.3}_{-0.7}$ & $ 0.560 ^{+0.004}_{-0.005}$ & $0.022^{+0.037}_{-0.022}$ & $ 66.8 ^{+0.9}_{-0.8}$ & $ 0.52 ^{+0.04}_{-0.05}$ \\ 
Abell~2204 & $ 0.92 ^{+0.30}_{-0.15}$ & $ 120 ^{+13}_{-18}$   & $ 2.01 ^{+0.12}_{-0.09} \times 10^{-1}$ & $ 7.5 ^{+0.3}_{-0.3}$ & $ 0.710 ^{+0.031}_{-0.025}$ & $ 0.960 ^{+0.003}_{-0.004}$ & $ 67.4 ^{+2.0}_{-1.8}$ & $ 0.61 ^{+0.06}_{-0.07}$ \\ 
Abell~2218 & $ 1.02 ^{+0.70}_{-0.60}$ & $ 110 ^{+35}_{-22}$ & $ 7.02 ^{+0.66}_{-0.66}$ $ \times 10^{-3}$ & -- & $ 0.739 ^{+0.014}_{-0.017}$ & -- & $ 68.3 ^{+1.7}_{-2.1}$ & $ 0.66 ^{+0.14}_{-0.11}$ \\ 
RX~J1716.4+6708  & $ 0.34 ^{+3.38}_{-0.30}$ &$ 146 ^{+545}_{-106}$  & $ 1.94 ^{+0.61}_{-0.40}$ $ \times 10^{-2}$ & -- & $ 0.589 ^{+0.042}_{-0.035}$ & -- & $ 12.3 ^{+2.0}_{-1.7}$ & $ 1.04 ^{+0.51}_{-0.43}$ \\ 
Abell~2259 & $ 0.65 ^{+1.15}_{-0.54}$ & $141 ^{+155}_{-56}$ & $ 9.29 ^{+2.97}_{-1.71}$ $ \times 10^{-3}$ & -- & $ 0.560 ^{+0.025}_{-0.024}$ & -- & $ 41.0 ^{+3.9}_{-2.8}$ & $ 0.58 ^{+0.29}_{-0.25}$ \\ 
Abell~2261 & $ 1.36 ^{+0.85}_{-0.85}$ &$ 68 ^{+25}_{-15}$  & $ 4.16 ^{+0.54}_{-0.63}$ $ \times 10^{-2}$ & $ 10.0 ^{+1.9}_{-1.7}$ & $ 0.628 ^{+0.025}_{-0.022}$ & $ 0.77 ^{+0.04}_{-0.05}$ & $ 37.8 ^{+6.5}_{-5.2}$ & $ 0.73 ^{+0.20}_{-0.13}$  \\ 
MS~2053.7-0449  & $ 0.26 ^{+1.41}_{-0.22}$ &$ 40 ^{+64}_{-22}$ & $ 9.22 ^{+1.08}_{-0.92}$ $ \times 10^{-3}$ & -- & $ 0.522 ^{+0.048}_{-0.042}$ & -- & $ 10.8 ^{+1.9}_{-1.7}$ & $ 2.48 ^{+0.41}_{-0.44}$ \\ 
MACS~J2129.4-0741  & $ 6.05 ^{+17.17}_{-5.15}$ &$ 20 ^{+23}_{-8}$ & $ 1.71 ^{+0.21}_{-0.19}$ $ \times 10^{-2}$ & -- & $ 0.626 ^{+0.027}_{-0.029}$ & -- & $ 19.7 ^{+1.5}_{-1.5}$ & $ 1.33 ^{+0.37}_{-0.28}$ \\ 
RX~J2129.7+0005 & $ 3.04 ^{+1.66}_{-1.41}$ & $ 84 ^{+21}_{-15}$  & $ 1.78 ^{+0.22}_{-0.21}i \times 10^{-1}$ & $ 3.6 ^{+0.5}_{-0.4}$ & $ 0.588 ^{+0.012}_{-0.015}$ & $ 0.91 ^{+0.01}_{-0.01}$ & $ 26.1 ^{+3.0}_{-2.9}$ & $ 0.46 ^{+0.11}_{-0.08}$ \\ 
MACS~J2214.9-1359  & $ 0.66 ^{+1.40}_{-0.51}$ &$ 64 ^{+62}_{-32}$  & $ 1.35 ^{+0.13}_{-0.13}$ $ \times 10^{-2}$ & -- & $ 0.615 ^{+0.016}_{-0.020}$ & -- & $ 22.8 ^{+1.2}_{-1.3}$ & $ 1.44 ^{+0.27}_{-0.23}$ \\
MACS~J2228.5+2036  & $ 0.41 ^{+1.12}_{-0.32}$ & $ 101 ^{+108}_{-45}$& $ 1.24 ^{+0.14}_{-0.11}$ $ \times 10^{-2}$ & -- & $ 0.519 ^{+0.014}_{-0.013}$ & -- & $ 21.7 ^{+1.4}_{-1.3}$ & $ 1.22 ^{+0.24}_{-0.23}$ \\ 
\enddata
\end{deluxetable}

\clearpage

\begin{deluxetable}{lcc}
\centering
\tablecaption{Sources of uncertainty in the measurement of $D_A$ \label{table-sys}} 
\tablewidth{350pt}
\tablehead{Source& Effect on $D_A$ & Reference}
\startdata
\multicolumn{3}{c}{STATISTICAL CONTRIBUTIONS}\\
Galactic $N_H$ & $\leq\pm$1\% &  (1)\\
Cluster asphericity & $\pm$15\% & (2) \\
SZE point sources & $\pm$8\% & (3) \\
Kinetic SZE effect & $\pm$8\% & (4)\\
CMB anisotropy & $\leq$1\% & (4)\\
X-ray background & $\pm2\%$ & (5)\\
\hline
\multicolumn{3}{c}{SYSTEMATIC CONTRIBUTIONS}\\
Presence of radio halos & +3\% & (4)\\
X-ray absolute flux calibration ($S_X$) & $\pm5$\% & (6) \\
X-ray temperature calibration ($T_e$) & $\pm7.5$\% & (7) \\
SZE calibration & $\pm8$\% & (4)\\
\enddata
\\
(1) This paper, Section \ref{nh}.\\
(2) \citet{sulkanen1999}.\\
(3)\citet{laroque2006}.\\
(4)\citet{reese2002}.\\
(5) This paper, Section \ref{xray-background}.\\
(6)  http://asc.harvard.edu/cal/.\\
(7) This paper, Section \ref{xray-cal}.\\
\end{deluxetable}
\clearpage

\begin{deluxetable}{lcccccccc}
\tablewidth{650pt}
\tablecaption{Results of the $r<$100 kpc-cut isothermal $\beta$ model of Section \ref{nocenter} \label{table:nocenter}}
\tabletypesize{\scriptsize}
\tablehead{ Cluster  &$ S_{X0} $  &$ r_c $  &$ \beta $  &$ \Delta T_0 $  &$ kT $  &metallicity  &$ \Lambda $  &$ D_A $ \\ 
 &  (cnt cm$^{-2}$ arcmin$^{-2}$)   & (arcsec)   &     & (mK)   & (keV)   & (Solar)   & (cnt cm$^{3}$ s$^{-1}$)   & (Gpc)  }
\rotate
\startdata
CL~0016+1609  & $ 24.2 ^{+1.0}_{-0.9}$ & $ 42.9 ^{+2.6}_{-2.4}$ & $ 0.744 ^{+0.029}_{-0.026}$ & $ -1.36 ^{+0.08}_{-0.08}$ & $ 10.7 ^{+0.6}_{-0.7}$ & $ 0.31 ^{+0.08}_{-0.08}$ & $ 2.48 ^{+0.03}_{-0.03}$ $ \times 10^{-15}$ & $ 1.22 ^{+0.22}_{-0.19}$ \\ 
Abell~0068 &  $ 4.3 ^{+0.4}_{-0.3}$ & $ 59.3 ^{+7.3}_{-5.6}$ & $ 0.790 ^{+0.067}_{-0.048}$ & $ -0.72 ^{+0.10}_{-0.11}$ & $ 9.1 ^{+1.0}_{-0.9}$ & $ 0.44 ^{+0.19}_{-0.21}$ & $ 2.98 ^{+0.09}_{-0.10}$ $ \times 10^{-15}$ & $ 0.68 ^{+0.27}_{-0.22}$ \\ 
Abell~0267 &  $ 6.8 ^{+1.0}_{-0.9}$ & $ 31.2 ^{+4.0}_{-2.8}$ & $ 0.656 ^{+0.029}_{-0.023}$ & $ -0.75 ^{+0.08}_{-0.09}$ & $ 5.6 ^{+0.7}_{-0.5}$ & $ 0.24 ^{+0.19}_{-0.14}$ & $ 2.54 ^{+0.10}_{-0.08}$ $ \times 10^{-15}$ & $ 1.14 ^{+0.37}_{-0.28}$ \\ 
Abell~0370 &  $ 10.9 ^{+0.3}_{-0.3}$ & $ 63.6 ^{+3.3}_{-3.6}$ & $ 0.829 ^{+0.036}_{-0.037}$ & $ -0.89 ^{+0.09}_{-0.07}$ & $ 8.6 ^{+0.5}_{-0.5}$ & $ 0.46 ^{+0.11}_{-0.11}$ & $ 2.19 ^{+0.04}_{-0.04}$ $ \times 10^{-15}$ & $ 1.83 ^{+0.41}_{-0.38}$ \\ 
MS~0451.6-0305 &  $ 21.5 ^{+0.9}_{-0.9}$ & $ 36.0 ^{+1.9}_{-1.6}$ & $ 0.795 ^{+0.026}_{-0.021}$ & $ -1.45 ^{+0.08}_{-0.07}$ & $ 9.7 ^{+0.5}_{-0.6}$ & $ 0.42 ^{+0.09}_{-0.12}$ & $ 2.04 ^{+0.03}_{-0.04}$ $ \times 10^{-15}$ & $ 1.47 ^{+0.27}_{-0.23}$ \\ 
MACS~J0647.7+7015 & $ 25.8 ^{+3.0}_{-2.3}$ & $ 22.0 ^{+2.1}_{-2.2}$ & $ 0.654 ^{+0.029}_{-0.027}$ & $ -1.36 ^{+0.14}_{-0.13}$ & $ 12.5 ^{+1.4}_{-1.2}$ & $ 0.18 ^{+0.11}_{-0.09}$ & $ 2.57 ^{+0.04}_{-0.04}$ $ \times 10^{-15}$ & $ 0.73 ^{+0.20}_{-0.17}$ \\ 
Abell~0586 &  $ 9.1 ^{+1.0}_{-0.7}$ & $ 47.3 ^{+3.9}_{-4.5}$ & $ 0.737 ^{+0.031}_{-0.033}$ & $ -0.65 ^{+0.09}_{-0.08}$ & $ 6.3 ^{+0.4}_{-0.3}$ & $ 0.59 ^{+0.20}_{-0.14}$ & $ 2.92 ^{+0.12}_{-0.08}$ $ \times 10^{-15}$ & $ 0.74 ^{+0.17}_{-0.22}$ \\ 
MACS~J0744.8+3927 & $ 17.0 ^{+1.5}_{-1.4}$ & $ 26.2 ^{+2.9}_{-2.2}$ & $ 0.733 ^{+0.048}_{-0.035}$ & $ -1.28 ^{+0.13}_{-0.15}$ & $ 8.1 ^{+0.5}_{-0.6}$ & $ 0.39 ^{+0.11}_{-0.11}$ & $ 2.35 ^{+0.06}_{-0.06}$ $ \times 10^{-15}$ & $ 1.83 ^{+0.43}_{-0.41}$ \\ 
Abell~0611  & $ 31.1 ^{+4.4}_{-3.6}$ & $ 23.9 ^{+2.4}_{-2.2}$ & $ 0.618 ^{+0.017}_{-0.015}$ & $ -0.77 ^{+0.08}_{-0.08}$ & $ 6.9 ^{+0.4}_{-0.4}$ & $ 0.39 ^{+0.11}_{-0.10}$ & $ 2.28 ^{+0.05}_{-0.04}$ $ \times 10^{-15}$ & $ 0.83 ^{+0.22}_{-0.19}$ \\ 
Abell~0665  & $ 29.3 ^{+1.9}_{-1.5}$ & $ 45.2 ^{+3.6}_{-3.5}$ & $ 0.567 ^{+0.020}_{-0.018}$ & $ -0.93 ^{+0.10}_{-0.10}$ & $ 7.7 ^{+0.4}_{-0.4}$ & $ 0.44 ^{+0.10}_{-0.08}$ & $ 3.00 ^{+0.05}_{-0.04}$ $ \times 10^{-15}$ & $ 0.76 ^{+0.16}_{-0.15}$ \\ 
Abell~0697  & $ 14.0 ^{+0.6}_{-0.6}$ & $ 43.2 ^{+2.1}_{-2.0}$ & $ 0.607 ^{+0.012}_{-0.013}$ & $ -1.22 ^{+0.12}_{-0.13}$  & $ 10.0 ^{+0.7}_{-0.6}$ & $ 0.32 ^{+0.10}_{-0.09}$ & $ 3.01 ^{+0.04}_{-0.04}$ $ \times 10^{-15}$ & $ 0.77 ^{+0.21}_{-0.17}$ \\ 
Abell~0773  & $ 14.6 ^{+1.2}_{-1.1}$ & $ 37.4 ^{+3.0}_{-2.4}$ & $ 0.588 ^{+0.014}_{-0.012}$ & $ -1.13 ^{+0.12}_{-0.10}$  & $ 7.4 ^{+0.5}_{-0.4}$ & $ 0.66 ^{+0.11}_{-0.10}$ & $ 3.18 ^{+0.06}_{-0.06}$ $ \times 10^{-15}$ & $ 1.56 ^{+0.36}_{-0.35}$ \\ 
ZW~3146 &   $ 85.6 ^{+2.3}_{-2.4}$ & $ 32.4 ^{+0.8}_{-0.7}$ & $ 0.745 ^{+0.007}_{-0.008}$ & $ -1.16 ^{+0.15}_{-0.13}$ & $ 7.9 ^{+0.3}_{-0.3}$ & $ 0.27 ^{+0.05}_{-0.05}$ & $ 2.67 ^{+0.03}_{-0.02}$ $ \times 10^{-15}$ & $ 0.76 ^{+0.19}_{-0.18}$ \\ 
MS~1054-0321 &  $ 8.8 ^{+0.2}_{-0.2}$ & $ 70.5 ^{+6.5}_{-6.9}$ & $ 1.083 ^{+0.129}_{-0.132}$ & $ -1.11 ^{+0.09}_{-0.09}$  & $ 9.7 ^{+1.1}_{-0.9}$ & $ 0.12 ^{+0.08}_{-0.07}$ & $ 1.80 ^{+0.03}_{-0.03}$ $ \times 10^{-15}$ & $ 1.58 ^{+0.42}_{-0.32}$ \\ 
MS~1137.5+6625 & $ 17.8 ^{+3.0}_{-2.0}$ & $ 20.5 ^{+2.2}_{-2.8}$ & $ 0.833 ^{+0.057}_{-0.060}$ & $ -0.80 ^{+0.10}_{-0.10}$ & $ 4.5 ^{+0.6}_{-0.5}$ & $ 0.79 ^{+0.44}_{-0.32}$ & $ 2.06 ^{+0.19}_{-0.15}$ $ \times 10^{-15}$ & $ 5.07 ^{+1.96}_{-1.43}$ \\ 
MACS~J1149.5+2223  & $ 9.5 ^{+0.4}_{-0.4}$ & $ 47.2 ^{+4.1}_{-2.7}$ & $ 0.695 ^{+0.040}_{-0.024}$ & $ -1.14 ^{+0.13}_{-0.12}$ & $ 8.7 ^{+0.5}_{-0.5}$ & $ 0.21 ^{+0.08}_{-0.08}$ & $ 2.61 ^{+0.04}_{-0.04}$ $ \times 10^{-15}$ & $ 1.56 ^{+0.40}_{-0.32}$ \\ 
Abell~1413 &  $ 25.2 ^{+2.9}_{-2.5}$ & $ 36.4 ^{+3.7}_{-3.4}$ & $ 0.532 ^{+0.015}_{-0.013}$ & $ -1.03 ^{+0.14}_{-0.14}$ & $ 7.5 ^{+0.4}_{-0.3}$ & $ 0.37 ^{+0.06}_{-0.06}$ & $ 3.05 ^{+0.03}_{-0.03}$ $ \times 10^{-15}$ & $ 0.62 ^{+0.19}_{-0.16}$ \\ 
CL~J1226.9+3332 &  $ 21.9 ^{+4.4}_{-4.1}$ & $ 16.4 ^{+3.9}_{-2.1}$ & $ 0.734 ^{+0.082}_{-0.042}$ & $ -1.69 ^{+0.19}_{-0.16}$ & $ 14.0 ^{+2.1}_{-1.8}$ & $ 0.17 ^{+0.13}_{-0.10}$ & $ 2.45 ^{+0.05}_{-0.05}$ $ \times 10^{-15}$ & $ 0.81 ^{+0.28}_{-0.22}$ \\ 
MACS~J1311.0-0310  & $ 65.5 ^{+35.8}_{-35.3}$ & $ 7.43 ^{+3.2}_{-1.2}$ & $ 0.633 ^{+0.029}_{-0.022}$ & $ -1.53 ^{+0.26}_{-0.25}$ & $ 6.8 ^{+1.4}_{-1.0}$ & $ 0.38 ^{+0.16}_{-0.19}$ & $ 2.67 ^{+0.11}_{-0.14}$ $ \times 10^{-15}$ & $ 1.50 ^{+0.76}_{-0.50}$ \\ 
Abell~1689 &  $ 36.1 ^{+1.4}_{-1.3}$ & $ 48.0 ^{+1.5}_{-1.7}$ & $ 0.686 ^{+0.010}_{-0.010}$ & $ -1.66 ^{+0.13}_{-0.14}$ & $ 10.1 ^{+0.5}_{-0.6}$ & $ 0.29 ^{+0.08}_{-0.10}$ & $ 2.96 ^{+0.03}_{-0.04}$ $ \times 10^{-15}$ & $ 0.90 ^{+0.16}_{-0.19}$ \\ 
RX~J1347.5-1145  & $236.2 ^{+11.7}_{-13.4}$  & $ 17.2 ^{+0.6}_{-0.6}$ & $ 0.633 ^{+0.005}_{-0.005}$ & $ -2.75 ^{+0.28}_{-0.30}$ & $ 16.1 ^{+1.0}_{-0.9}$ & $ 0.32 ^{+0.08}_{-0.09}$ & $ 2.79 ^{+0.03}_{-0.03}$ $ \times 10^{-15}$ & $ 0.51 ^{+0.12}_{-0.11}$ \\ 
MS~1358.4+6245  & $ 18.7 ^{+1.0}_{-0.9}$ & $ 31.9 ^{+1.2}_{-1.5}$ & $ 0.658 ^{+0.010}_{-0.012}$ & $ -0.69 ^{+0.10}_{-0.10}$ & $ 8.5 ^{+0.7}_{-0.6}$ & $ 0.54 ^{+0.16}_{-0.13}$ & $ 2.39 ^{+0.06}_{-0.05}$ $ \times 10^{-15}$ & $ 0.81 ^{+0.28}_{-0.23}$ \\ 
Abell~1835 &  $ 62.8 ^{+3.0}_{-2.7}$ & $ 32.4 ^{+1.4}_{-1.1}$ & $ 0.670 ^{+0.012}_{-0.009}$ & $ -1.70 ^{+0.10}_{-0.11}$ & $ 10.9 ^{+0.7}_{-0.5}$ & $ 0.38 ^{+0.09}_{-0.08}$ & $ 2.35 ^{+0.03}_{-0.03}$ $ \times 10^{-15}$ & $ 0.69 ^{+0.16}_{-0.09}$ \\ 
MACS~J1423+2404  & $ 156.5 ^{+19.2}_{-18.2}$ & $ 11.2 ^{+0.9}_{-0.7}$ & $ 0.607 ^{+0.011}_{-0.009}$ & $ -1.39 ^{+0.24}_{-0.21}$  & $ 7.4 ^{+0.4}_{-0.4}$ & $ 0.36 ^{+0.08}_{-0.08}$ & $ 2.19 ^{+0.03}_{-0.04}$ $ \times 10^{-15}$ & $ 1.71 ^{+0.65}_{-0.57}$ \\ 
Abell~1914 &  $ 78.7 ^{+2.4}_{-2.7}$ & $ 45.3 ^{+1.5}_{-1.1}$ & $ 0.742 ^{+0.011}_{-0.008}$ & $ -1.55 ^{+0.15}_{-0.13}$ & $ 9.6 ^{+0.3}_{-0.3}$ & $ 0.24 ^{+0.05}_{-0.06}$ & $ 3.10 ^{+0.03}_{-0.03}$ $ \times 10^{-15}$ & $ 0.67 ^{+0.12}_{-0.13}$ \\ 
Abell~1995 &  $ 24.9 ^{+0.4}_{-0.4}$ & $ 50.4 ^{+1.4}_{-1.5}$ & $ 0.923 ^{+0.021}_{-0.023}$ & $ -0.92 ^{+0.05}_{-0.05}$ & $ 9.1 ^{+0.5}_{-0.5}$ & $ 0.45 ^{+0.13}_{-0.11}$ & $ 2.35 ^{+0.05}_{-0.04}$ $ \times 10^{-15}$ & $ 1.20 ^{+0.21}_{-0.16}$ \\ 
Abell~2111 &  $ 2.3 ^{+0.2}_{-0.2}$ & $ 58.8 ^{+7.1}_{-6.6}$ & $ 0.648 ^{+0.043}_{-0.038}$ & $ -0.57 ^{+0.11}_{-0.11}$ & $ 8.2 ^{+0.8}_{-0.8}$ & $ 0.19 ^{+0.13}_{-0.12}$ & $ 2.76 ^{+0.06}_{-0.05}$ $ \times 10^{-15}$ & $ 0.72 ^{+0.35}_{-0.28}$ \\ 
Abell~2163\tablenotemark{a} &  $ 69.2 ^{+0.7}_{-0.7}$ & $ 78.8 ^{+0.6}_{-0.6}$ & $ 0.700$ & $ -1.55 ^{+0.15}_{-0.15}$ & $ 13.8 ^{+0.8}_{-0.7}$ & $ 0.23 ^{+0.04}_{-0.04}$ & $ 2.52 ^{+0.01}_{-0.01}$ $ \times 10^{-15}$ & $ 0.73 ^{+0.27}_{-0.22}$ \\ 
Abell~2204 &  $ 27.0 ^{+8.0}_{-4.9}$ & $ 35.9 ^{+8.4}_{-7.4}$ & $ 0.623 ^{+0.066}_{-0.040}$ & $ -1.62 ^{+0.21}_{-0.27}$ & $ 11.2 ^{+0.8}_{-0.7}$ & $ 0.46 ^{+0.15}_{-0.12}$ & $ 2.36 ^{+0.05}_{-0.040}$ $ \times 10^{-15}$ & $ 0.46 ^{+0.11}_{-0.10}$ \\ 
Abell~2218 &  $ 20.9 ^{+0.3}_{-0.3}$ & $ 70.4 ^{+1.7}_{-1.6}$ & $ 0.767 ^{+0.015}_{-0.012}$ & $ -0.87 ^{+0.07}_{-0.07}$ & $ 7.8 ^{+0.4}_{-0.4}$ & $ 0.35 ^{+0.08}_{-0.08}$ & $ 3.01 ^{+0.040}_{-0.04}$ $ \times 10^{-15}$ & $ 1.18 ^{+0.24}_{-0.22}$ \\ 
RX~J1716.4+6708 & $ 11.0 ^{+3.2}_{-1.8}$ & $ 14.7 ^{+3.5}_{-3.5}$ & $ 0.624 ^{+0.085}_{-0.070}$ & $ -0.70 ^{+0.15}_{-0.19}$ & $ 5.8 ^{+0.7}_{-0.7}$ & $ 0.87 ^{+0.35}_{-0.30}$ & $ 2.18 ^{+0.16}_{-0.11}$ $ \times 10^{-15}$ & $ 2.09 ^{+1.07}_{-0.80}$ \\ 
Abell~2259 &  $ 5.4 ^{+0.5}_{-0.3}$ & $ 48.1 ^{+3.0}_{-4.5}$ & $ 0.611 ^{+0.022}_{-0.026}$ & $ -0.45 ^{+0.14}_{-0.14}$ & $ 5.7 ^{+0.5}_{-0.4}$ & $ 0.35 ^{+0.12}_{-0.12}$ & $ 3.01 ^{+0.09}_{-0.08}$ $ \times 10^{-15}$ & $ 0.51 ^{+0.37}_{-0.29}$ \\ 
Abell~2261 &  $ 14.4 ^{+2.2}_{-2.3}$ & $ 29.2 ^{+4.8}_{-2.9}$ & $ 0.628 ^{+0.030}_{-0.020}$ & $ -1.18 ^{+0.12}_{-0.12}$  & $ 7.9 ^{+0.8}_{-1.1}$ & $ 0.55 ^{+0.42}_{-0.29}$ & $ 2.80 ^{+0.21}_{-0.12}$ $ \times 10^{-15}$ & $ 0.95 ^{+0.30}_{-0.26}$ \\ 
MS~2053.7-0449 &  $ 9.4 ^{+1.6}_{-1.1}$ & $ 24.4 ^{+3.1}_{-3.2}$ & $ 0.775 ^{+0.050}_{-0.055}$ & $ -0.44 ^{+0.07}_{-0.08}$ & $ 4.5 ^{+0.6}_{-0.5}$ & $ 0.46 ^{+0.24}_{-0.22}$ & $ 2.14 ^{+0.12}_{-0.12}$ $ \times 10^{-15}$ & $ 3.58 ^{+1.62}_{-1.24}$ \\ 
MACS~J2129.4-0741  & $ 41.1 ^{+10.6}_{-7.2}$ & $ 14.2 ^{+1.9}_{-2.0}$ & $ 0.605 ^{+0.020}_{-0.018}$ & $ -1.40 ^{+0.19}_{-0.15}$ & $ 8.3 ^{+0.8}_{-0.7}$ & $ 0.65 ^{+0.18}_{-0.14}$ & $ 2.70 ^{+0.08}_{-0.06}$ $ \times 10^{-15}$ & $ 1.22 ^{+0.34}_{-0.28}$ \\ 
RX~J2129.7+0005  & $ 13.3 ^{+2.3}_{-1.9}$ & $ 26.9 ^{+2.9}_{-3.2}$ & $ 0.617 ^{+0.017}_{-0.017}$ & $ -0.75 ^{+0.10}_{-0.11}$  & $ 6.9 ^{+0.7}_{-0.7}$ & $ 0.46 ^{+0.23}_{-0.18}$ & $ 2.83 ^{+0.13}_{-0.10}$ $ \times 10^{-15}$ & $ 0.56 ^{+0.21}_{-0.16}$ \\ 
MACS~J2214.9-1359  & $ 19.3 ^{+2.2}_{-1.9}$ & $ 30.5 ^{+3.1}_{-3.3}$ & $ 0.700 ^{+0.038}_{-0.038}$ & $ -1.45 ^{+0.12}_{-0.13}$ & $ 9.9 ^{+1.1}_{-0.7}$ & $ 0.29 ^{+0.13}_{-0.13}$ & $ 2.84 ^{+0.06}_{-0.06}$ $ \times 10^{-15}$ & $ 1.86 ^{+0.42}_{-0.34}$ \\ 
MACS~J2228.5+2036  & $ 18.2 ^{+3.5}_{-2.8}$ & $ 17.8 ^{+3.1}_{-2.6}$ & $ 0.532 ^{+0.024}_{-0.022}$ & $ -1.75 ^{+0.21}_{-0.19}$  & $ 8.4 ^{+0.8}_{-0.8}$ & $ 0.35 ^{+0.13}_{-0.10}$ & $ 2.73 ^{+0.07}_{-0.05}$ $ \times 10^{-15}$ & $ 1.99 ^{+0.47}_{-0.44}$ \\ 
\enddata
\tablenotetext{a}{NOTE: The disturbed morphology
of Abell~2163 (see Appendix A1) required us to fix the value of the
$\beta$ parameter to a fiducial value of 0.7.} 
\end{deluxetable}

\clearpage

\begin{deluxetable}{lccccccccc}
\tablewidth{650pt}
\tablecaption{Results of the isothermal $\beta$ model of Section \ref{nocenter} \label{table:center}}
\tabletypesize{\scriptsize}
\tablehead{ Cluster   &$ S_{X0} $  &$ r_c $  &$ \beta $  &$ \Delta T_0 $  &$ kT $  & metallicity  &$ \Lambda $  &$ D_A $ \\ 
 &   (cnt cm$^{-2}$ arcmin$^{-2}$)   & (arcsec)   &     & (mK)   & (keV)   & (Solar)   & (cnt cm$^{3}$ s$^{-1}$)   & (Gpc)  }
\rotate
\startdata
CL~0016+1609 &  $ 27.4 ^{+0.7}_{-0.6}$ & $ 36.7 ^{+1.4}_{-1.5}$ & $ 0.686 ^{+0.017}_{-0.017}$ & $ -1.44 ^{+0.09}_{-0.09}$ & $ 10.3 ^{+0.5}_{-0.5}$ & $ 0.37 ^{+0.07}_{-0.07}$ & $ 2.50 ^{+0.03}_{-0.03}$ $ \times 10^{-15}$ & $ 1.30 ^{+0.21}_{-0.19}$ \\ 
Abell~0068 &  $ 5.0 ^{+0.2}_{-0.2}$ & $ 49.6 ^{+3.6}_{-3.1}$ & $ 0.721 ^{+0.035}_{-0.029}$ & $ -0.75 ^{+0.10}_{-0.11}$ & $ 10.0 ^{+1.1}_{-0.9}$ & $ 0.42 ^{+0.19}_{-0.18}$ & $ 2.98 ^{+0.08}_{-0.08}$ $ \times 10^{-15}$ & $ 0.54 ^{+0.20}_{-0.16}$ \\ 
Abell~0267 &  $ 4.6 ^{+0.2}_{-0.2}$ & $ 41.3 ^{+2.8}_{-2.6}$ & $ 0.712 ^{+0.030}_{-0.027}$ & $ -0.70 ^{+0.08}_{-0.07}$ & $ 6.0 ^{+0.6}_{-0.5}$ & $ 0.24 ^{+0.17}_{-0.13}$ & $ 2.57 ^{+0.09}_{-0.07}$ $ \times 10^{-15}$ & $ 1.13 ^{+0.34}_{-0.27}$ \\ 
Abell~0370 &  $ 11.7 ^{+0.2}_{-0.2}$ & $ 57.0 ^{+2.5}_{-2.6}$ & $ 0.768 ^{+0.029}_{-0.027}$ & $ -0.89 ^{+0.09}_{-0.10}$ & $ 8.9 ^{+0.4}_{-0.4}$ & $ 0.47 ^{+0.09}_{-0.09}$ & $ 2.20 ^{+0.03}_{-0.03}$ $ \times 10^{-15}$ & $ 1.64 ^{+0.40}_{-0.34}$ \\ 
MS~0451.6-0305 & $ 23.3 ^{+0.5}_{-0.5}$ & $ 33.5 ^{+1.2}_{-1.2}$ & $ 0.767 ^{+0.018}_{-0.018}$ & $ -1.48 ^{+0.09}_{-0.09}$ & $ 10.4 ^{+0.6}_{-0.7}$ & $ 0.45 ^{+0.10}_{-0.10}$ & $ 2.06 ^{+0.03}_{-0.03}$ $ \times 10^{-15}$ & $ 1.26 ^{+0.22}_{-0.19}$ \\ 
MACS~J0647.7+7015 & $ 31.0 ^{+1.1}_{-1.1}$ & $ 19.7 ^{+0.8}_{-0.8}$ & $ 0.645 ^{+0.012}_{-0.012}$ & $ -1.39 ^{+0.12}_{-0.12}$ & $ 13.8 ^{+1.6}_{-1.3}$ & $ 0.25 ^{+0.14}_{-0.12}$ & $ 2.61 ^{+0.05}_{-0.05}$ $ \times 10^{-15}$ & $ 0.59 ^{+0.17}_{-0.14}$ \\ 
Abell~0586 & $ 13.7 ^{+0.5}_{-0.5}$ & $ 32.6 ^{+1.6}_{-1.5}$ & $ 0.639 ^{+0.015}_{-0.014}$ & $ -0.72 ^{+0.09}_{-0.09}$ & $ 6.6 ^{+0.4}_{-0.4}$ & $ 0.54 ^{+0.12}_{-0.12}$ & $ 2.90 ^{+0.07}_{-0.07}$ $ \times 10^{-15}$ & $ 0.61 ^{+0.18}_{-0.15}$ \\ 
MACS~J0744.8+3927 & $ 86.4 ^{+6.2}_{-5.9}$ & $ 6.7 ^{+0.5}_{-0.4}$ & $ 0.516 ^{+0.008}_{-0.007}$ & $ -2.17 ^{+0.23}_{-0.23}$ & $ 8.2 ^{+0.6}_{-0.6}$ & $ 0.19 ^{+0.09}_{-0.09}$ & $ 2.28 ^{+0.04}_{-0.04}$ $ \times 10^{-15}$ & $ 1.64 ^{+0.41}_{-0.36}$ \\ 
Abell~0611 & $ 39.9 ^{+1.2}_{-1.1}$ & $ 19.9 ^{+0.7}_{-0.7}$ & $ 0.592 ^{+0.008}_{-0.007}$ & $ -0.80 ^{+0.09}_{-0.09}$ & $ 6.8 ^{+0.3}_{-0.3}$ & $ 0.39 ^{+0.08}_{-0.08}$ & $ 2.28 ^{+0.03}_{-0.03}$ $ \times 10^{-15}$ & $ 0.76 ^{+0.18}_{-0.17}$ \\ 
Abell~0665 & $ 23.1 ^{+0.4}_{-0.4}$ & $ 49.4 ^{+1.5}_{-1.5}$ & $ 0.536 ^{+0.008}_{-0.008}$ & $ -1.04 ^{+0.10}_{-0.10}$ & $ 7.3 ^{+0.2}_{-0.2}$ & $ 0.35 ^{+0.06}_{-0.06}$ & $ 2.95 ^{+0.03}_{-0.03}$ $ \times 10^{-15}$ & $ 1.01 ^{+0.22}_{-0.18}$ \\ 
Abell~0697 &  $ 14.2 ^{+0.4}_{-0.4}$ & $ 42.9 ^{+1.6}_{-1.5}$ & $ 0.607 ^{+0.011}_{-0.010}$ & $ -1.22 ^{+0.12}_{-0.12}$ & $ 9.9 ^{+0.6}_{-0.6}$ & $ 0.41 ^{+0.09}_{-0.09}$ & $ 3.05 ^{+0.04}_{-0.04}$ $ \times 10^{-15}$ & $ 0.77 ^{+0.20}_{-0.17}$ \\ 
Abell~0773 & $ 12.4 ^{+0.3}_{-0.3}$ & $ 43.3 ^{+1.4}_{-1.4}$ & $ 0.613 ^{+0.010}_{-0.010}$ & $ -1.08 ^{+0.11}_{-0.11}$ & $ 7.6 ^{+0.5}_{-0.4}$ & $ 0.60 ^{+0.09}_{-0.09}$ & $ 3.14 ^{+0.06}_{-0.05}$ $ \times 10^{-15}$ & $ 1.51 ^{+0.37}_{-0.32}$ \\ 
ZW~3146 &  $ 571.5 ^{+9.6}_{-9.5}$ & $ 9.0 ^{+0.1}_{-0.1}$ & $ 0.567 ^{+0.002}_{-0.002}$ & $ -2.02 ^{+0.25}_{-0.25}$ & $ 6.6 ^{+0.1}_{-0.1}$ & $ 0.38 ^{+0.03}_{-0.03}$ & $ 2.67 ^{+0.02}_{-0.02}$ $ \times 10^{-15}$ & $ 0.98 ^{+0.26}_{-0.23}$ \\ 
MS~1054-0321 & $ 8.9 ^{+0.2}_{-0.2}$ & $ 68.7 ^{+10.6}_{-8.4}$ & $ 1.045 ^{+0.213}_{-0.146}$ & $ -1.12 ^{+0.09}_{-0.11}$ & $ 9.8 ^{+1.1}_{-1.1}$ & $ 0.12 ^{+0.10}_{-0.07}$ & $ 1.81 ^{+0.03}_{-0.04}$ $ \times 10^{-15}$ & $ 1.60 ^{+0.43}_{-0.37}$ \\ 
MS~1137.5+6625 & $ 24.8 ^{+1.4}_{-1.4}$ & $ 15.5 ^{+1.2}_{-1.1}$ & $ 0.739 ^{+0.034}_{-0.031}$ & $ -0.90 ^{+0.10}_{-0.10}$ & $ 5.3 ^{+0.5}_{-0.5}$ & $ 0.54 ^{+0.26}_{-0.22}$ & $ 2.04 ^{+0.10}_{-0.09}$ $ \times 10^{-15}$ & $ 3.65 ^{+1.25}_{-0.97}$ \\ 
MACS~J1149.5+2223 & $ 10.9 ^{+0.4}_{-0.3}$ & $ 39.4 ^{+1.8}_{-1.7}$ & $ 0.633 ^{+0.015}_{-0.014}$ & $ -1.21 ^{+0.12}_{-0.12}$ & $ 9.8 ^{+0.7}_{-0.7}$ & $ 0.24 ^{+0.10}_{-0.09}$ & $ 2.66 ^{+0.05}_{-0.04}$ $ \times 10^{-15}$ & $ 1.26 ^{+0.31}_{-0.27}$ \\ 
Abell~1413 & $ 44.8 ^{+1.1}_{-1.1}$ & $ 21.1 ^{+0.7}_{-0.6}$ & $ 0.476 ^{+0.004}_{-0.004}$ & $ -1.24 ^{+0.18}_{-0.18}$ & $ 7.3 ^{+0.2}_{-0.2}$ & $ 0.46 ^{+0.07}_{-0.06}$ & $ 3.09 ^{+0.03}_{-0.03}$ $ \times 10^{-15}$ & $ 0.62 ^{+0.20}_{-0.16}$ \\ 
CL~J1226.9+3332 & $ 22.9 ^{+1.5}_{-1.4}$ & $ 15.3 ^{+1.4}_{-1.3}$ & $ 0.701 ^{+0.041}_{-0.036}$ & $ -1.73 ^{+0.17}_{-0.17}$ & $ 12.7 ^{+2.0}_{-1.6}$ & $ 0.21 ^{+0.18}_{-0.12}$ & $ 2.44 ^{+0.07}_{-0.05}$ $ \times 10^{-15}$ & $ 0.98 ^{+0.35}_{-0.27}$ \\ 
MACS~J1311.0-0310 & $ 33.7 ^{+2.4}_{-2.2}$ & $ 9.5 ^{+0.7}_{-0.7}$ & $ 0.624 ^{+0.020}_{-0.019}$ & $ -1.33 ^{+0.20}_{-0.21}$ & $ 6.4 ^{+0.6}_{-0.5}$ & $ 0.54 ^{+0.18}_{-0.19}$ & $ 2.72 ^{+0.10}_{-0.11}$ $ \times 10^{-15}$ & $ 1.97 ^{+0.75}_{-0.60}$ \\ 
Abell~1689 & $ 102.4 ^{+1.7}_{-1.8}$ & $ 20.7 ^{+0.4}_{-0.4}$ & $ 0.554 ^{+0.003}_{-0.003}$ & $ -2.06 ^{+0.17}_{-0.16}$ & $ 10.0 ^{+0.3}_{-0.3}$ & $ 0.37 ^{+0.05}_{-0.05}$ & $ 3.00 ^{+0.02}_{-0.02}$ $ \times 10^{-15}$ & $ 0.70 ^{+0.13}_{-0.11}$ \\ 
RX~J1347.5-1145 & $ 1837.0 ^{+30.0}_{-30.0}$ & $ 4.8 ^{+0.1}_{-0.1}$ & $ 0.542 ^{+0.001}_{-0.001}$ & $ -5.15 ^{+0.58}_{-0.60}$ & $ 13.5 ^{+0.5}_{-0.5}$ & $ 0.37 ^{+0.05}_{-0.05}$ & $ 2.80 ^{+0.02}_{-0.02}$ $ \times 10^{-15}$ & $ 0.76 ^{+0.20}_{-0.17}$ \\ 
MS~1358.4+6245 & $ 113.1 ^{+5.0}_{-4.5}$ & $ 6.6 ^{+0.3}_{-0.3}$ & $ 0.483 ^{+0.003}_{-0.003}$ & $ -1.36 ^{+0.18}_{-0.18}$ & $ 8.3 ^{+0.6}_{-0.6}$ & $ 0.76 ^{+0.17}_{-0.16}$ & $ 2.48 ^{+0.07}_{-0.07}$ $ \times 10^{-15}$ & $ 1.11 ^{+0.38}_{-0.30}$ \\ 
Abell~1835 & $ 524.9 ^{+7.1}_{-7.4}$ & $ 8.1 ^{+0.1}_{-0.1}$ & $ 0.543 ^{+0.001}_{-0.001}$ & $ -2.90 ^{+0.21}_{-0.20}$ & $ 8.4 ^{+0.2}_{-0.2}$ & $ 0.42 ^{+0.05}_{-0.05}$ & $ 2.36 ^{+0.02}_{-0.02}$ $ \times 10^{-15}$ & $ 0.98 ^{+0.16}_{-0.14}$ \\ 
MACS~J1423.8+2504 & $ 1219.0 ^{+12.2}_{-14.4}$ & $ 3.59 ^{+0.01}_{-0.01}$ & $ 0.550 ^{+0.001}_{-0.001}$ & $ -2.41 ^{+0.38}_{-0.41}$ & $ 5.8 ^{+0.2}_{-0.1}$ & $ 0.56 ^{+0.05}_{-0.05}$ & $ 2.19 ^{+0.02}_{-0.02}$ $ \times 10^{-15}$ & $ 2.52 ^{+0.86}_{-0.77}$ \\ 
Abell~1914 & $ 48.2 ^{+0.5}_{-0.5}$ & $ 68.8 ^{+1.1}_{-1.1}$ & $ 0.903 ^{+0.013}_{-0.012}$ & $ -1.36 ^{+0.11}_{-0.12}$ & $ 9.9 ^{+0.3}_{-0.3}$ & $ 0.25 ^{+0.05}_{-0.05}$ & $ 3.11 ^{+0.02}_{-0.02}$ $ \times 10^{-15}$ & $ 0.68 ^{+0.13}_{-0.12}$ \\ 
Abell~1995 & $ 25.0 ^{+0.4}_{-0.4}$ & $ 50.1 ^{+1.6}_{-1.5}$ & $ 0.918 ^{+0.024}_{-0.023}$ & $ -0.90 ^{+0.05}_{-0.06}$  & $ 8.7 ^{+0.4}_{-0.4}$ & $ 0.46 ^{+0.09}_{-0.09}$ & $ 2.35 ^{+0.03}_{-0.03}$ $ \times 10^{-15}$ & $ 1.30 ^{+0.21}_{-0.18}$ \\ 
Abell~2111 & $ 2.6 ^{+0.1}_{-0.1}$ & $ 51.4 ^{+4.5}_{-4.2}$ & $ 0.613 ^{+0.031}_{-0.028}$ & $ -0.59^{+0.12}_{-0.12}$ & $ 8.1 ^{+0.9}_{-0.8}$ & $ 0.14 ^{+0.12}_{-0.08}$ & $ 2.74 ^{+0.05}_{-0.04}$ $ \times 10^{-15}$ & $ 0.72 ^{+0.36}_{-0.28}$ \\ 
Abell~2163 & $ 64.6 ^{+0.5}_{-0.5}$ & $ 68.8 ^{+1.1}_{-1.0}$ & $ 0.576 ^{+0.006}_{-0.005}$ & $ -1.89 ^{+0.17}_{-0.17}$ & $ 14.8 ^{+0.4}_{-0.3}$ & $ 0.34 ^{+0.04}_{-0.04}$ & $ 2.56 ^{+0.01}_{-0.01}$ $ \times 10^{-15}$ & $ 0.42 ^{+0.08}_{-0.07}$ \\ 
Abell~2204 & $ 428.3 ^{+9.1}_{-8.9}$ & $ 5.2 ^{+0.1}_{-0.1}$ & $ 0.483 ^{+0.002}_{-0.002}$ & $ -3.22 ^{+0.30}_{-0.32}$ & $ 6.5 ^{+0.2}_{-0.2}$ & $ 0.64 ^{+0.06}_{-0.06}$ & $ 2.44 ^{+0.02}_{-0.02}$ $ \times 10^{-15}$ & $ 1.08 ^{+0.23}_{-0.20}$ \\ 
Abell~2218  & $ 20.8 ^{+0.3}_{-0.3}$ & $ 70.4 ^{+1.7}_{-1.6}$ & $ 0.766 ^{+0.014}_{-0.012}$ & $ -0.87 ^{+0.08}_{-0.08}$  & $ 8.2 ^{+0.4}_{-0.4}$ & $ 0.33 ^{+0.07}_{-0.07}$ & $ 3.01 ^{+0.04}_{-0.03}$ $ \times 10^{-15}$ & $ 1.07 ^{+0.22}_{-0.20}$ \\ 
RX~J1716.4+6708 & $ 11.4 ^{+1.3}_{-1.0}$ & $ 12.4 ^{+1.8}_{-1.8}$ & $ 0.577 ^{+0.037}_{-0.033}$ & $ -0.76 ^{+0.17}_{-0.17}$ & $ 7.7 ^{+1.2}_{-1.0}$ & $ 0.66 ^{+0.25}_{-0.25}$ & $ 2.24 ^{+0.11}_{-0.11}$ $ \times 10^{-15}$ & $ 1.31 ^{+0.75}_{-0.53}$ \\ 
Abell~2259 & $ 5.9 ^{+0.2}_{-0.2}$ & $ 42.4 ^{+3.1}_{-2.6}$ & $ 0.579 ^{+0.021}_{-0.018}$ & $ -0.46 ^{+0.16}_{-0.16}$ & $ 5.6 ^{+0.3}_{-0.3}$ & $ 0.49 ^{+0.12}_{-0.11}$ & $ 3.09 ^{+0.08}_{-0.08}$ $ \times 10^{-15}$ & $ 0.52 ^{+0.43}_{-0.30}$ \\ 
Abell~2261 &  $ 25.1 ^{+0.9}_{-0.9}$ & $ 18.4 ^{+0.8}_{-0.7}$ & $ 0.559 ^{+0.008}_{-0.008}$ & $ -1.36 ^{+0.13}_{-0.14}$ & $ 7.2 ^{+0.4}_{-0.4}$ & $ 0.44 ^{+0.13}_{-0.12}$ & $ 2.74 ^{+0.06}_{-0.05}$ $ \times 10^{-15}$ & $ 0.99 ^{+0.25}_{-0.22}$ \\ 
MS~2053.7-0449 & $ 15.0 ^{+1.2}_{-1.0}$ & $ 15.3 ^{+1.6}_{-1.4}$ & $ 0.639 ^{+0.033}_{-0.029}$ & $ -0.52 ^{+0.09}_{-0.09}$ & $ 4.7 ^{+0.5}_{-0.4}$ & $ 0.28 ^{+0.16}_{-0.14}$ & $ 2.07 ^{+0.09}_{-0.09}$ $ \times 10^{-15}$ & $ 3.11 ^{+1.27}_{-0.99}$ \\ 
MACS~J2129.4-0741 & $ 21.9 ^{+1.0}_{-0.9}$ & $ 22.0 ^{+1.3}_{-1.3}$ & $ 0.678 ^{+0.023}_{-0.021}$ & $ -1.22 ^{+0.13}_{-0.14}$ & $ 8.6 ^{+0.7}_{-0.6}$ & $ 0.69 ^{+0.13}_{-0.13}$ & $ 2.72 ^{+0.06}_{-0.06}$ $ \times 10^{-15}$ & $ 1.39 ^{+0.39}_{-0.33}$ \\ 
RX~J2129.7+0005 & $ 66.5 ^{+3.6}_{-3.5}$ & $ 8.0 ^{+0.4}_{-0.4}$ & $ 0.507 ^{+0.005}_{-0.005}$ & $ -1.21 ^{+0.19}_{-0.19}$ & $ 5.9 ^{+0.3}_{-0.3}$ & $ 0.53 ^{+0.09}_{-0.10}$ & $ 2.82 ^{+0.05}_{-0.06}$ $ \times 10^{-15}$ & $ 0.76 ^{+0.27}_{-0.21}$ \\ 
MACS~J2214.9-1359 & $ 24.9 ^{+1.0}_{-0.9}$ & $ 22.7 ^{+1.3}_{-1.2}$ & $ 0.618 ^{+0.017}_{-0.016}$ & $ -1.65 ^{+0.13}_{-0.14}$ & $ 9.8 ^{+0.8}_{-0.7}$ & $ 0.25 ^{+0.10}_{-0.10}$ & $ 2.82 ^{+0.05}_{-0.05}$ $ \times 10^{-15}$ & $ 1.97 ^{+0.43}_{-0.38}$ \\ 
MACS~J2228.5+2036 & $ 12.5 ^{+0.6}_{-0.6}$ & $ 22.4 ^{+1.3}_{-1.3}$ & $ 0.532 ^{+0.011}_{-0.011}$ & $ -1.68 ^{+0.16}_{-0.16}$  & $ 9.1 ^{+0.8}_{-0.7}$ & $ 0.41 ^{+0.12}_{-0.13}$ & $ 2.78 ^{+0.05}_{-0.06}$ $ \times 10^{-15}$ & $ 1.85 ^{+0.47}_{-0.39}$ \\ 

\enddata
\end{deluxetable}

\clearpage

\section*{Appendix 1: X-ray and SZE images}
\begin{figure}[!h]
\caption{\chandra\/ images of the background-subtracted X-ray surface brightness
in 0.7-7 keV band (color) in units of counts pixel$^{-1}$
(1.97$^{\prime\prime}$ pixels).
 Overlaid are the SZE decrement contours,
with contour levels (+1,-1,-2,-3,-4,...) times the rms noise in each image.
The
full-width at half maximum of the synthesized beams (effective psf) of
these deconvolutions are shown in the lower left-hand corners.}
\begin{center}
\vspace{-0cm}
\includegraphics[angle=-90,width=5.5in]{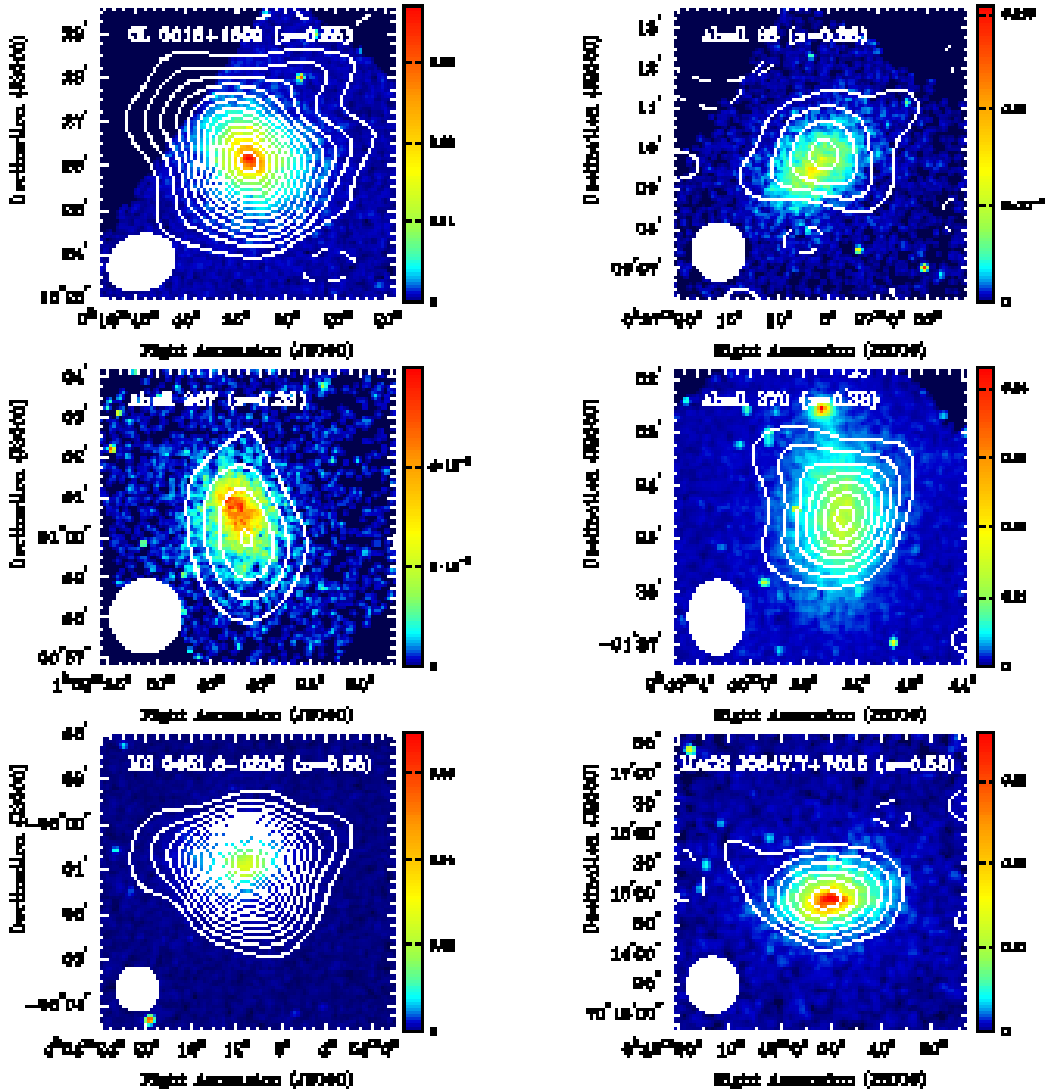}
\end{center}
\end{figure}
\clearpage
\begin{center}
\includegraphics[angle=-90,width=6in]{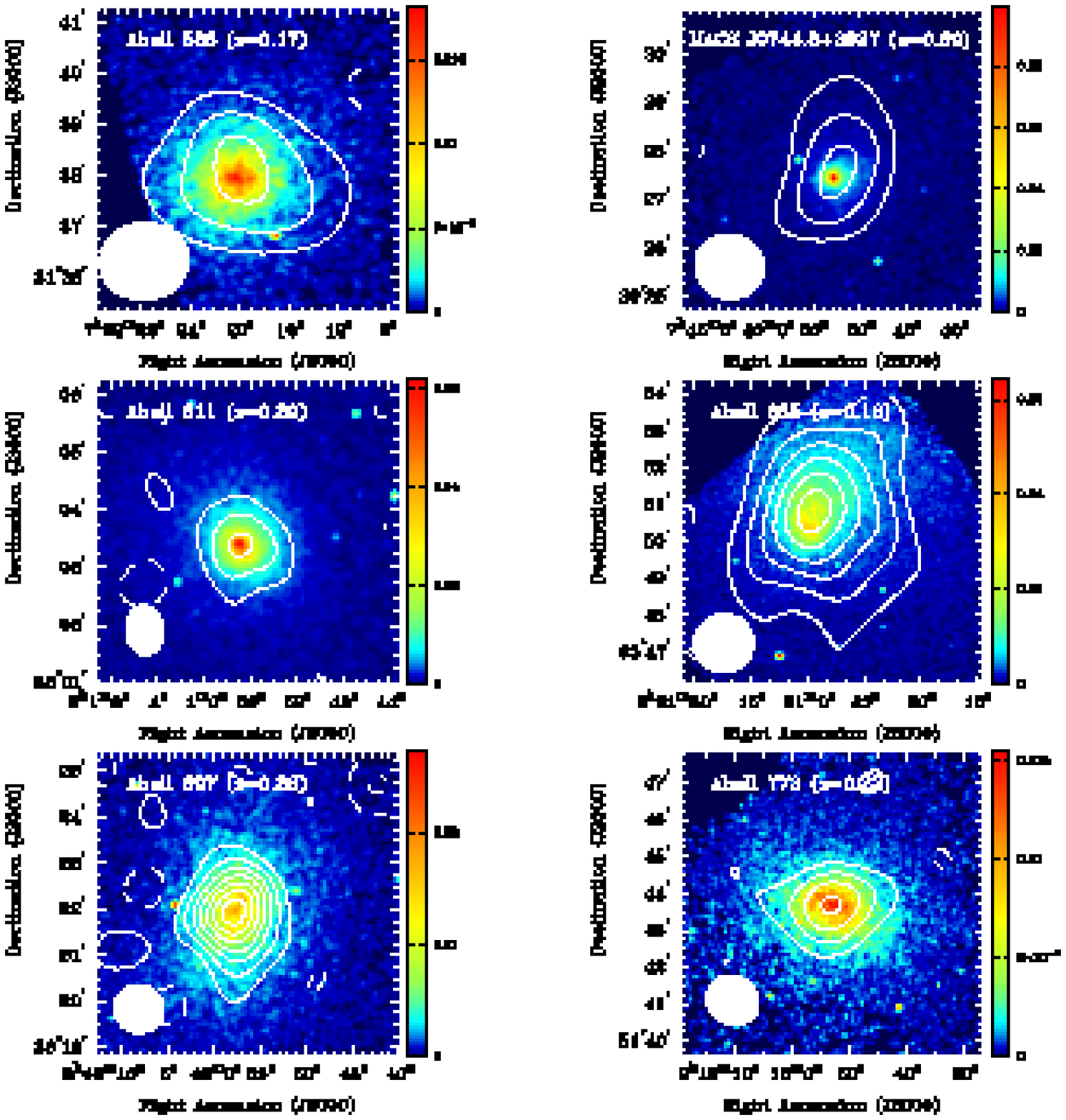}
\end{center}
\begin{center}
\includegraphics[angle=-90,width=6in]{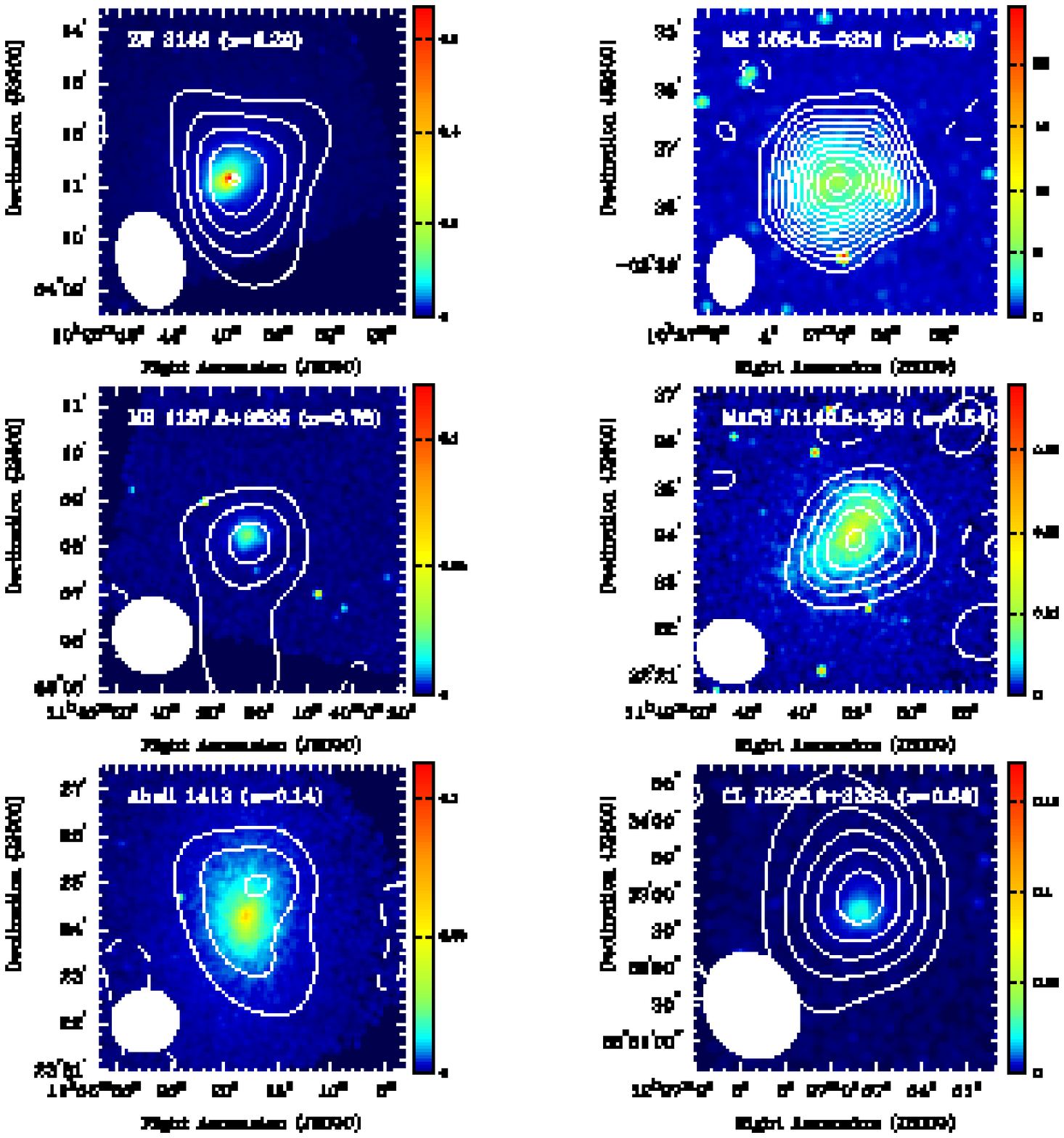}
\end{center}
\begin{center}
\includegraphics[angle=-90,width=6in]{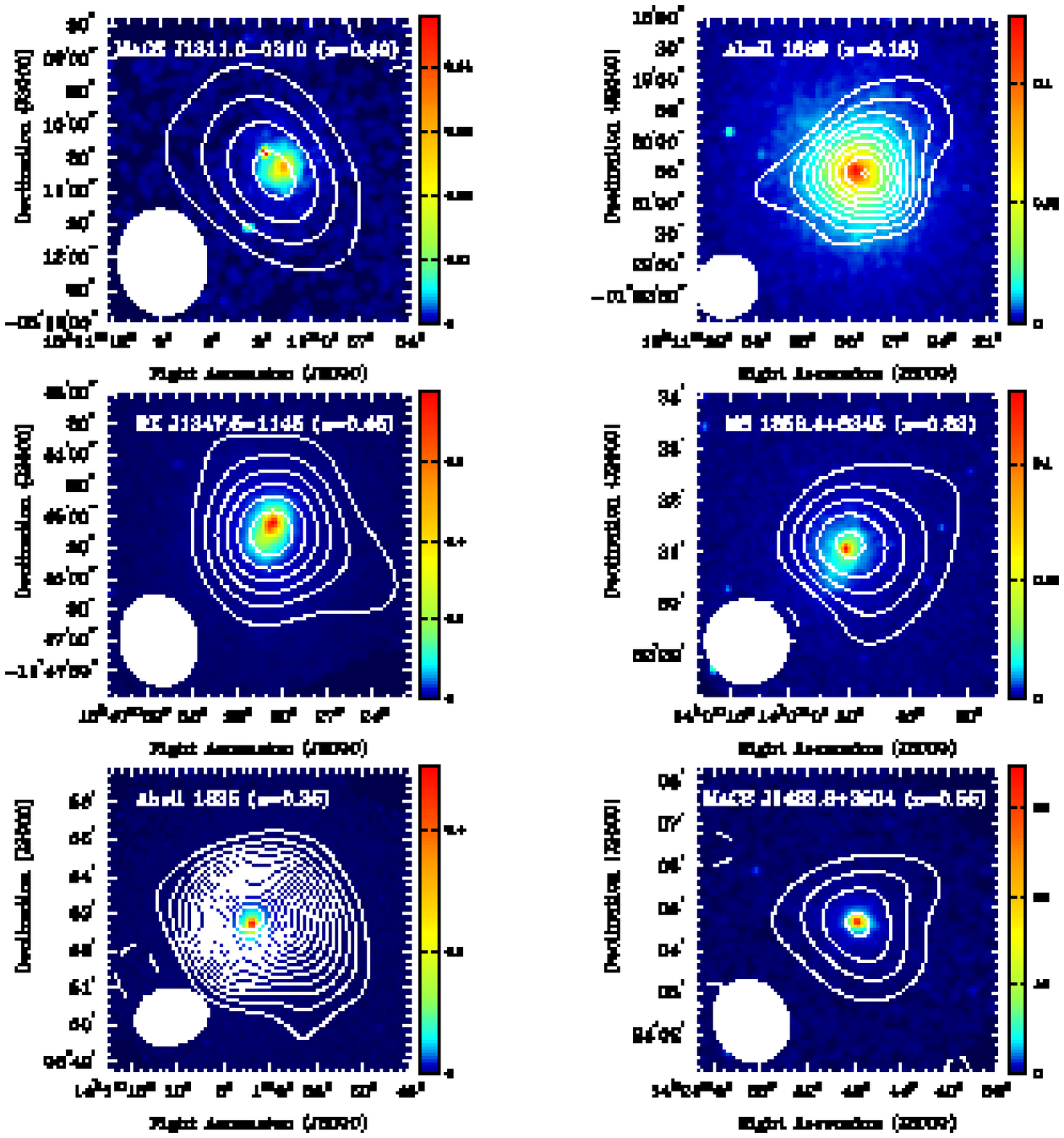}
\end{center}
\begin{center}
\includegraphics[angle=-90,width=6in]{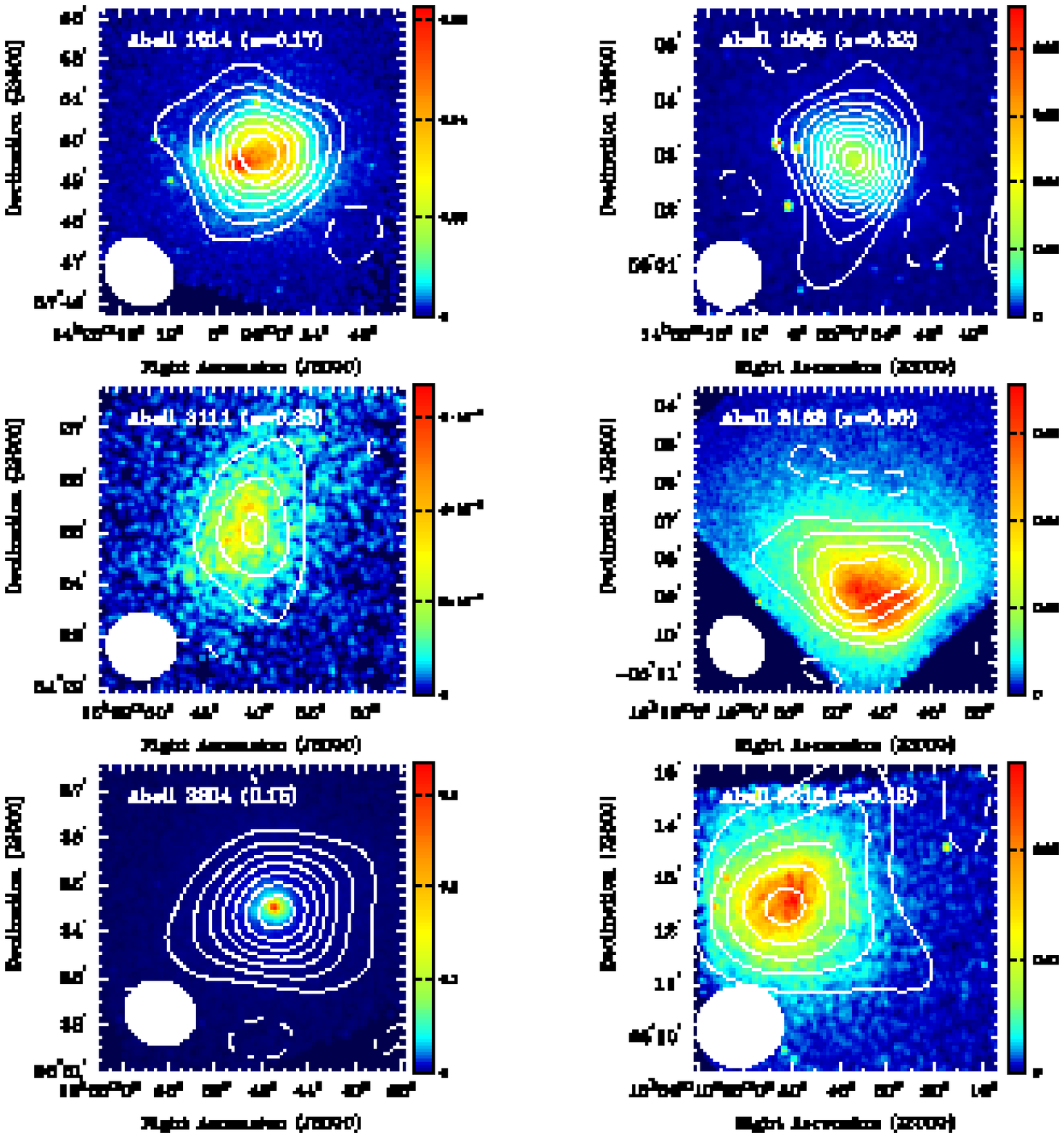}
\end{center}
\begin{center}
\includegraphics[angle=-90,width=6in]{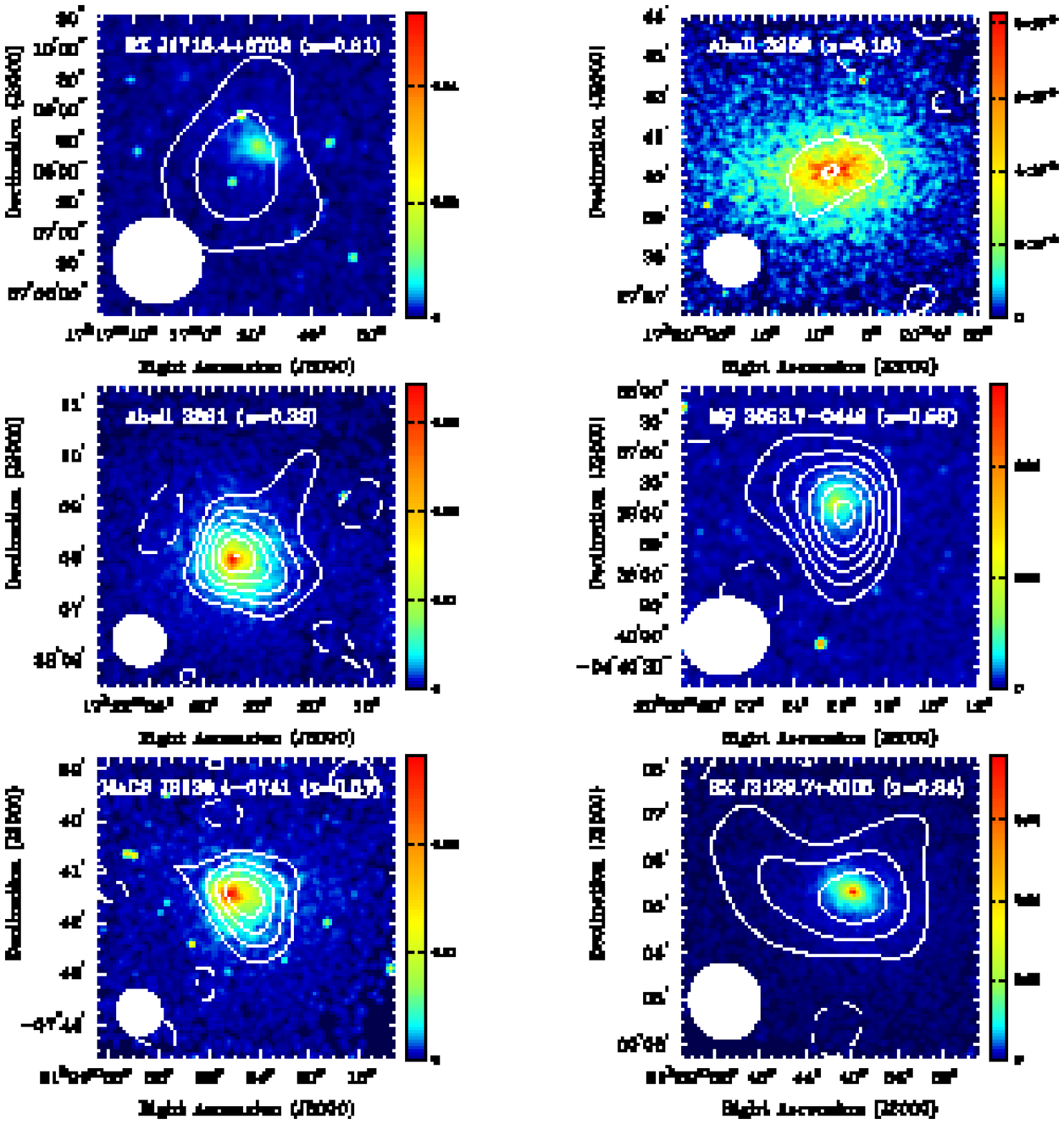}
\end{center}
\begin{center}
\includegraphics[angle=-90,width=6in]{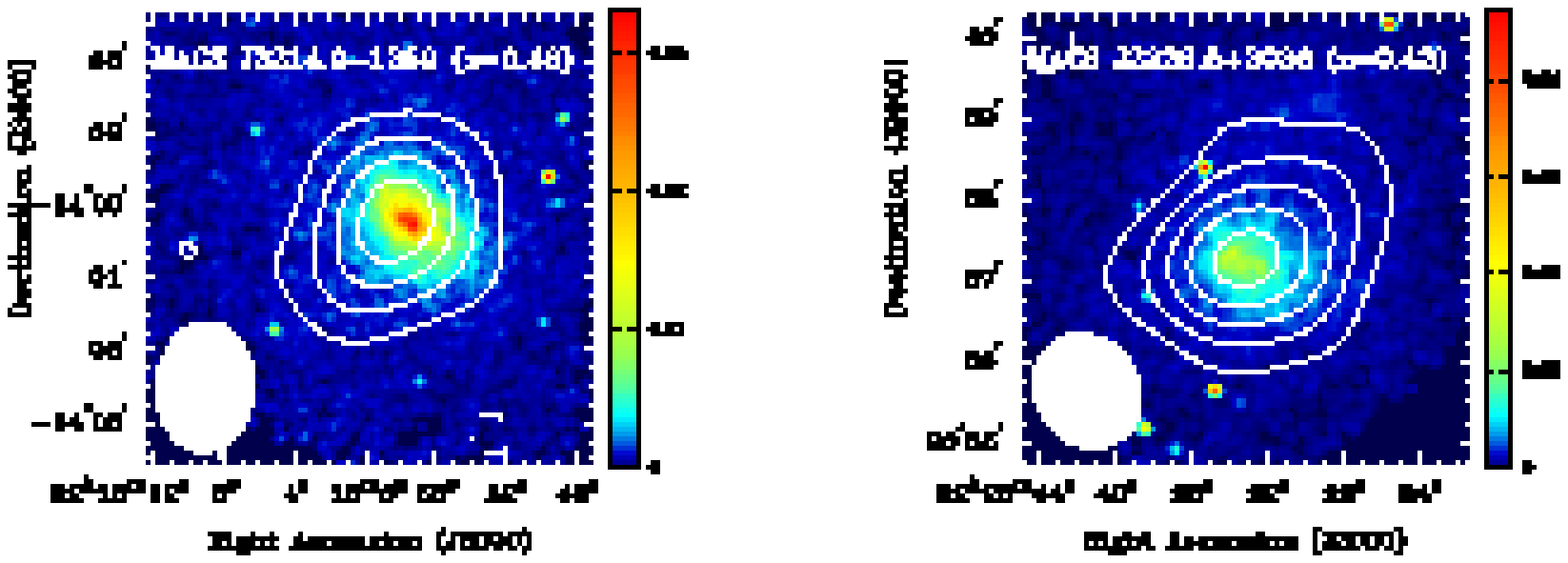}
\end{center}
\vspace{-5cm}
\clearpage

\section*{Appendix 2: Surface brightness profiles}
\begin{figure}[h!]
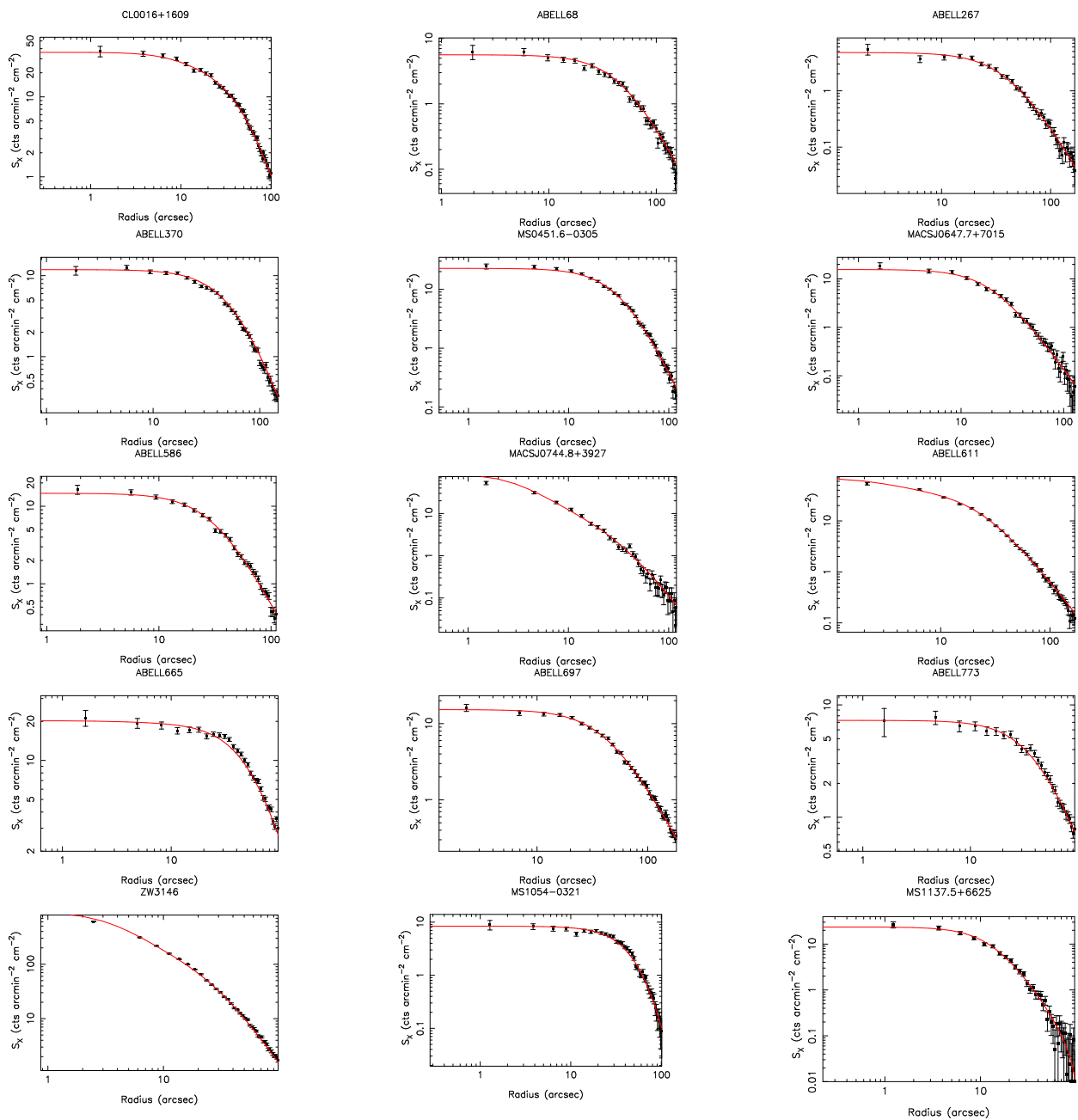

\caption{
Background subtracted X-ray surface brightness
profiles of the \nclu\/ clusters in our sample, with best-fit curves
from the hydrostatic model results.  
\label{sbprofiles}}
\includegraphics[angle=-90,width=1.65in]{f6a.eps}
\includegraphics[angle=-90,width=1.65in]{f6b.eps}
\includegraphics[angle=-90,width=1.65in]{f6c.eps}
\includegraphics[angle=-90,width=1.65in]{f6d.eps}
\includegraphics[angle=-90,width=1.65in]{f6e.eps}
\includegraphics[angle=-90,width=1.65in]{f6f.eps}
\includegraphics[angle=-90,width=1.65in]{f6g.eps}
\includegraphics[angle=-90,width=1.65in]{f6h.eps}
\includegraphics[angle=-90,width=1.65in]{f6i.eps}
\includegraphics[angle=-90,width=1.65in]{f6j.eps}
\includegraphics[angle=-90,width=1.65in]{f6k.eps}
\includegraphics[angle=-90,width=1.65in]{f6l.eps}
\includegraphics[angle=-90,width=1.65in]{f6m.eps}
\hspace{1.6cm}
\includegraphics[angle=-90,width=1.65in]{f6n.eps}
\hspace{1.6cm}
\includegraphics[angle=-90,width=1.75in]{f6o.eps}
\end{figure}

\vspace{-0cm}
In Figure \ref{sbprofiles} we show the radial profiles of the X-ray surface brightness for the 
\nclu\ clusters in our sample. 
The $\chi^2$ of the 
hydrostatic equilibrium model fits to the surface brightness profiles
are shown in Table \ref{tab:results}.

\clearpage

\begin{figure}
\includegraphics[angle=-90,width=1.75in]{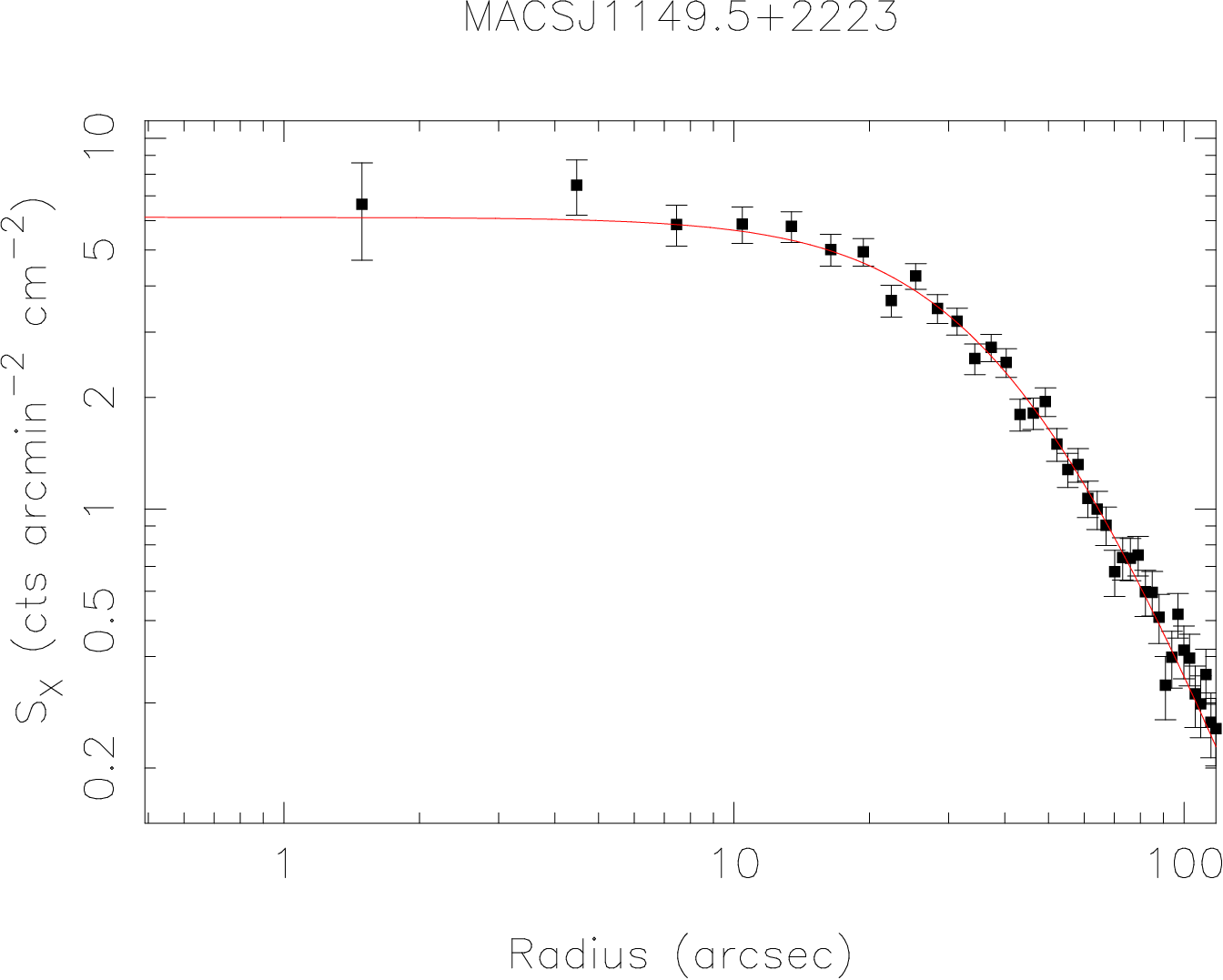}
\includegraphics[angle=-90,width=1.75in]{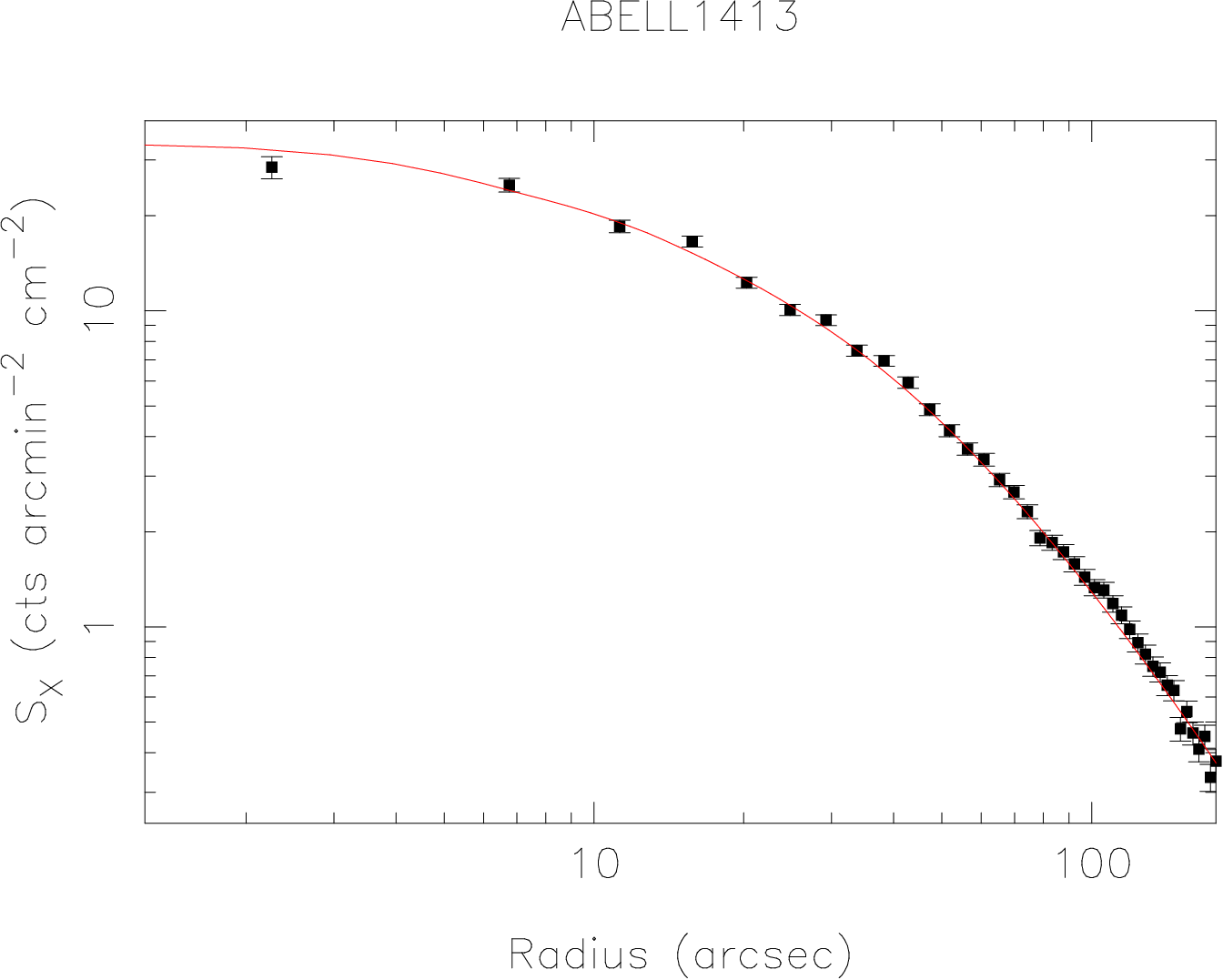}
\includegraphics[angle=-90,width=1.75in]{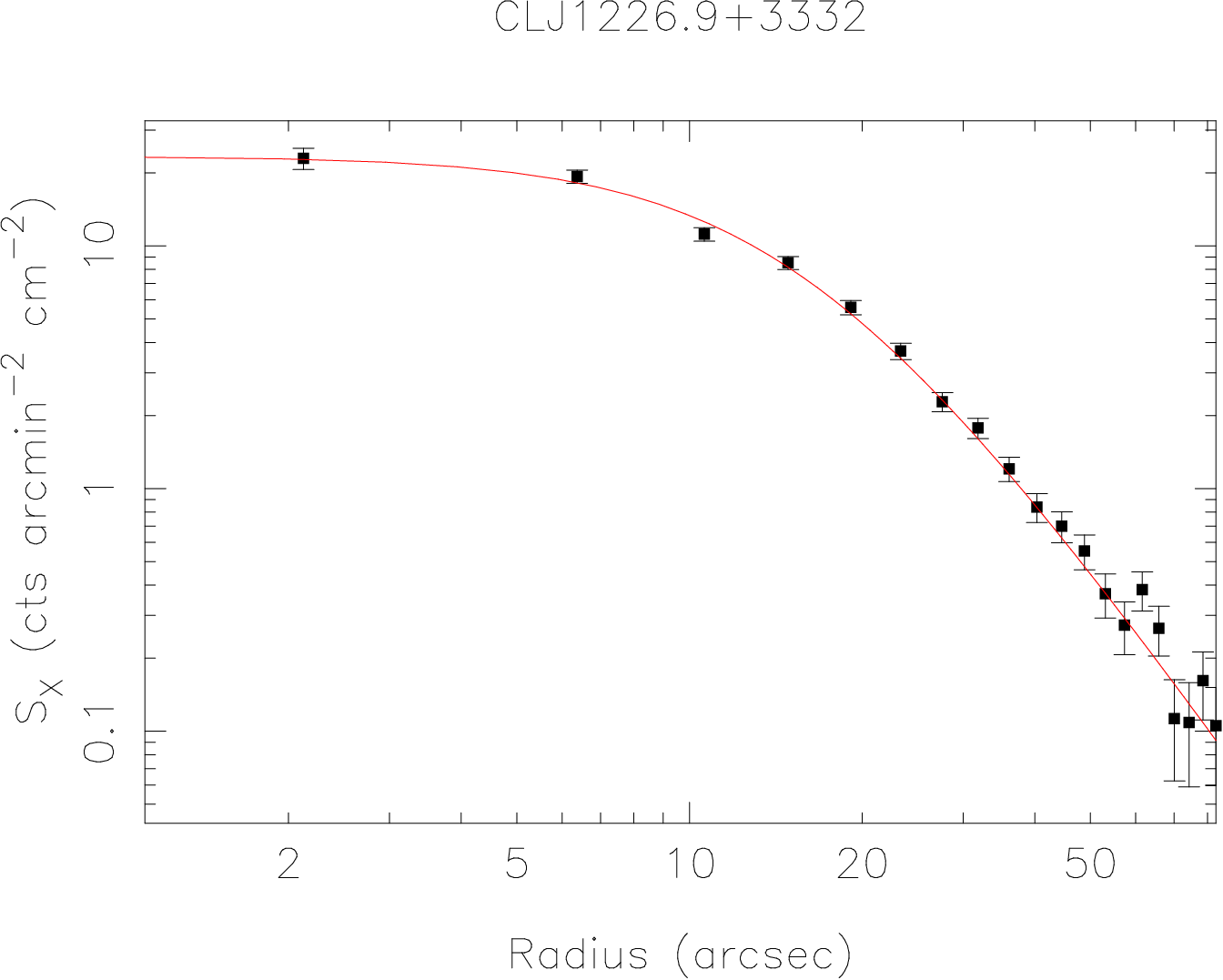}
\includegraphics[angle=-90,width=1.75in]{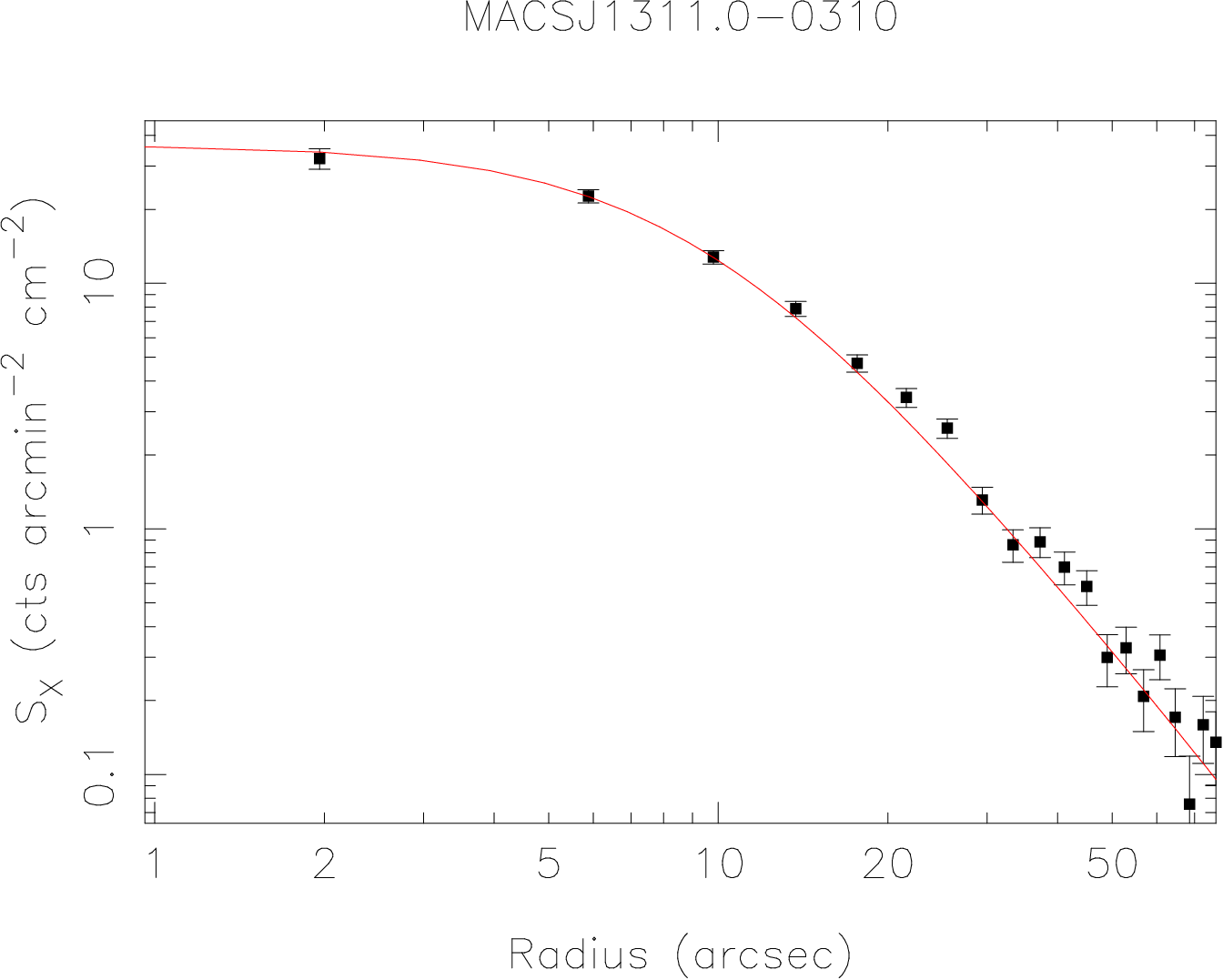}
\includegraphics[angle=-90,width=1.75in]{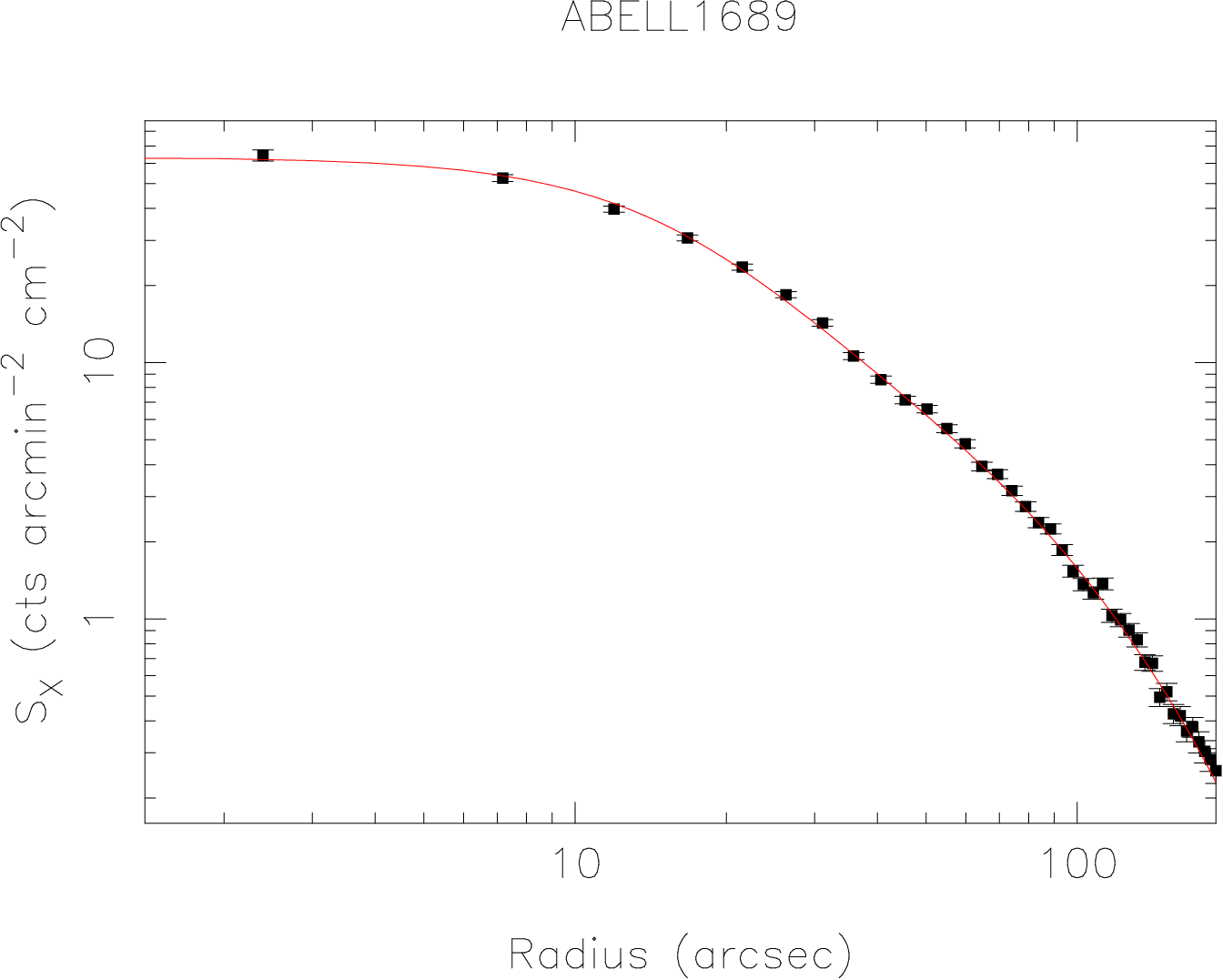}
\includegraphics[angle=-90,width=1.75in]{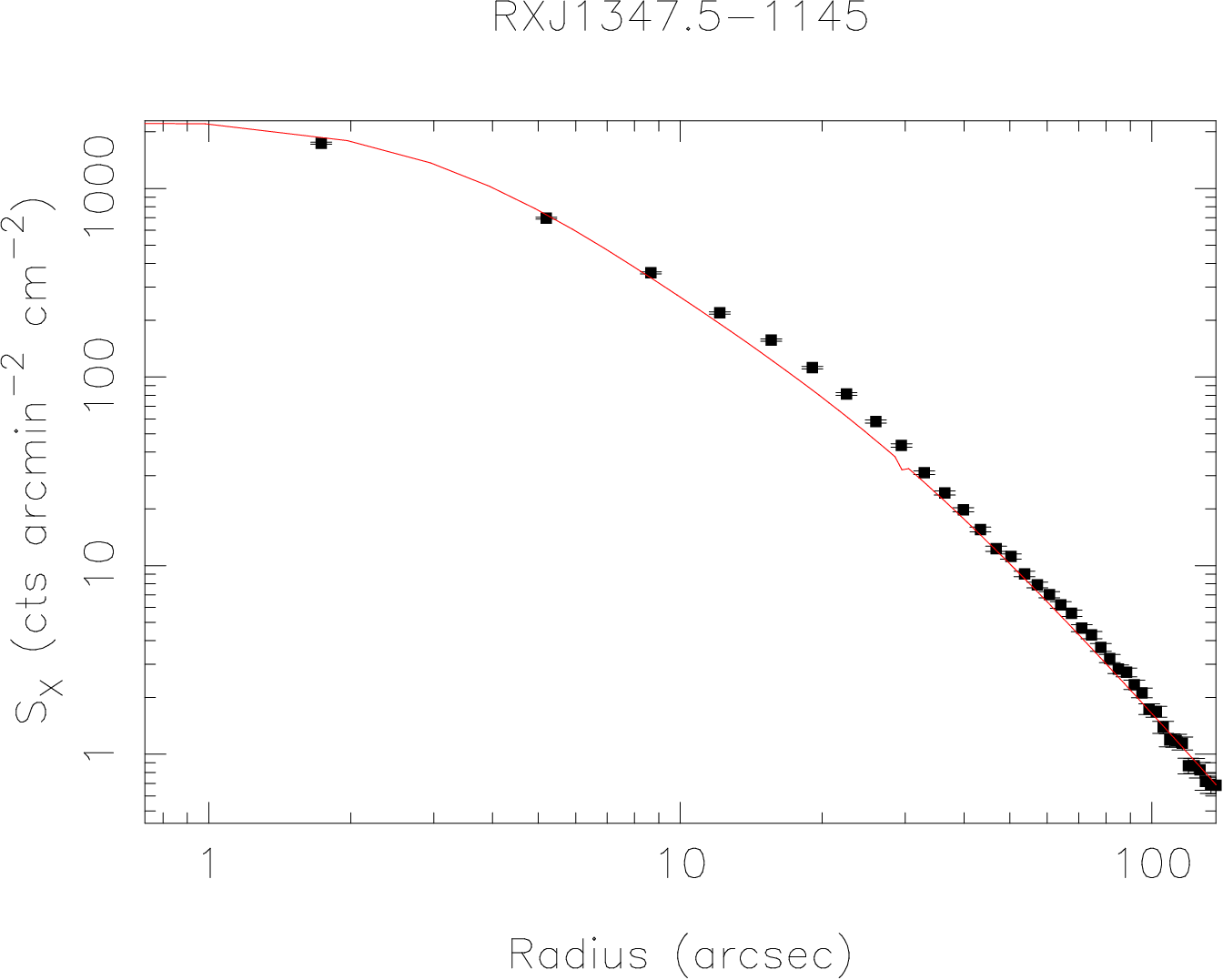}
\includegraphics[angle=-90,width=1.75in]{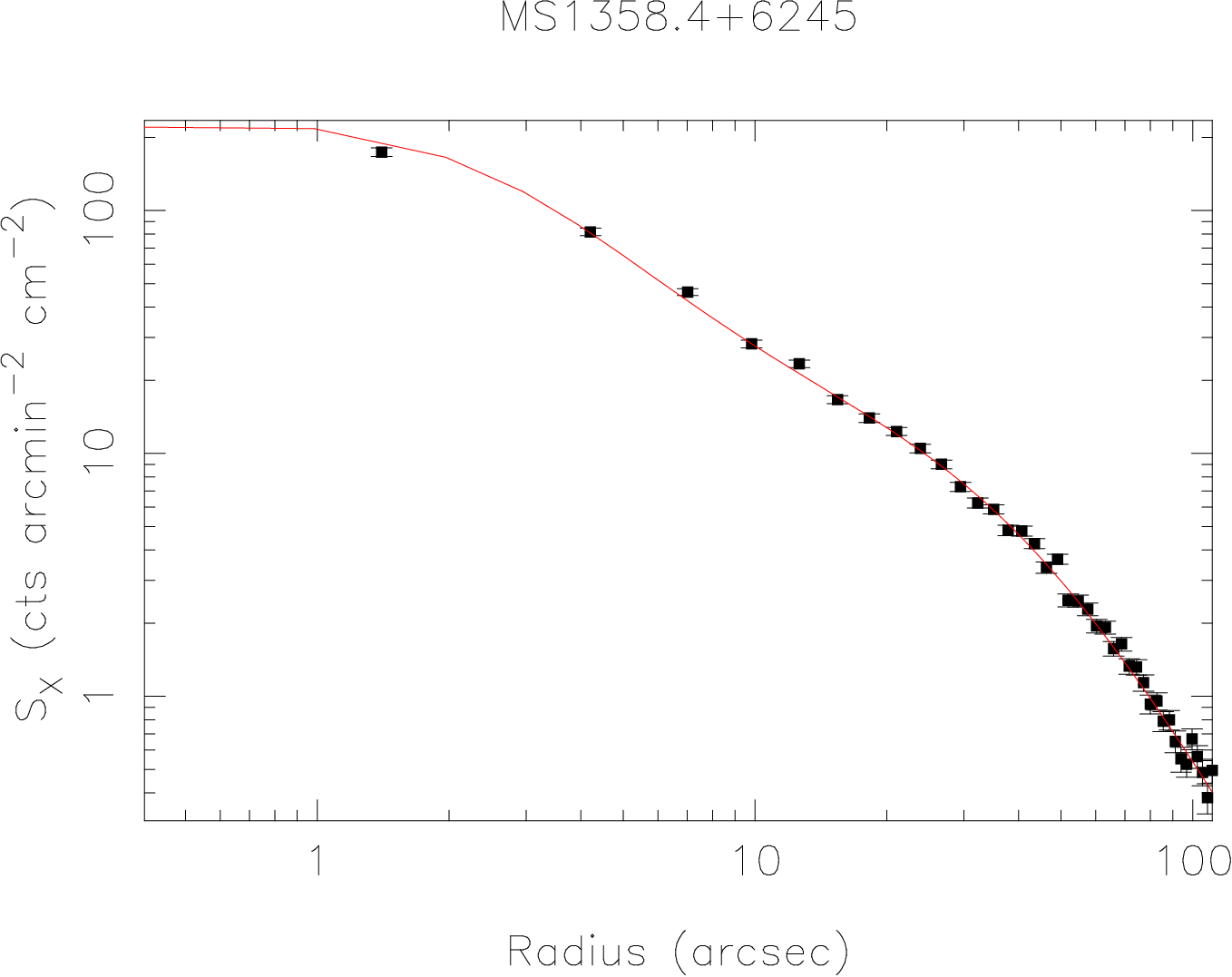}
\includegraphics[angle=-90,width=1.75in]{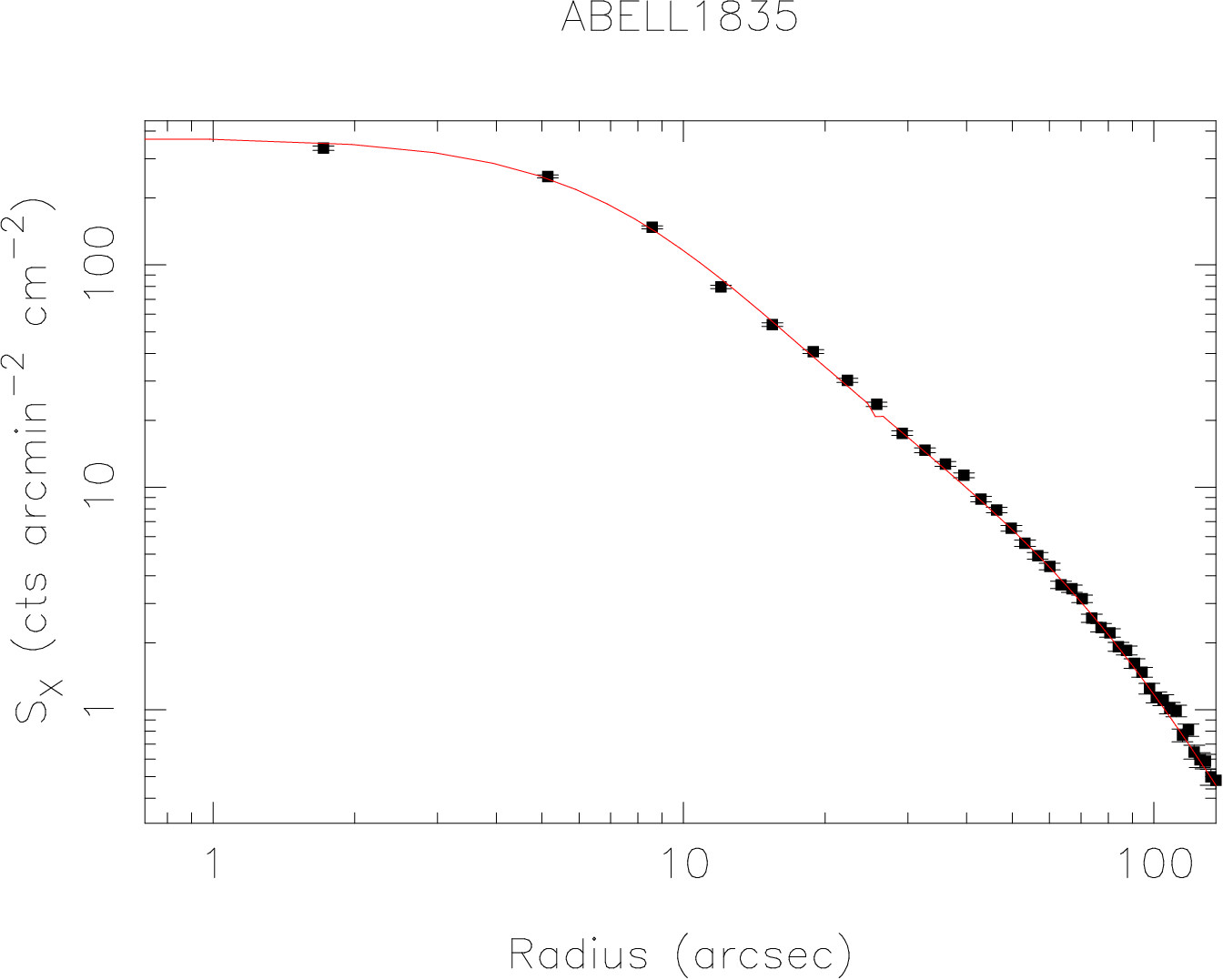}
\includegraphics[angle=-90,width=1.75in]{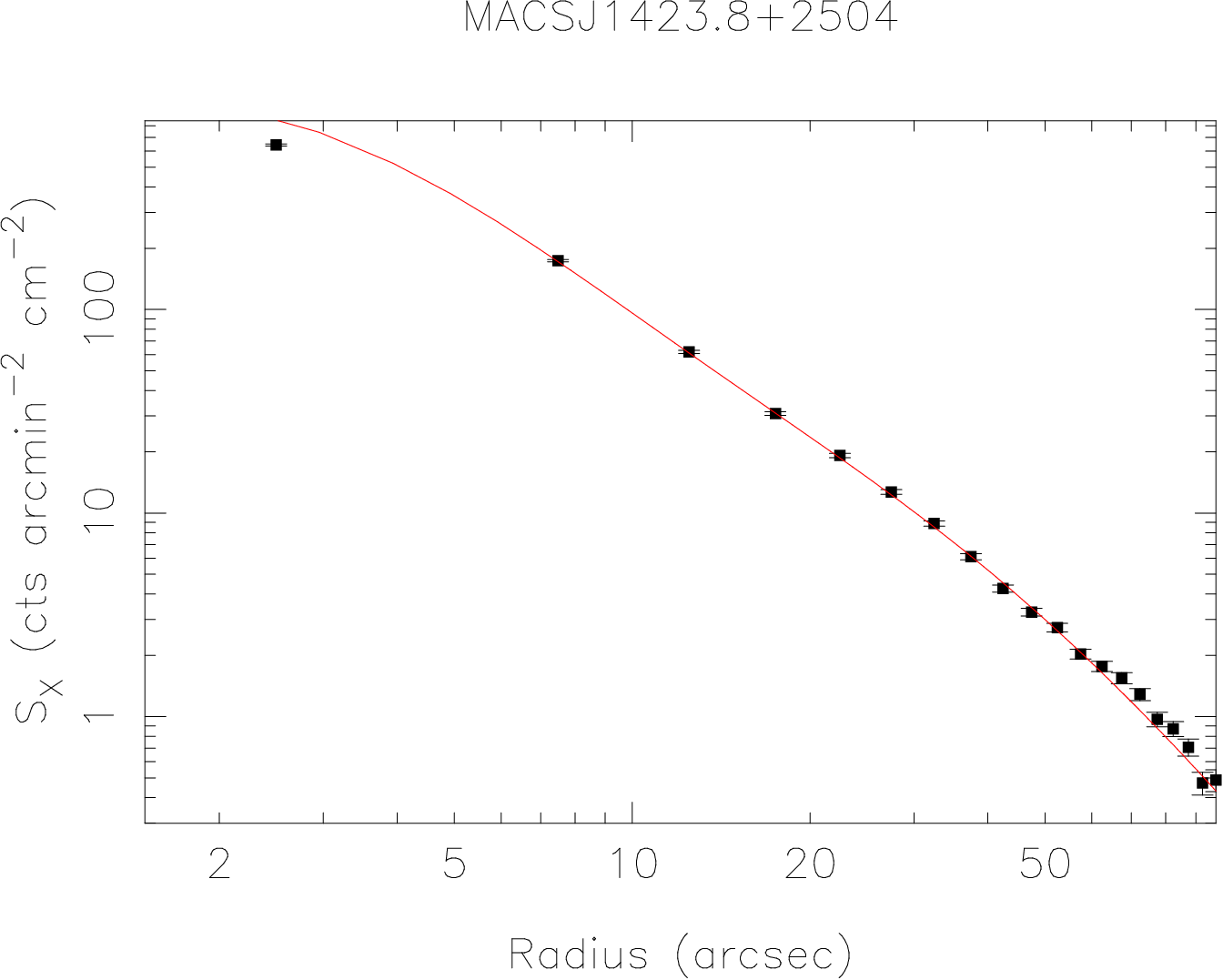}
\includegraphics[angle=-90,width=1.75in]{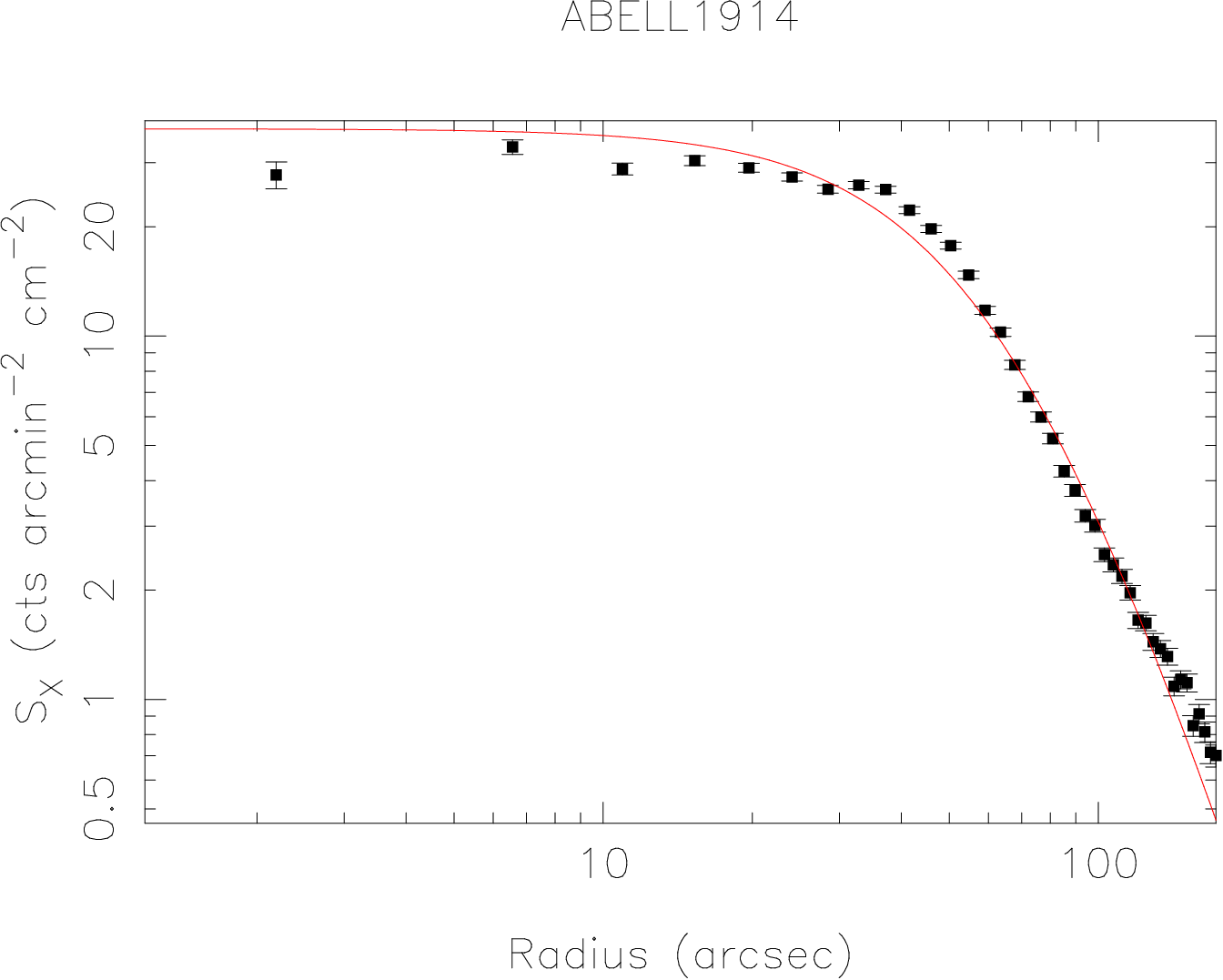}
\includegraphics[angle=-90,width=1.75in]{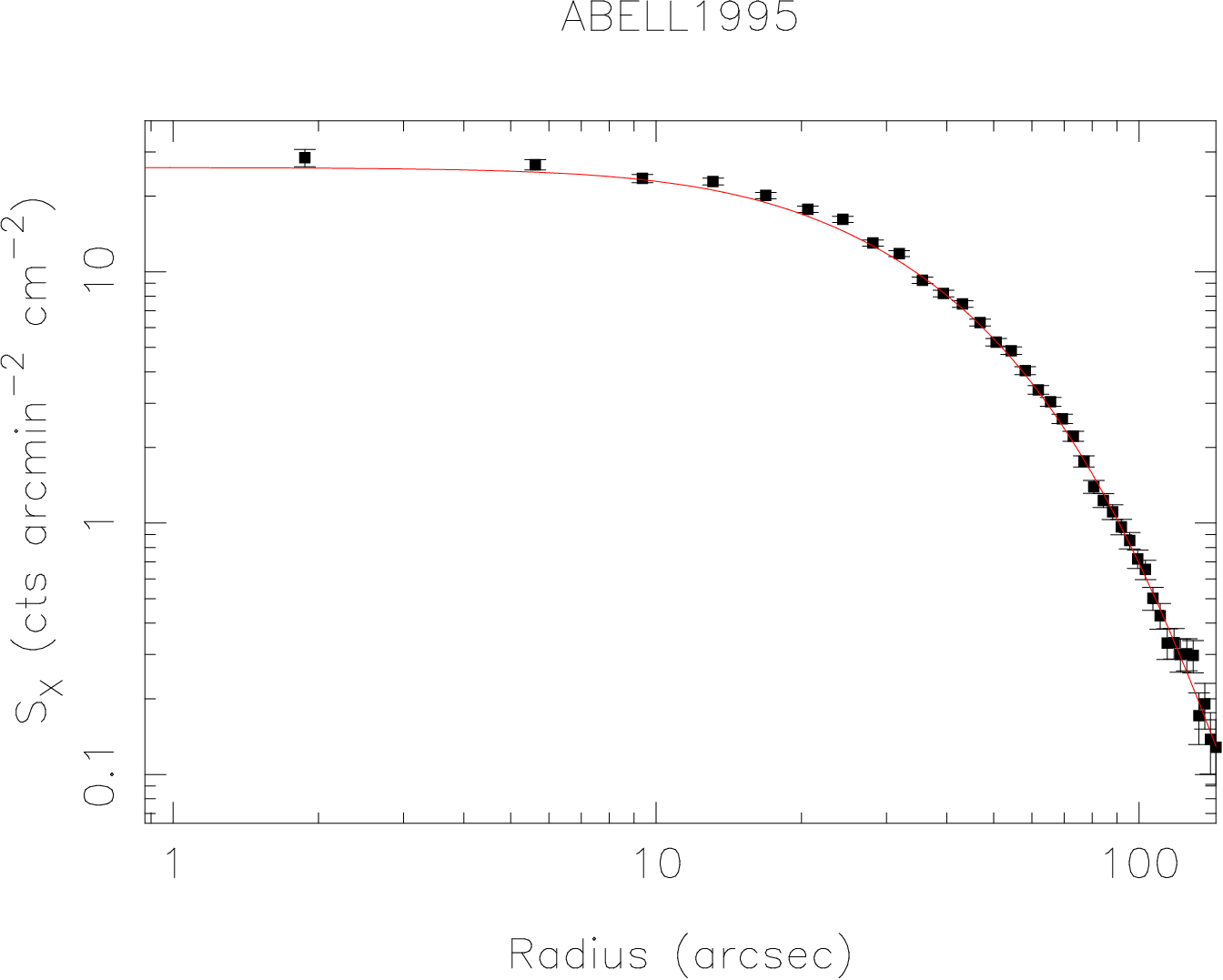}
\includegraphics[angle=-90,width=1.75in]{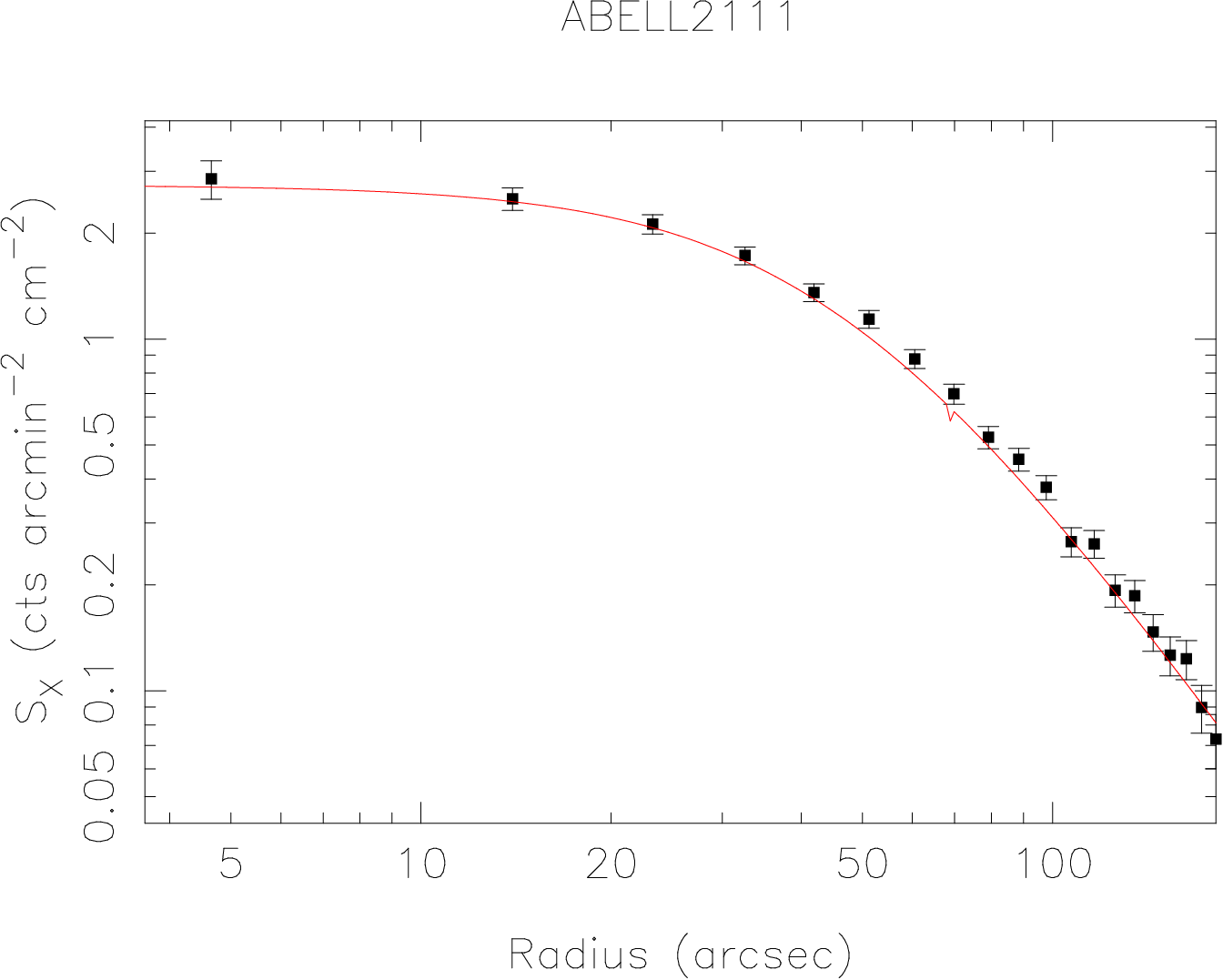}
\hspace{1.2cm}
\includegraphics[angle=-90,width=1.75in]{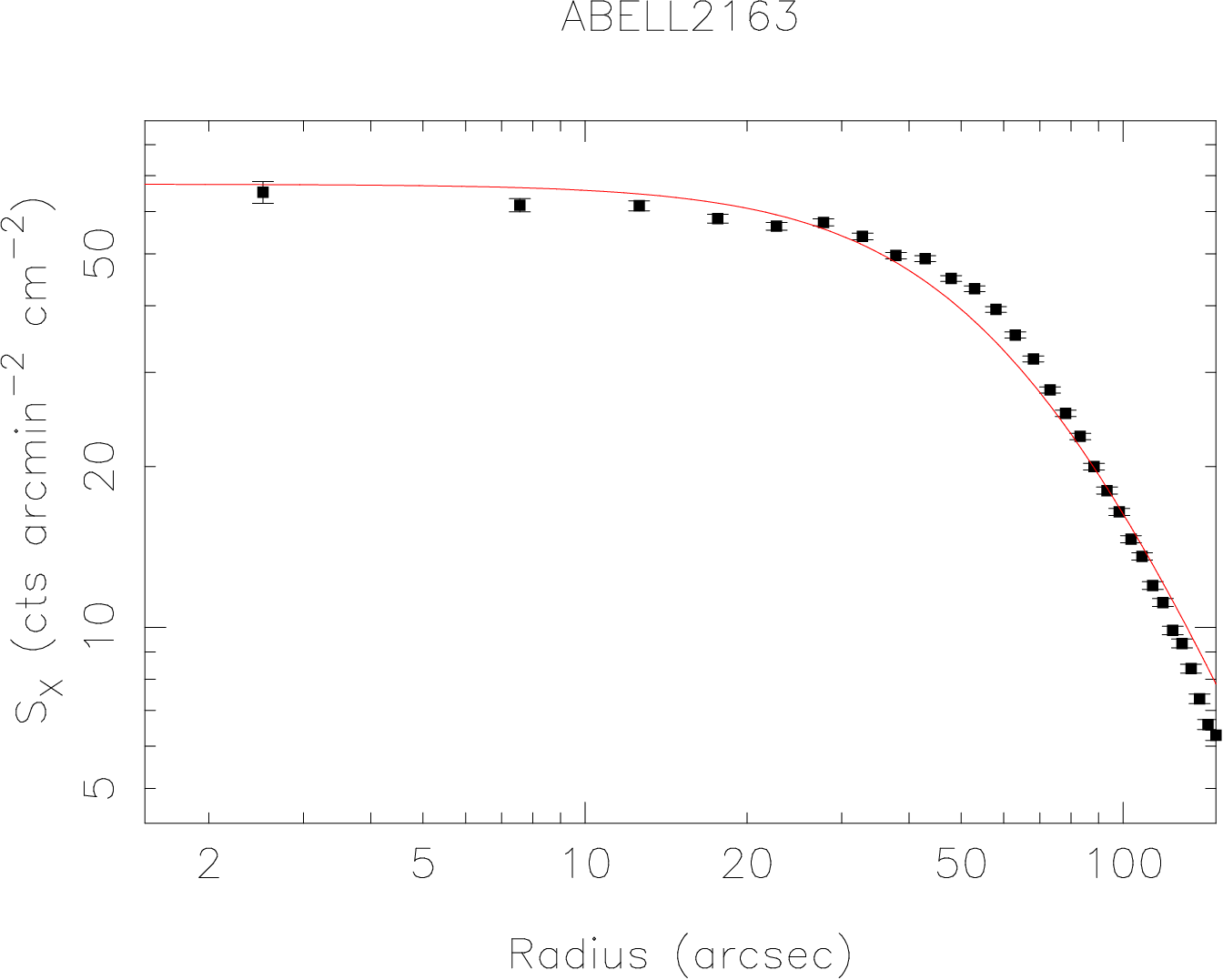}
\hspace{1.2cm}
\includegraphics[angle=-90,width=1.75in]{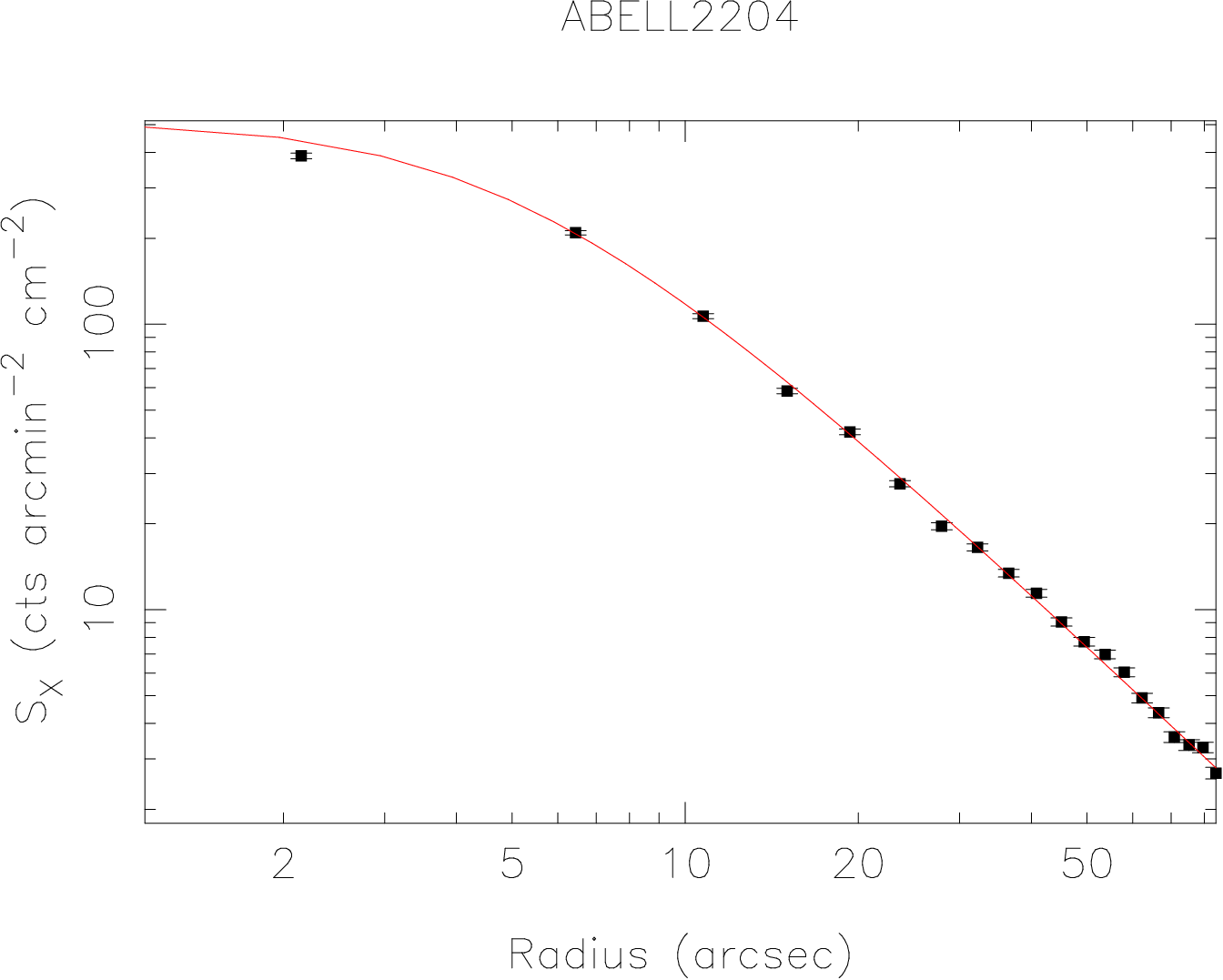}
\includegraphics[angle=-90,width=1.75in]{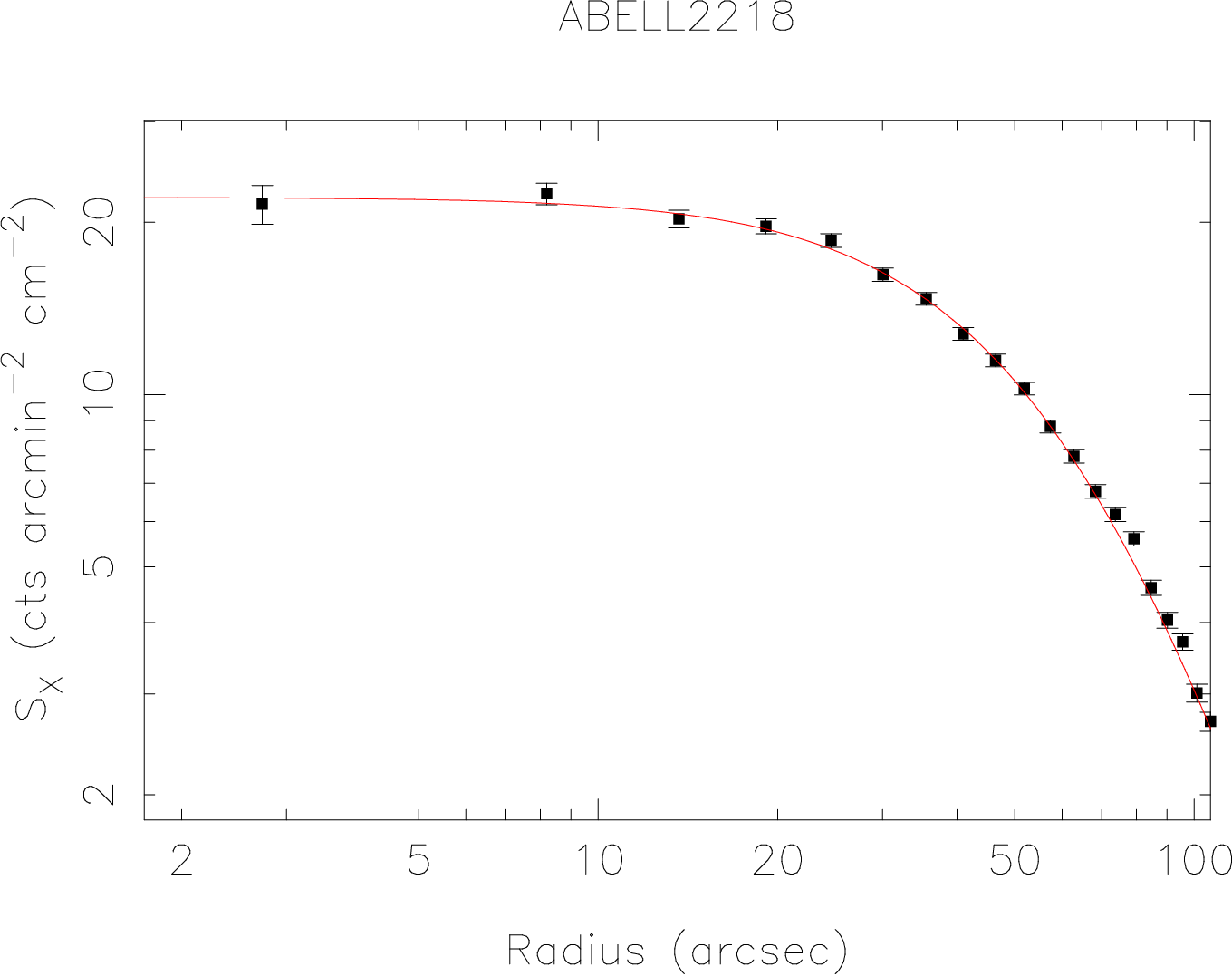}
\end{figure}

\clearpage

\begin{figure}
\includegraphics[angle=-90,width=1.75in]{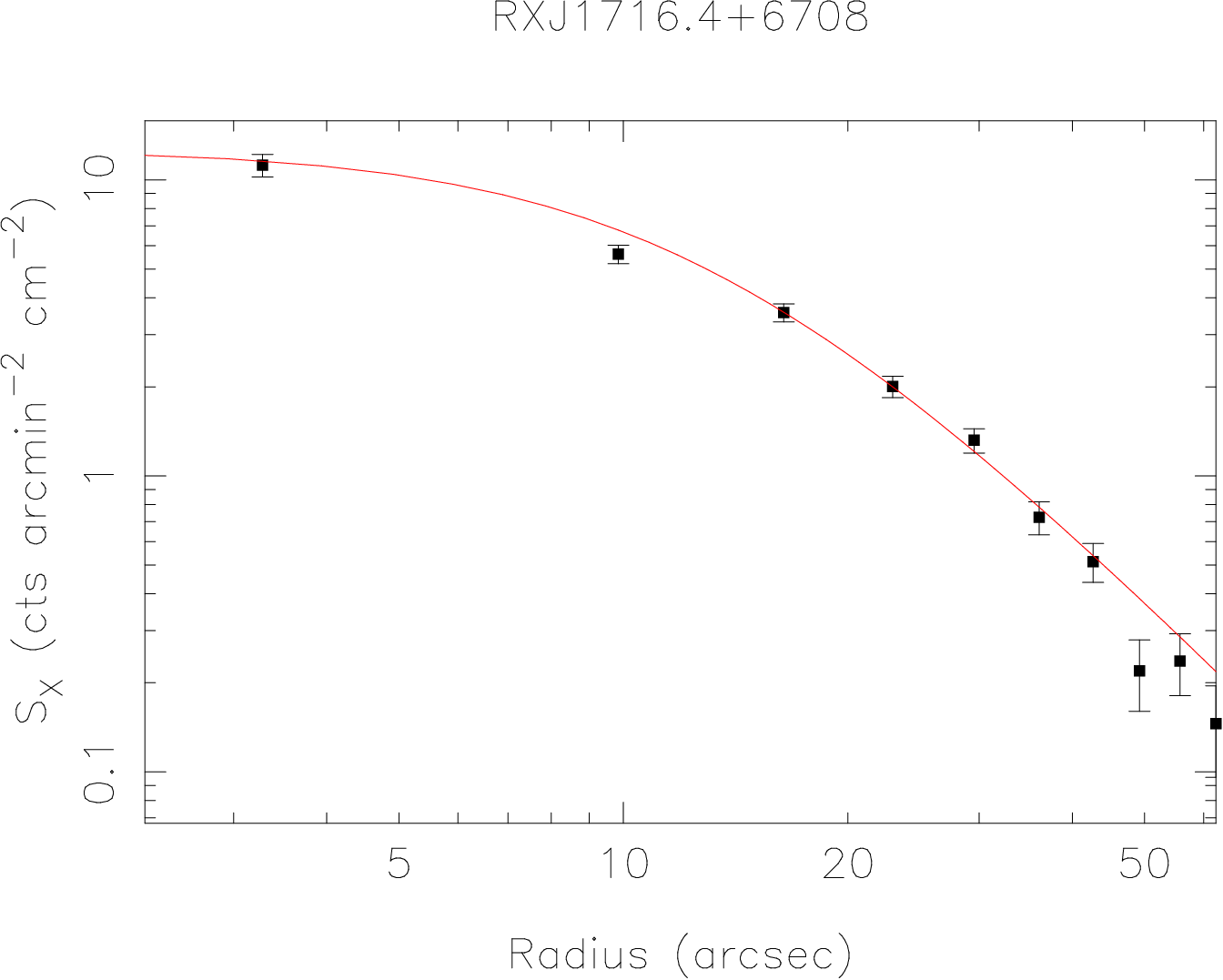}
\includegraphics[angle=-90,width=1.75in]{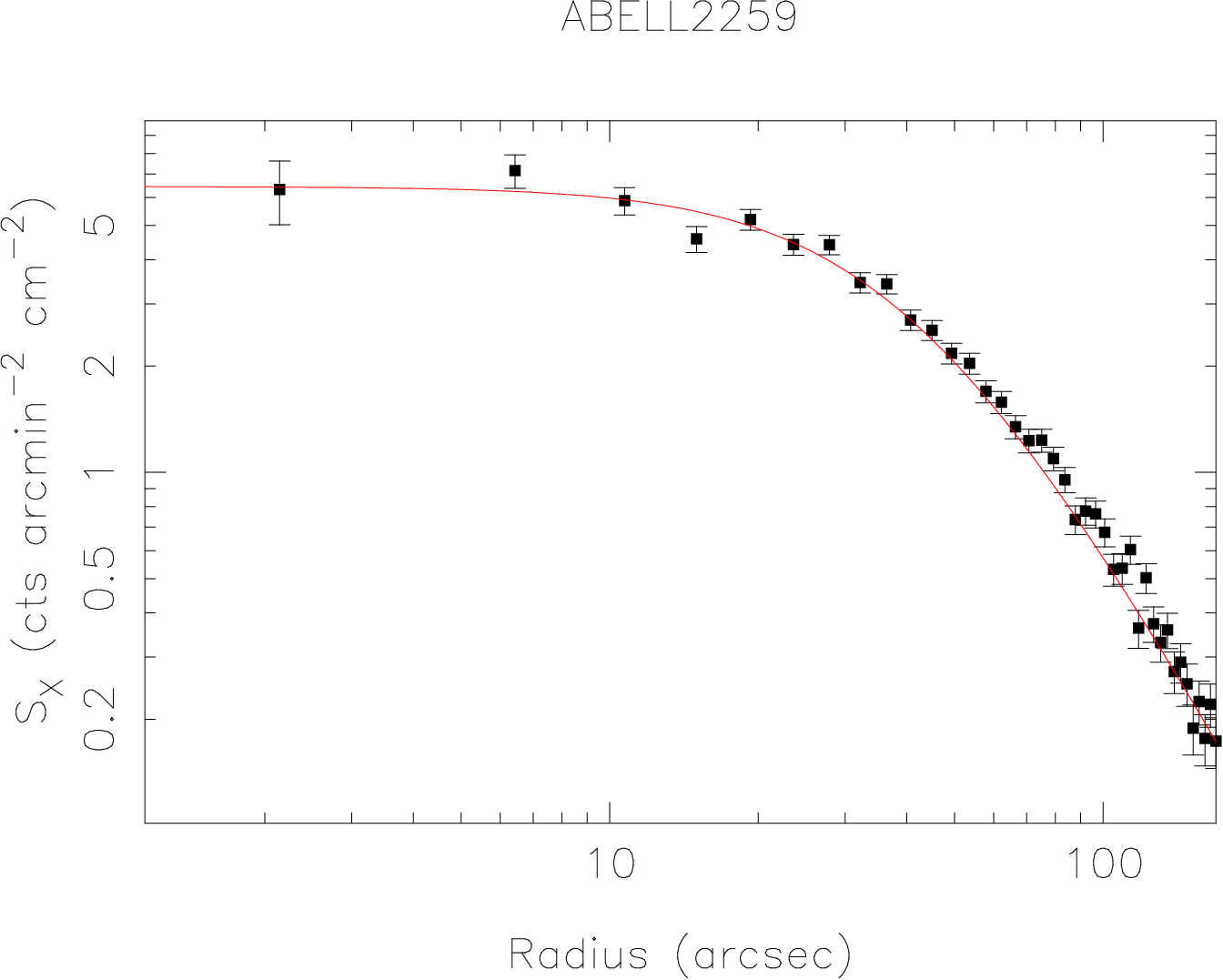}
\includegraphics[angle=-90,width=1.75in]{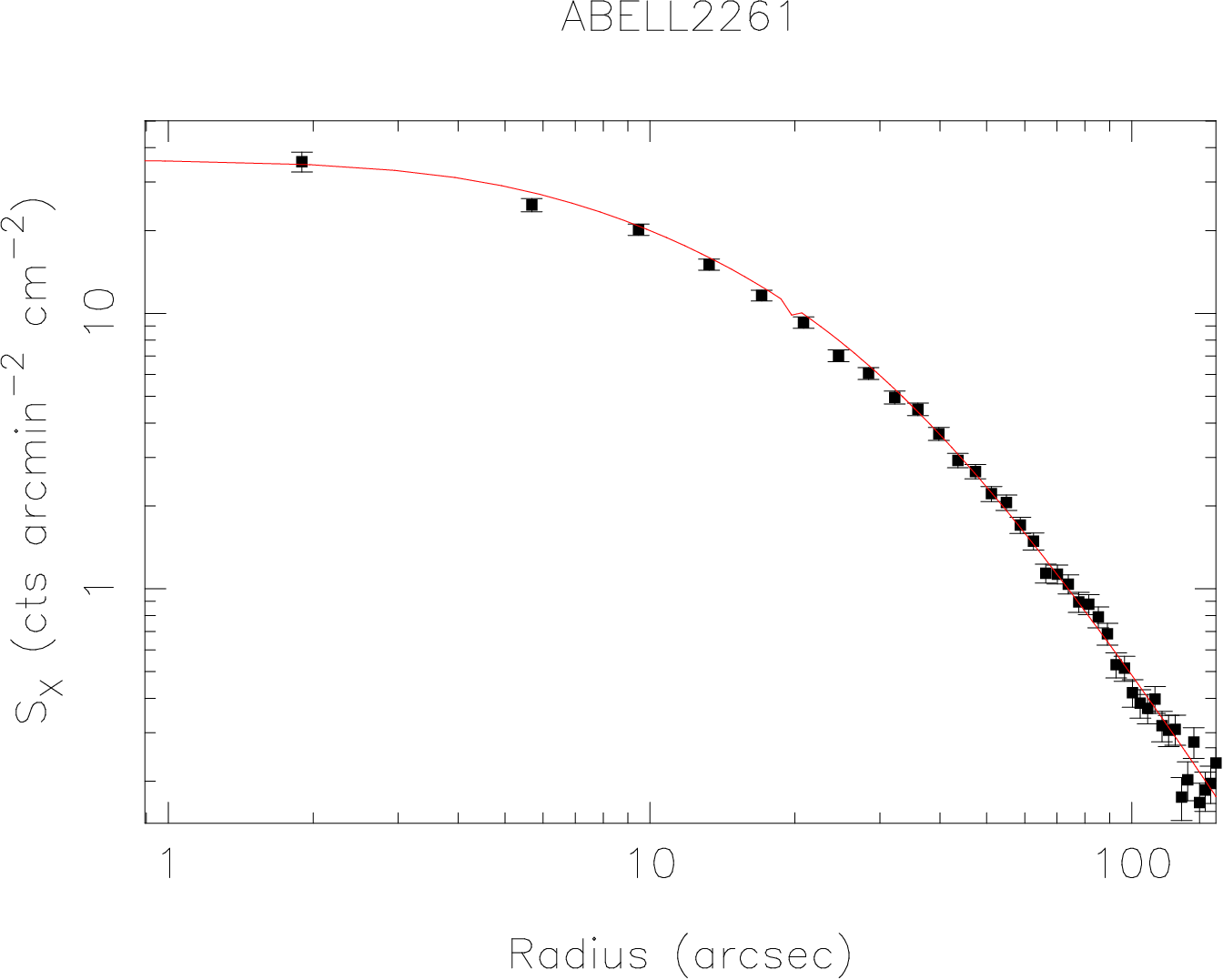}
\includegraphics[angle=-90,width=1.75in]{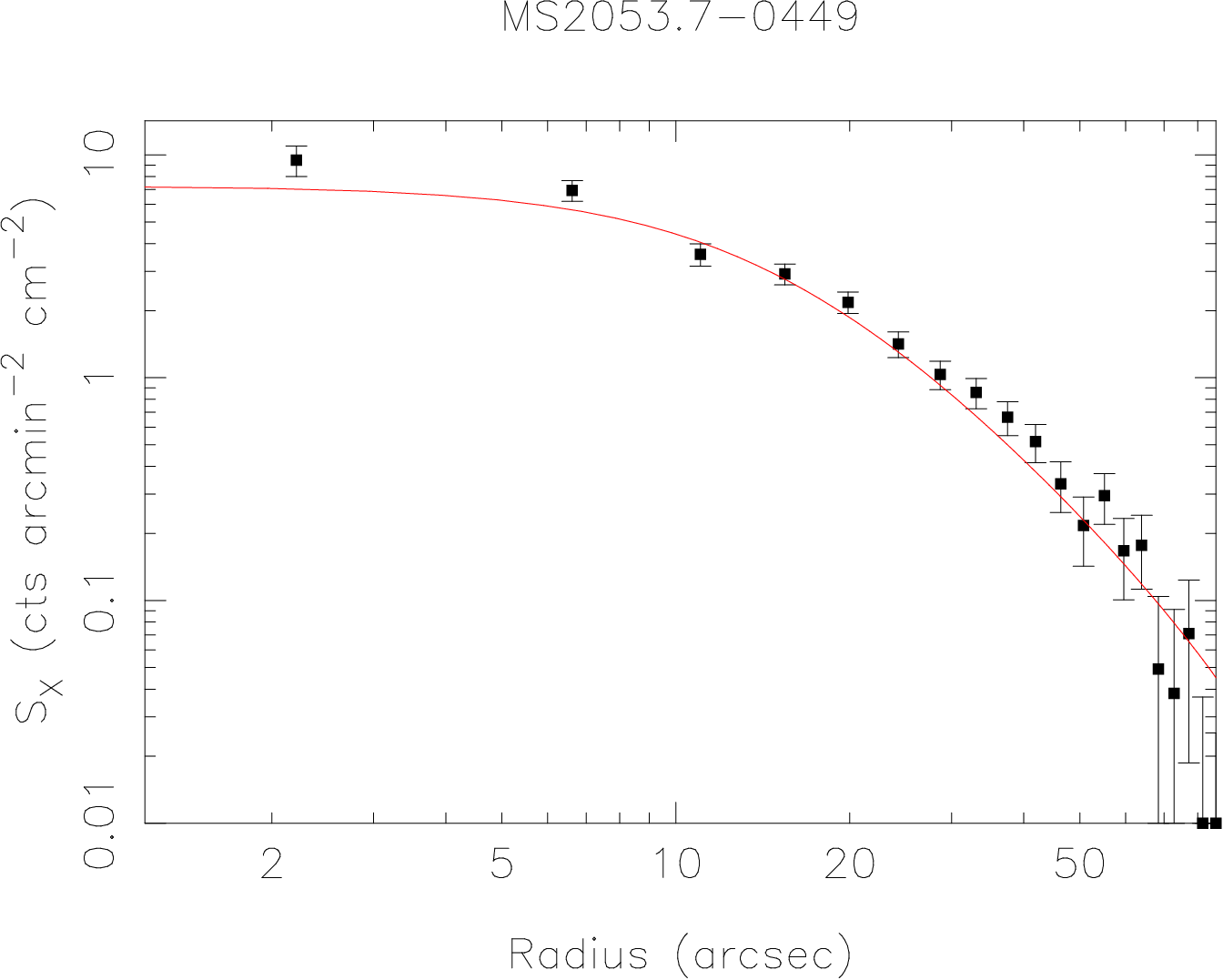}
\includegraphics[angle=-90,width=1.75in]{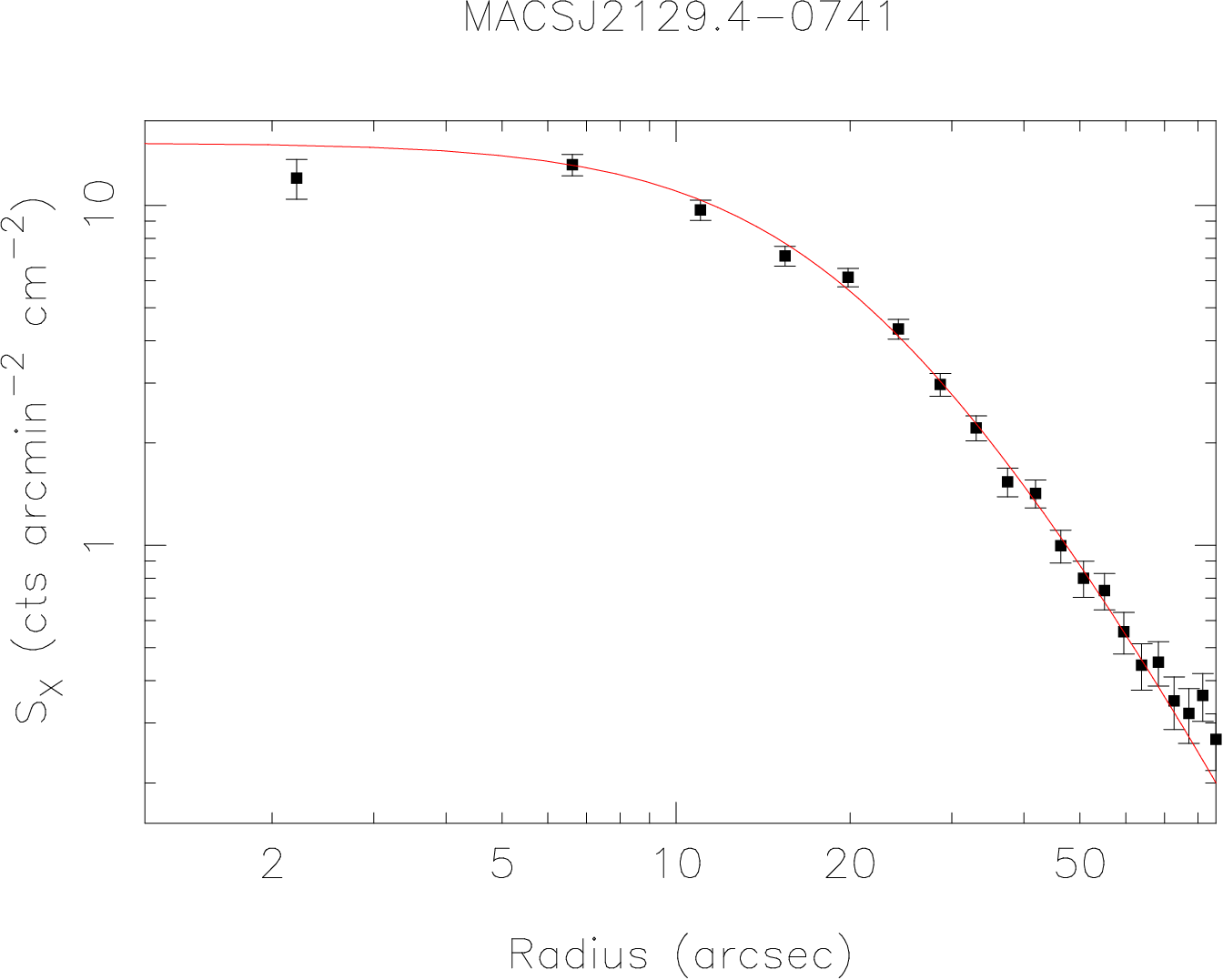}
\includegraphics[angle=-90,width=1.75in]{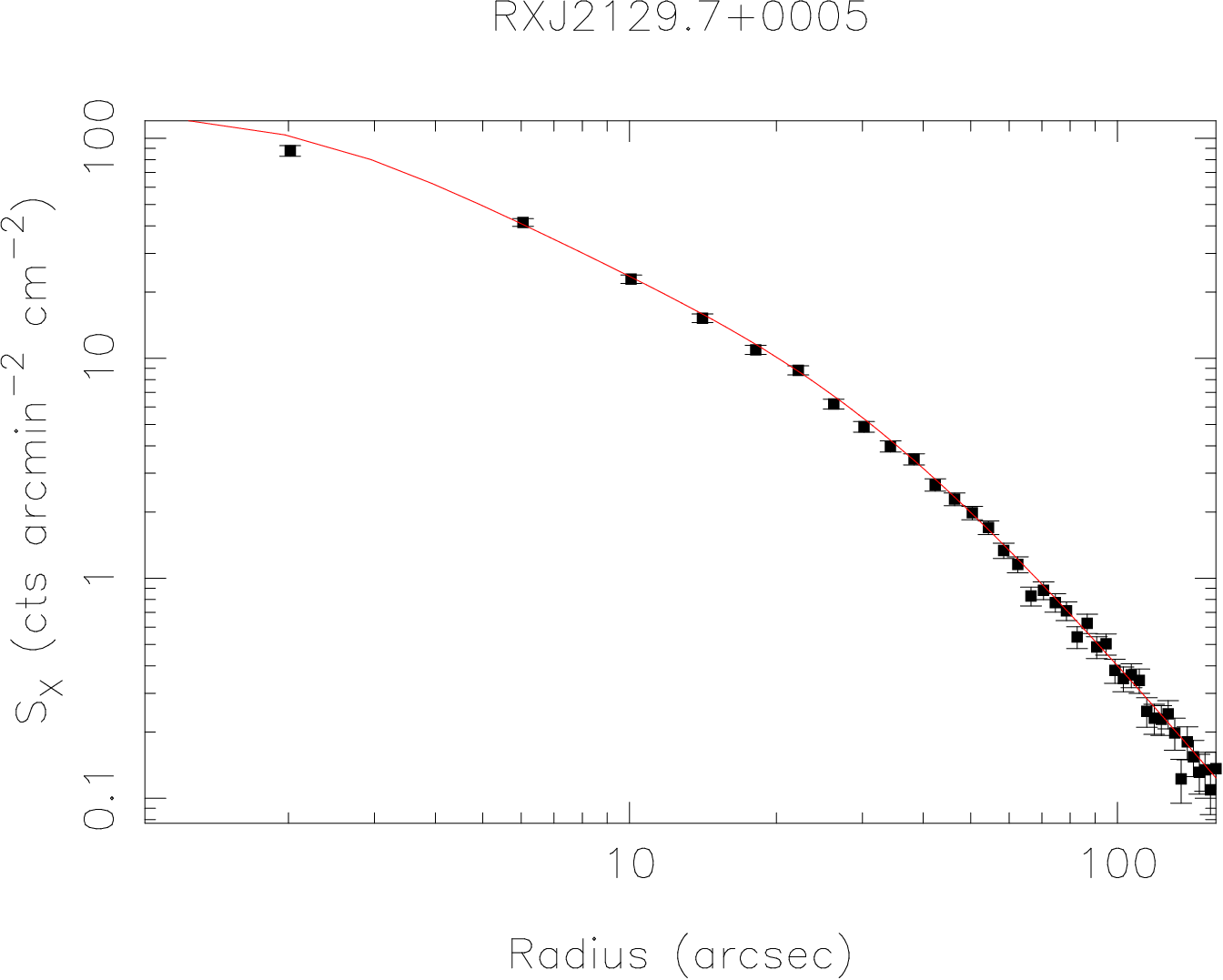}
\hspace{1.2cm}
\includegraphics[angle=-90,width=1.75in]{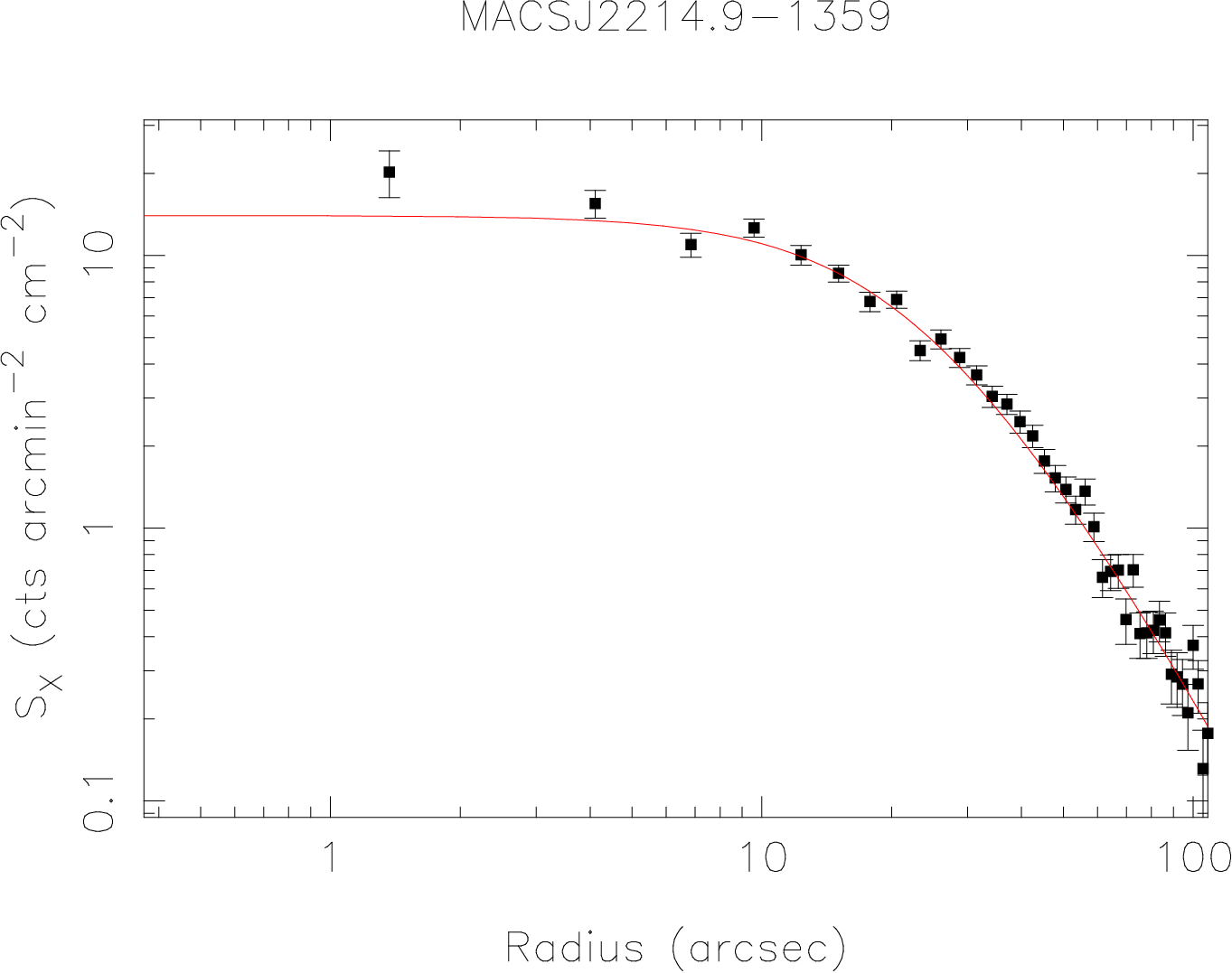}
\hspace{1.2cm}
\includegraphics[angle=-90,width=1.75in]{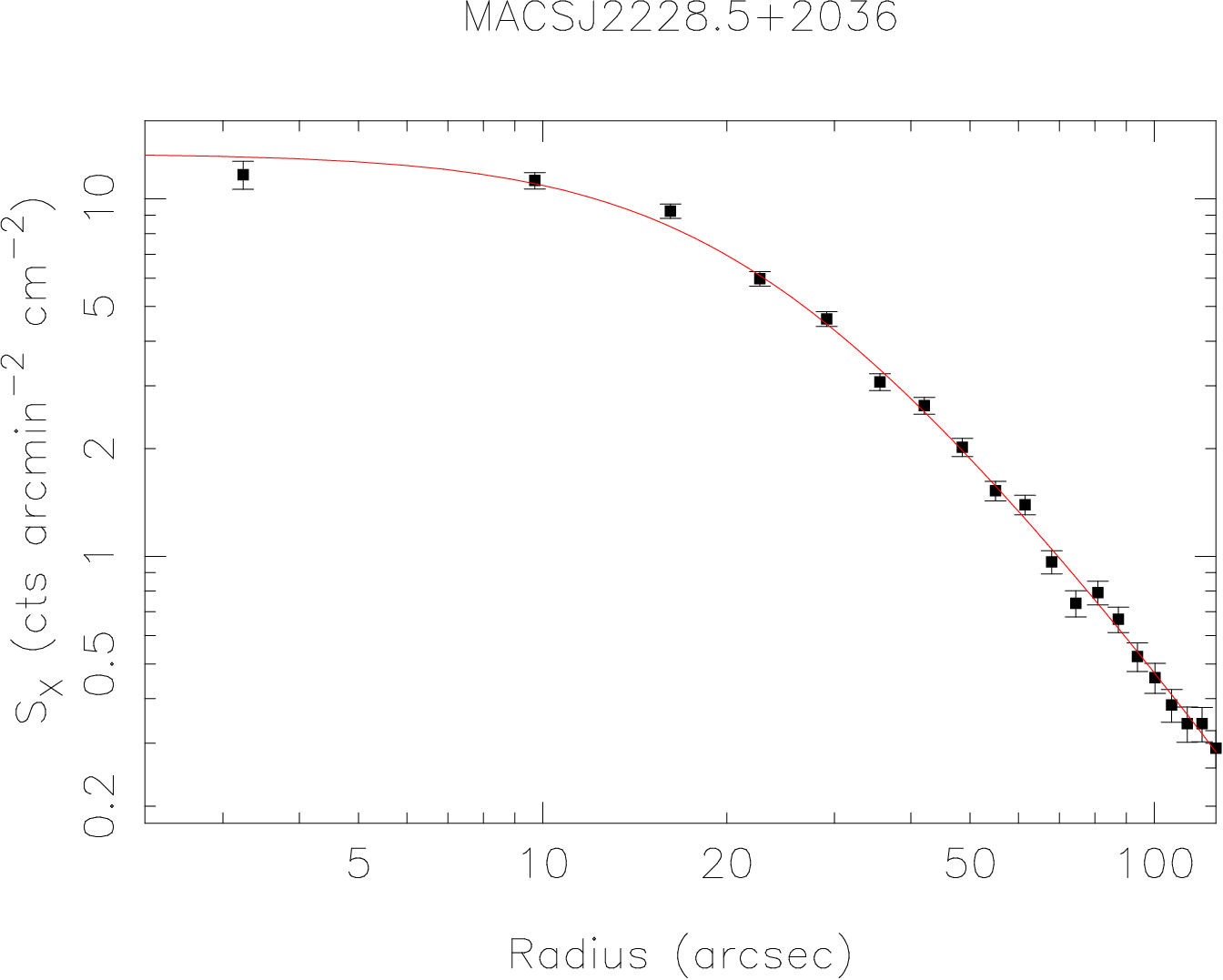}
\end{figure}

\vspace{-1cm}
\begin{deluxetable}{lclclclc}
\tabletypesize{\scriptsize}
\tablecaption{Reduced $\chi^2$ values of the hydrostatic equilibrium model fits to 
X-ray surface brightness profiles \label{table_sbprofiles}}
\tablehead{Cluster & $\chi^2_r$ & Cluster & $\chi^2_r$ & Cluster & $\chi^2_r$  & Cluster & $\chi^2_r$}
\startdata
CL0016+1609 & 1.38 & Abell~697 & 1.12 & RX~J1347.5-1145 & 18.24 &  RX~J1716.4+6708 & 0.97 \\
Abell~68 & 0.90    & Abell~773 & 0.98 & MS~1358.4+6245 & 2.52 & Abell~2259 & 1.47 \\
Abell~267 & 0.74   & ZW~3146 & 2.43 & Abell~1835 & 3.31 & Abell~2261 & 1.06 \\
Abell~370 & 1.44   & MS~1054-0321 & 1.40 & MACS~J1423+2404 & 17.60 & MS~2053.7-0449 & 0.80\\
MS~0451.6-0305 & 1.28 & MS~1137.5+6625 & 0.71 & Abell~1914 & 15.29 & MACS~J2129.4-0741 & 1.24 \\
MACS~J0647.7+7015& 1.07 & MACS~J1149.5+2223 & 1.14 & Abell~1995 & 1.23 & RX~J2129.7+0005 & 1.15 \\
Abell~586 & 1.30   & Abell~1413 & 1.47 & Abell~2111 & 1.39 & MACS~J2214.9-1359 & 1.38\\
MACS~J0744.8+3927 & 1.18 &CL~J1226.9+3332 & 1.34 & Abell~2163 & 41.62 & MACS~J2228.5+2036 & 1.11\\
Abell~611 & 1.40 & MACS~J1311.0-0310 & 1.47 & Abell~2204 & 3.04 & & \\
Abell~665 & 11.84 & Abell~1689 & 1.25 & Abell~2218 & 1.22 & \\
\enddata
\end{deluxetable}

\clearpage

\section*{Appendix 3: Temperature profiles}
\begin{center}
\end{center}
\begin{deluxetable}{cccc}
\tablecolumns{4}
\tablewidth{0pc}
\tablecaption{Temperature profiles \label{table_tprof}}
\tablehead{ \colhead{r (arcsec)} & \colhead{Counts} & \colhead{kT (keV)} & \colhead{$\chi^2$ (dof)}} 
\startdata
\multicolumn{4}{c}{CL~0016+1609 -- Obs. 520}\\ 
0-17 & 2198.3 & 10.2 $\pm$ 2.2 & 73.5 (68) \\ 
17-24 & 1773.6 & 9.3 $\pm$ 2.2 & 64.1 (58)\\ 
24-32 & 1671.8 & 9.8 $\pm$ 2.3 & 68.7 (54)\\ 
32-42 & 2224.6 & 10.1 $\pm$ 2.7 & 86.4 (71)\\ 
42-52 & 2070.9 & 8.8 $\pm$ 1.7 & 84.9 (66)\\ 
52-65 & 1946.2 & 9.1 $\pm$ 2.0 & 64.6 (67)\\ 
65-83 & 1934.0 & 11.1 $\pm$ 3.0 & 64.6 (70)\\ 
83-101 & 1355.0 & 10.1 $\pm$ 3.1 & 39.6 (58)\\ 
\multicolumn{4}{c}{ }\\ 
\multicolumn{4}{c}{Abell~0068 -- Obs. 3250}\\ 
0-49 & 2036.9 &10.8 $\pm$ 2.8 & 70.5 (67)\\ 
49-111 & 1910.0 & 7.5 $\pm$ 1.8 & 67.0 (65) \\ 
\multicolumn{4}{c}{ }\\ 
\multicolumn{4}{c}{Abell~0267 -- Obs. 1448}\\ 
0-53 & 1998.4 & 6.0 $\pm$ 0.9 & 71.1 (60)\\ 
53-166 & 1658.0 & 5.5 $\pm$ 1.2 & 50.6 (65)  \\ 
\multicolumn{4}{c}{ }\\ 
\multicolumn{4}{c}{Abell~0370 -- Obs. 515}\\ 
0-13 & 1071.9 & 9.1 $\pm$ 3.1 & 23.4 (37) \\ 
13-21 & 1434.9 & 13.3 $\pm$ 5.2 & 38.6 (47)  \\ 
21-28 & 1654.3 & 8.5 $\pm$ 2.0 & 67.8 (55)\\ 
28-36 & 1842.9 & 9.1 $\pm$ 2.3  & 59.4 (59)\\ 
36-40 & 965.5 & 8.6 $\pm$ 2.6 & 33.6 (34)\\ 
40-47 & 1876.2 & 13.1 $\pm$ 4.5 & 75.3 (63) \\ 
47-55 & 1718.9 & 10.1 $\pm$ 3.2 & 72.1 (60)\\ 
55-62 & 1591.6 & 7.1 $\pm$ 1.7 & 41.3 (57)\\ 
62-70 & 1381.4 & 8.5 $\pm$ 2.7 & 58.0 (53)\\ 
70-81 & 1856.6 & 7.3 $\pm$ 1.9 & 63.2 (76)\\ 
81-89 & 1004.7 & 6.1 $\pm$ 2.3 & 62.3 (46)\\ 
89-107 & 1942.2 & 6.8 $\pm$ 1.7 & 76.0 (93)\\ 
107-134 & 1880.2 & 5.4 $\pm$ 1.5 & 129.8 (116) \\ 
\multicolumn{4}{c}{ }\\ 
\multicolumn{4}{c}{MS~0451.6-0305 -- Obs. 529}\\ 
0-44 & 2002.3 & 7.9 $\pm$ 1.5 & 63.3 (62)\\ 
44-119 & 1317.6 & 6.8 $\pm$ 2.0 & 51.2 (57)\\ 
\multicolumn{4}{c}{ }\\ 
\multicolumn{4}{c}{MS~0451.6-0305 -- Obs. 902}\\ 
0-14 & 1922.4 & 14.3 $\pm$ 4.6 & 41.2 (59)\\ 
14-23 & 2226.5 & 9.3 $\pm$ 1.9 & 77.5 (67)\\ 
23-29 & 1476.1 & 8.3 $\pm$ 1.8 & 34.1 (49)\\ 
29-38 & 1985.2 & 9.6 $\pm$ 2.0 & 73.7 (64)\\ 
38-50 & 2171.4 & 10.3 $\pm$ 2.4 & 72.8 (71)\\ 
50-68 & 2123.6 & 9.1 $\pm$ 2.2 & 60.7 (73)\\ 
68-101 & 2008.7 & 10.8 $\pm$ 4.7 & 116.3 (87)\\ 
\multicolumn{4}{c}{ }\\ 
\multicolumn{4}{c}{MACS~J0647.7+7015 -- Obs. 3196}\\ 
0-37 & 1973.8 & 12.4 $\pm$ 3.1 & 56.6 (65) \\ 
37-128 & 1323.0 & 10.5 $\pm$ 4.5 & 66.3 (65) \\ 
\multicolumn{4}{c}{ }\\ 
\multicolumn{4}{c}{MACS~J0647.7+7015 -- Obs. 3584}\\ 
0-34 & 1885.6 & 10.7 $\pm$ 2.4 & 57.3 (60) \\ 
34-160 & 1790.8 & 19.1 $\pm$ 10.8 & 64.2 (83)\\ 
\multicolumn{4}{c}{ }\\ 
\multicolumn{4}{c}{Abell~0586 -- Obs. 530}\\ 
0-24 & 1909.4 & 7.8 $\pm$ 1.5 & 47.2 (60) \\ 
24-43 & 1925.4 & 7.7 $\pm$ 1.6 & 75.2 (62)\\ 
43-70 & 1939.5 & 6.9 $\pm$ 1.2 & 58.8 (63)\\ 
70-130 & 2062.0 & 6.4 $\pm$ 1.2 & 74.3 (73)\\ 
\multicolumn{4}{c}{ }\\ 
\multicolumn{4}{c}{MACS~J0744.8+3927 -- Obs. 3197}\\ 
0-36 & 1948.1 & 8.47 $\pm$ 1.42 & 64.6 (62) \\ 
36-125 & 1108.9 & 9.15 $\pm$ 4.76 & 59.8 (57)\\ 
\multicolumn{4}{c}{ }\\ 
\multicolumn{4}{c}{MACS~J0744.8+3927 -- Obs. 3585}\\ 
0-41 & 2019.9  & 8.1 $\pm$ 1.3 & 96.2  (65) \\ 
41-60 & 359.2  & 6.0 $\pm$ 2.3 & 18.2 (13)\\ 
\multicolumn{4}{c}{ }\\ 
\multicolumn{4}{c}{Abell~0611 -- Obs. 3194}\\ 
0-11 & 1916.2 &  6.0 $\pm$ 0.8 & 73.9 (60)  \\ 
11-19 & 1268.4 & 7.3 $\pm$ 1.3 & 87.1 (79)\\ 
19-24 & 1351.4 & 6.0 $\pm$ 1.2 & 44.3 (44)\\ 
24-32 & 2341.6 & 6.7 $\pm$ 1.3 & 96.8 (74)\\ 
32-41 & 1850.1 & 6.0 $\pm$ 1.1 & 67.8 (57)\\ 
41-49 & 1487.9 & 7.1 $\pm$ 1.4 & 52.6 (51)\\ 
49-66 & 2279.9 & 5.9 $\pm$ 1.0 & 91.4 (78)\\ 
66-88 & 1857.0 & 6.0 $\pm$ 1.1 & 49.1 (71)\\ 
88-126 & 1982.7 & 6.8 $\pm$ 2.1 & 121.5 (102)\\ 
\multicolumn{4}{c}{ }\\ 
\multicolumn{4}{c}{Abell~0665 -- Obs. 3586}\\ 
0-19 & 2240.2 &  7.8 $\pm$ 1.4 & 72.3 (69) \\ 
19-26 & 1442.0 & 9.2 $\pm$ 2.7 & 42.1 (49)\\ 
26-37 & 2277.0 & 8.5 $\pm$ 1.7 & 94.1 (70)\\ 
37-47 & 2140.0 & 7.8 $\pm$ 1.4 & 68.2 (67)\\ 
47-58 & 2085.3 & 7.7 $\pm$ 1.5 & 53.3 (66)\\ 
58-69 & 1958.1 & 9.0 $\pm$ 2.3 & 80.0 (64)\\ 
69-79 & 1950.6 & 9.2 $\pm$ 2.3 & 66.5 (66)\\ 
79-90 & 1838.9 &7.4 $\pm$ 1.7 & 44.5 (61)\\ 
90-100 & 1825.5 & 7.3 $\pm$ 1.6 & 48.7 (61) \\ 
100-111 & 1727.1 & 9.7 $\pm$ 2.9 & 63.4 (59)\\ 
111-125 & 2149.2 & 8.3 $\pm$ 1.8 & 88.8 (74)\\ 
125-139 & 1989.6 & 8.7 $\pm$ 2.2 & 78.0 (68) \\ 
\multicolumn{4}{c}{ }\\ 
\multicolumn{4}{c}{Abell~0665 -- Obs. 531}\\ 
0-37 & 2029.8 & 6.7 $\pm$ 1.1 & 58.7 (62)\\ 
37-67 & 1904.4 & 8.8 $\pm$ 2.2 & 60.3 (60)\\ 
67-103 & 2006.3 & 11.9 $\pm$ 3.6 & 65.8 (66)\\ 
\multicolumn{4}{c}{ }\\ 
\multicolumn{4}{c}{Abell~0697 -- Obs. 4217}\\ 
0-25 & 2176.0 & 10.1 $\pm$ 2.3 & 67.7 (67) \\ 
25-39 & 2031.6 & 10.2 $\pm$ 2.6 & 76.6 (63)\\ 
39-48 & 1397.5 & 9.1 $\pm$ 2.6 & 44.3 (46)\\ 
48-67 & 2213.1 & 11.2 $\pm$ 2.9 & 74.1 (70)\\ 
67-90 & 2117.5 & 10.8 $\pm$ 2.8 & 61.4 (70)\\ 
90-117 & 2029.7 & 12.6 $\pm$ 4.0 & 66.6 (73)\\ 
117-154 & 1914.1 & 9.9 $\pm$ 2.8 & 81.8 (72)\\ 
\multicolumn{4}{c}{ }\\ 
\multicolumn{4}{c}{Abell~0773 -- Obs. 3588}\\ 
0-41 & 1930.5 & 6.0 $\pm$ 1.0 & 75.2 (61) \\ 
41-80 & 1934.3 & 6.2 $\pm$ 1.2 & 89.4 (63)\\ 
80-157 & 2066.8 & 6.4 $\pm$ 1.3 & 72.3 (73)\\ 
\multicolumn{4}{c}{ }\\ 
\multicolumn{4}{c}{Abell~0773 -- Obs. 533}\\ 
0-36 & 1990.2 & 7.2 $\pm$ 1.4 & 66.9 (61)  \\ 
36-62 & 1900.2 & 8.6 $\pm$ 2.1 & 61.3 (61)\\ 
62-106 & 2019.5 & 7.1 $\pm$ 1.3 & 67.0 (67)\\ 
\multicolumn{4}{c}{ }\\ 
\multicolumn{4}{c}{ZW~3146 -- Obs. 909}\\ 
0-5 & 3491.4 & 3.9 $\pm$ 0.2 & 133 .6 (130) \\ 
5-11 & 7557.8 &  4.9 $\pm$ 0.2 & 213.8 (154) \\ 
11-18 & 8887.2 & 6.5 $\pm$ 0.3 & 218.0 (176)\\ 
18-25 & 6361.8 & 7.0 $\pm$ 0.3 & 180.7 (150) \\ 
25-35 & 6237.6 & 8.4 $\pm$ 0.6 & 158.3 (150) \\ 
35-46 & 5598.2 & 8.2 $\pm$ 0.6 & 109.3 (140)\\ 
46-60 & 4756.4 & 9.1 $\pm$ 0.6 & 153.1 (134)\\ 
60-76 & 3401.7 & 9.7 $\pm$ 1.2 & 83.5 (109) \\ 
76-101 &3252.7 &  8.7 $\pm$ 0.9 & 91.8 (115)\\ 
101-140 & 2238.8 & 7.7 $\pm$ 1.2 & 101.7 (89) \\ 
\multicolumn{4}{c}{ }\\ 
\multicolumn{4}{c}{MS~1054-0321 -- Obs. 512}\\ 
0-14 & 740.3 & 13.8 $\pm$ 6.3 & 27.7 (26)\\ 
14-22 & 937.0 & 7.7 $\pm$ 2.6 & 26.1 (33)\\ 
22-35 & 1755.5 & 9.6 $\pm$ 2.1 & 66.7 (62)\\ 
35-50 & 1718.1 & 13.2 $\pm$ 4.7 & 65.1 (68)\\ 
50-83 & 1751.9 & 8.7 $\pm$ 3.5 & 131.7 (109)\\ 
\multicolumn{4}{c}{ }\\ 
\multicolumn{4}{c}{MS~1137.5+6625 -- Obs. 536}\\ 
0-25 & 1585.5 & 6.4 $\pm$ 1.2 & 56.1 (52) \\ 
25-75 & 897.7 & 4.6 $\pm$ 1.4 & 43.5 (55)\\ 
\multicolumn{4}{c}{ }\\ 
\multicolumn{4}{c}{MACS~J1149.5+2223 -- Obs. 1656}\\ 
0-48 & 1924.1 & 9.30 $\pm$ 1.97 & 68.3 (63) \\ 
48-108 & 1702.6 &  7.71 $\pm$ 1.90 & 50.4 (65)\\ 
\multicolumn{4}{c}{ }\\ 
\multicolumn{4}{c}{MACS~J1149.5+2223 -- Obs. 3589}\\ 
0-46 & 1985.5 & 10.8 $\pm$ 2.5 & 64.67 (66)\\ 
46-112 & 1911.5 & 7.0 $\pm$ 1.6 & 57.6 (71)\\ 
\multicolumn{4}{c}{ }\\ 
\multicolumn{4}{c}{Abell~1413 -- Obs. 1661}\\ 
0-20 & 2190.8 &  6.0 $\pm$ 0.9 & 68.1 (67)\\ 
20-34 & 2040.8 & 9.0 $\pm$ 2.0 & 63.6 (64)\\ 
34-47 & 2005.8 & 8.3 $\pm$ 1.5 & 57.0 (64)\\ 
47-61 & 1640.5 & 9.2 $\pm$ 2.4 & 55.7 (54)\\ 
61-79 & 1841.1 & 7.4 $\pm$ 1.6 & 57.5 (60)\\ 
79-106 & 2156.7 & 6.0 $\pm$ 1.0 & 78.4 (68)\\ 
106-133 & 1789.5 & 6.2 $\pm$ 1.0 & 61.7 (59)\\ 
133-178 & 1982.4 & 7.4 $\pm$ 1.7 & 76.6 (74)\\ 
\multicolumn{4}{c}{ }\\ 
\multicolumn{4}{c}{Abell~1413 -- Obs. 537}\\ 
0-18 & 1908.9 &  7.1 $\pm$ 1.5 & 55.5 (59) \\ 
18-31 & 1973.0 & 8.2 $\pm$ 1.8 & 62.3 (62)\\ 
31-46 & 2129.7 & 6.2 $\pm$ 1.0 & 79.1 (63)\\ 
46-62 & 1823.0 & 7.0 $\pm$ 1.3 & 49.6 (58)\\ 
62-81 & 2045.6 & 9.2 $\pm$ 2.3 & 85.3 (65)\\ 
\multicolumn{4}{c}{ }\\ 
\multicolumn{4}{c}{CL~J1226.9+3332 -- Obs. 3180}\\ 
0-82 & 1962.2 & 12.1 $\pm$ 2.6 & 58.8 (63) \\ 
\multicolumn{4}{c}{ }\\ 
\multicolumn{4}{c}{CL~J1226.9+3332 -- Obs. 932}\\ 
0-58 &  1003.9 & 15.4 $\pm$ 7.5 & 26.9 (35) \\ 
\multicolumn{4}{c}{ }\\ 
\multicolumn{4}{c}{MACS~J1311.0-0310 -- Obs. 3258}\\ 
0-73 & 1963.2 & 6.7 $\pm$ 1.1 & 76.3 (65)\\ 
\multicolumn{4}{c}{Abell~1689 -- Obs. 1663}\\ 
0-12 & 2136.8 & 9.4 $\pm$ 2.1 & 76.3 (67) \\ 
12-17 & 1409.5 & 9.5 $\pm$ 2.9 & 47.0 (45)\\ 
17-26 & 2928.5 & 9.47 $\pm$ 1.81 & 94.7 (87)\\ 
26-31 & 1271.1 & 16.57 $\pm$ 9.01 & 41.0 (42)\\ 
31-41 & 2159.3 & 9.41 $\pm$ 2.09 & 65.9 (67)\\ 
41-50 & 1922.3 & 9.77 $\pm$ 2.69 & 72.0 (63)\\ 
50-60 & 1687.4 & 8.72 $\pm$ 2.18 & 59.7 (54)\\ 
60-74 & 2150.8 & 9.78 $\pm$ 2.28 & 75.1 (68)\\ 
74-93 & 2244.1 & 14.10 $\pm$ 4.95 & 81.6 (72)\\ 
93-117 & 2012.4 & 9.22 $\pm$ 2.14 & 59.6 (68)\\ 
117-146 & 1759.2 & 10.55 $\pm$ 3.27 & 56.7 (61)\\ 
146-189 & 1532.9 & 6.53 $\pm$ 1.57 & 51.2 (59)\\ 
\multicolumn{4}{c}{ }\\ 
\multicolumn{4}{c}{Abell~1689 -- Obs. 540}\\ 
0-10 & 1601.5 & 11.9 $\pm$ 3.8 & 64.8 (51) \\ 
10-18 & 2431.2 & 7.8 $\pm$ 1.3 & 97.7 (71)\\ 
18-24 & 1706.8 & 8.0 $\pm$ 1.6 & 47.5 (53)\\ 
24-32 & 2236.1 & 13.1 $\pm$ 3.8 & 59.1 (68)\\ 
32-41 & 1798.6 & 13.4 $\pm$ 5.0 & 57.5 (59)\\ 
41-52 & 2178.3 & 9.7 $\pm$ 2.1 & 82.2 (68)\\ 
52-63 & 1890.4 & 11.0 $\pm$ 3.0 & 66.6 (60)\\ 
63-78 & 2035.9 & 14.7 $\pm$ 5.2 & 59.5 (67)\\ 
78-94 & 1898.7 & 11.7 $\pm$ 4.9 & 61.4 (60)\\ 
94-111 & 1371.4 & 12.7 $\pm$ 4.7 & 53.0 (46)\\ 
\multicolumn{4}{c}{ }\\ 
\multicolumn{4}{c}{RX~J1347.5-1145 -- Obs. 3592}\\ 
0-2 & 2062.7 & 6.3 $\pm$ 0.9 & 95.9 (103) \\ 
2-5 & 7479.7 & 8.5 $\pm$ 0.7 & 208.3 (166)\\ 
5-9 & 6539.0 & 11.2 $\pm$ 1.3 & 192.2 (159)\\ 
9-12 & 5672.1 & 16.3 $\pm$ 3.1 & 199.6 (151)\\ 
12-16 & 4939.2 & 14.8 $\pm$ 2.9 & 166.5 (137)\\ 
16-19 & 4392.6 & 16.8 $\pm$ 3.4 & 159.7 (126)\\ 
19-23 & 4010.0 & 14.6 $\pm$ 3.0 & 106.2 (122)\\ 
23-29 & 5869.5 & 13.5 $\pm$ 2.0 & 182.8 (154)\\ 
29-36 & 4081.0 & 18.4 $\pm$ 4.2 & 115.7 (124)\\ 
36-47 & 4143.3 & 19.1 $\pm$ 4.3 & 158.9 (128)\\ 
47-68 & 5440.7 & 14.1 $\pm$ 2.5 & 171.7 (154)\\ 
68-102 & 4970.9 & 12.8 $\pm$ 2.5 & 161.3 (154)\\ 
\multicolumn{4}{c}{ }\\ 
\multicolumn{4}{c}{MS~1358.4+6245 -- Obs. 516}\\ 
0-7 & 2112.3 & 4.02 $\pm$ 0.38 & 61.4 (63) \\ 
7-13 & 1629.5 & 5.92 $\pm$ 1.18 & 39.3 (50)\\ 
13-21 & 2237.5 & 8.36 $\pm$ 1.71 & 65.9 (71)\\ 
21-29 & 2038.2 & 8.8 $\pm$ 1.8 & 63.4 (66)\\ 
29-41 & 2143.1 & 9.2 $\pm$ 2.2 & 96.1 (68)\\ 
41-52 & 1928.4 & 8.6 $\pm$ 2.1 & 66.2 (66)\\ 
52-66 & 1726.7 & 7.5 $\pm$ 2.0 & 50.0 (62)\\ 
66-88 & 1959.2 & 8.1 $\pm$ 2.3 & 94.3 (82)\\ 
88-111 & 1203.9 & 12.6 $\pm$ 9.9 & 46.7 (67)\\ 
\multicolumn{4}{c}{ }\\ 
\multicolumn{4}{c}{Abell~1835 -- Obs. 495}\\ 
0-5 & 4283.1 & 4.3 $\pm$ 0.4 & 121.3 (101) \\ 
5-9 & 4989.8 & 5.0 $\pm$ 0.4 & 117.7 (114)\\ 
9-12 & 4033.2 & 6.3 $\pm$ 0.7 & 117.6 (104)\\ 
12-19 & 6163.1 & 9.1 $\pm$ 1.1 & 142.9 (140)\\ 
19-26 & 4943.0 & 8.3 $\pm$ 1.1 & 143.5 (121)\\ 
26-36 & 5573.5 & 9.1 $\pm$ 1.2 & 150.2 (133)\\ 
36-46 & 4466.6 & 12.6 $\pm$ 2.4 & 146.1 (125)\\ 
46-64 & 5255.3 & 9.2 $\pm$ 1.3 & 116.4 (129)\\ 
64-91 & 5233.8 & 11.6 $\pm$ 2.2 & 162.7 (142)\\ 
901-136 & 4718.5 & 18.4 $\pm$ 7.2 & 135.3 (151)\\ 
\multicolumn{4}{c}{ }\\ 
\multicolumn{4}{c}{Abell~1835 -- Obs. 496}\\ 
0-11 & 5201.4 & 5.0 $\pm$ 0.4 & 95.6 (117)\\ 
11-20 & 3868.4 & 7.0 $\pm$ 0.9 & 111.6 (103)\\ 
20-42 & 6416.7 & 8.2 $\pm$ 0.9 & 173.9 (142)\\ 
42-68 & 4488.9 & 11.2 $\pm$ 2.5 & 110.7 (120)\\ 
68-138 & 4982.1 & 13.4 $\pm$ 4.0 & 139.0 (143) \\ 
\multicolumn{4}{c}{ }\\ 
\multicolumn{4}{c}{MACS~J1423.8+2404 -- Obs. 4195}\\ 
0-3 & 2988.9 & 4.2 $\pm$ 0.4 & 97.4 (87)\\ 
3-8 & 7992.4 & 4.8 $\pm$ 0.3 & 199.7 (146)\\ 
8-13 & 4588.5 & 6.6 $\pm$ 0.6 & 106.6 (118)\\ 
13-18 & 2869.9 & 7.4 $\pm$ 1.0 & 66.1 (87)\\ 
18-23 & 2169.4 & 7.4 $\pm$ 1.2 & 72.9 (66)\\ 
23-28 & 1731.4 & 7.2 $\pm$ 1.5 & 47.6 (54)\\ 
28-32 & 1459.8 & 8.7 $\pm$ 2.2 & 52.8 (49)\\ 
32-38 & 1140.1 & 6.1 $\pm$ 1.4 & 44.2 (41)\\ 
38-42 & 943.7 & 7.0 $\pm$ 2.2 & 28.9 (37)\\ 
42-47 & 739.6  & 6.9 $\pm$ 3.3 & 34.1 (29)\\ 
47-52 & 657.5 & 5.9 $\pm$ 2.1 & 19.5 (28)\\ 
52-57 & 553.7  & 7.4 $\pm$ 3.9 & 15.9 (25)\\ 
57-62 & 484.9 & 6.8 $\pm$ 4.2 & 14.5 (24)\\ 
62-67 & 477.5 & 7.0 $\pm$ 6.5 & 34.0 (24)\\ 
67-77 & 806.5 & 5.4$\pm$ 2.0 & 38.3 (47)\\ 
77-97 & 1075.9 & 4.6$\pm$ 1.4 & 73.0 (77)\\ 
\multicolumn{4}{c}{ }\\ 
\multicolumn{4}{c}{Abell~1914 -- Obs. 3593}\\ 
0-24 & 4677.0 &  13.1 $\pm$ 2.5 & 125.6 (138) \\ 
24-37 & 5716.9 & 11.3 $\pm$ 1.7 & 184.5 (154)\\ 
37-46 & 4416.5 & 11.0 $\pm$ 1.9 & 117.5 (130)\\ 
46-55 & 4180.6 & 11.4 $\pm$ 2.3 & 117.6 (126)\\ 
55-72 & 6022.7 & 8.5 $\pm$ 1.1  & 150.1 (152)\\ 
72-94 & 4535.4 & 11.5 $\pm$ 2.3 & 167.2 (132)\\ 
94-138 & 5294.2 &  8.9 $\pm$ 1.3 & 144.8 (154)\\ 
138-173 & 2636.7 & 8.2 $\pm$ 1.8 & 100.7 (93)\\ 
\multicolumn{4}{c}{ }\\ 
\multicolumn{4}{c}{Abell~1914 -- Obs. 542}\\ 
0-25 & 2276.3 & 10.8 $\pm$ 2.7 & 62.5 (69)\\ 
25-32 & 1345.1 & 13.4 $\pm$ 6.7 & 43.2 (44)\\ 
32-44 & 2278.0 & 10.8 $\pm$ 2.8 & 55.7 (69)\\ 
44-51 & 1613.1 & 16.7 $\pm$ 6.8 & 47.9 (53)\\ 
51-63 & 2160.2 & 13.6 $\pm$ 4.1 & 71.4 (66)\\ 
63-78 & 2050.4 & 10.9 $\pm$ 3.2 & 65.2 (64)\\ 
78-104 & 2186.0 & 8.8 $\pm$ 2.2 & 71.7 (69)\\ 
104-142 & 1849.5 & 8.4 $\pm$ 2.4 & 61.2 (66)\\ 
\multicolumn{4}{c}{ }\\ 
\multicolumn{4}{c}{Abell~1995 -- Obs. 906}\\ 
0-13 & 2060.1 & 7.8 $\pm$ 1.4 & 51.9 (64)\\ 
13-17 & 1120.6 & 10.2 $\pm$ 4.0 & 38.3 (37)\\ 
17-24 & 2658.1 & 7.4 $\pm$ 1.3 & 72.9 (78)\\ 
24-28 & 1461.6 & 9.3 $\pm$ 2.8 & 45.5 (47)\\ 
28-36 & 2715.0 & 9.9 $\pm$ 2.0 & 60.4 (84)\\ 
36-39 & 1293.1 & 13.0 $\pm$ 4.9 & 34.5 (43)\\ 
39-47 & 2449.4 & 9.1 $\pm$ 2.2 & 98.2 (75)\\ 
47-54 & 2228.6 & 10.0 $\pm$ 2.7 & 86.8 (73)\\ 
54-62 & 1814.6 & 7.1 $\pm$ 1.3 & 68.4 (61)\\ 
62-69 & 1557.0 & 7.2 $\pm$ 1.6 & 55.2 (55)\\ 
69-84 & 2209.6 & 8.2 $\pm$ 2.0 & 83.9 (85)\\ 
84-107 & 1992.8 & 11.1 $\pm$ 4.6 & 92.1 (92) \\ 
107-148 & 1805.4 & 8.4 $\pm$ 4.7 & 149.8 (131) \\ 
\multicolumn{4}{c}{ }\\ 
\multicolumn{4}{c}{Abell~2111 -- Obs. 544}\\ 
0-70 & 2114.6 & 7.5 $\pm$ 1.50 &  75.2 (66)\\ 
70-163 & 1932.3 & 9.3 $\pm$ 3.0 & 87.6 (70)\\ 
\multicolumn{4}{c}{ }\\ 
\multicolumn{4}{c}{Abell~2163 -- Obs. 1653}\\ 
0-23 & 8821.2 & 18.07 $\pm$ 4.57 & 210.6 (201)\\ 
23-33 & 9098.7 &  11.62 $\pm$ 2.20 &  243.6 (199)\\ 
33-43 & 11446.3 & 19.36 $\pm$ 4.33 & 256.4 (228)\\ 
43-53 & 12904.6 & 12.71 $\pm$ 1.89 & 260.6 (243)\\ 
53-58 & 6900.5 & 11.4 $\pm$ 2.3 & 173.0 (174) \\ 
58-68 & 13317.5 & 13.6 $\pm$ 1.9 & 231.4 (246)\\ 
68-73 & 6111.6 & 11.7 $\pm$ 2.1 & 177.8 (164)\\ 
73-83 & 11752.2 & 15.0 $\pm$ 2.8 & 249.0 (235)\\ 
83-93 & 10457.4 & 13.1 $\pm$ 2.2 & 231.6 (221)\\ 
93-103 &9532.6 &  16.3 $\pm$ 3.7 & 213.3 (212)\\ 
103-113 &8423.0 &  16.5 $\pm$ 4.0 & 205.4 (202)\\ 
113-129 & 10962.5 & 18.5 $\pm$ 4.5 & 257.2 (233)\\ 
129-144 & 8863.9 & 14.6 $\pm$ 3.1 & 191.3 (207)\\ 
144-164 & 10300.9 & 17.1 $\pm$ 4.7 & 231.2 (231) \\ 
164-184 & 9774.5 & 14.5 $\pm$ 3.0 & 275.3 (232)\\ 
184-199 & 6843.6  & 18.1 $\pm$ 6.1 & 173.2 (195)\\ 
\multicolumn{4}{c}{ }\\ 
\multicolumn{4}{c}{Abell~2204 -- Obs. 499}\\ 
0-7 & 4961.3 & 3.6 $\pm$ 0.2 & 105.5 (114)\\ 
7-11 & 4238.5 & 4.3 $\pm$ 0.3 &115.8 (108) \\ 
11-19 & 5979.5 & 5.7 $\pm$ 0.5 & 166.5 (136)\\ 
19-28 & 3962.0 & 8.1 $\pm$ 1.2 & 110.2 (111)\\ 
28-45 & 6013.9 & 9.4 $\pm$ 1.3 & 145.1 (144)\\ 
45-62 & 4447.9 & 9.4 $\pm$ 1.6 & 118.5 (122)\\ 
62-93 & 5059.2 & 10.8 $\pm$ 2.0 & 104.3 (137)\\ 
93-153 & 5317.6 & 13.4 $\pm$ 4.0 & 148.0 (151)\\ 
\multicolumn{4}{c}{ }\\ 
\multicolumn{4}{c}{Abell~2204 -- Obs. 6104}\\ 
0-6 & 3805.5 & 3.3 $\pm$ 0.2 & 125.1 (105) \\ 
6-15 & 5643.9 & 5.4 $\pm$ 0.4 & 196.7 (142)\\ 
15-28 & 5274.9 & 7.3 $\pm$ 0.8 & 192.0 (138)\\ 
28-45 & 4774.2 & 11.1 $\pm$ 1.9 & 130.7 (136)\\ 
45-75 & 5517.0 & 9.3 $\pm$ 1.2 & 142.9 (151)\\ 
\multicolumn{4}{c}{ }\\ 
\multicolumn{4}{c}{Abell~2218 -- Obs. 1666}\\ 
0-30 & 5103.0 & 9.2 $\pm$ 1.2 & 135.4 (142) \\ 
30-46 & 4969.6 &  8.9 $\pm$ 1.3 & 128.5 (141)\\ 
46-63 & 4936.2 & 8.0 $\pm$ 1.1 & 145.5 (140)\\ 
63-79 & 4515.1 & 8.3 $\pm$ 1.2 & 144.7 (136)\\ 
79-107 & 5582.4 & 6.5 $\pm$ 0.7 & 151.1 (153)\\ 
\multicolumn{4}{c}{ }\\ 
\multicolumn{4}{c}{RX~J1716.4+6708 -- Obs. 584}\\ 
0-62 & 1428.3 & 7.1 $\pm$ 1.6 & 54.2 (57) \\ 
\multicolumn{4}{c}{ }\\ 
\multicolumn{4}{c}{Abell~2259 -- Obs. 3245}\\ 
0-41 & 1933.3 & 5.3 $\pm$ 0.7 & 82.4 (62) \\ 
41-79 & 2113.1 & 5.0 $\pm$ 0.7 & 67.0 (65)\\ 
79-135 &1830.5 &  5.4 $\pm$ 0.9 & 83.3 (63)\\ 
\multicolumn{4}{c}{Abell~2261 -- Obs. 550}\\ 
0-17 & 1801.4 &  6.8 $\pm$ 1.4 & 73.1 (57) \\ 
17-36 & 2244.5 & 7.4 $\pm$ 1.4 & 94.4 (67)\\ 
36-59 & 1874.8 & 7.0 $\pm$ 1.5 & 50.0 (59)\\ 
59-100 & 1957.3 & 9.3 $\pm$ 2.2 & 66.9 (67)\\ 
100-150 & 1140.8 & 5.8 $\pm$ 1.6 & 38.2 (47)\\ 
\multicolumn{4}{c}{ }\\ 
\multicolumn{4}{c}{MS~2053.7-0449 -- Obs. 1667}\\ 
0-86 & 1095.1 & 3.8 $\pm$ 0.9 & 36.0 (57) \\ 
\multicolumn{4}{c}{ }\\ 
\multicolumn{4}{c}{MS~2053.7-0449 -- Obs. 551}\\ 
0-117 & 1188.4 &  6.7 $\pm$ 2.8 & 87.4 (79) \\ 
\multicolumn{4}{c}{ }\\ 
\multicolumn{4}{c}{MACS~J2129.4-0741 -- Obs. 3199}\\ 
0-46 & 2012.3 & 7.1 $\pm$ 1.1 & 63.4 (66) \\ 
46-90 & 651.2 & 5.5 $\pm$ 1.7 & 38.7 (30)\\ 
\multicolumn{4}{c}{ }\\ 
\multicolumn{4}{c}{MACS~J2129.4-0741 -- Obs. 3595}\\ 
0-42 & 1974.1 & 12.3 $\pm$ 2.9 & 55.6 (65) \\ 
42-90 & 875.8 & 8.7 $\pm$ 3.7 & 41.2 (37)\\ 
\multicolumn{4}{c}{ }\\ 
\multicolumn{4}{c}{RX~J2129.7+0005 -- Obs. 552}\\ 
0-14 & 1852.5 & 4.4 $\pm$ 0.5 & 57.2 (57) \\ 
14-34 & 2160.4 & 6.1 $\pm$ 0.9 & 65.9 (65) \\ 
34-67 & 1931.2 & 6.4 $\pm$ 1.0 & 75.4 (64)\\ 
67-159 & 1970.1 &  7.4 $\pm$ 1.6 & 85.2 (73)\\ 
\multicolumn{4}{c}{ }\\ 
\multicolumn{4}{c}{MACS~J2214.9-1359 -- Obs. 3259}\\ 
0-32 & 1882.6 & 9.5 $\pm$ 2.4 & 64.2 (60) \\ 
32-95 & 2099.3 &15.1 $\pm$ 5.0 & 82.1 (76)\\ 
\multicolumn{4}{c}{ }\\ 
\multicolumn{4}{c}{MACS~J2214.9-1359 -- Obs. 5011}\\ 
0-40 & 2030.9 & 10.6 $\pm$ 2.8 & 75.4 (64)\\ 
40-119 & 1691.7 & 8.6 $\pm$ 2.6 & 50.0 (67)\\ 
\multicolumn{4}{c}{ }\\ 
\multicolumn{4}{c}{MACS~J2228.5+2036 -- Obs. 3285}\\ 
0-29 & 1835.3 & 8.4 $\pm$ 1.7 & 58.5 (59) \\ 
29-68 & 2140.5 & 8.9$\pm$ 1.9 & 53.3 (68)\\ 
68-126 & 1605.8 & 10.5 $\pm$ 3.6 & 57.8 (67)\\ 
\enddata
\end{deluxetable}
\end{document}